\documentclass[11pt,a4paper]{article}
\usepackage{amsmath,amssymb,bm,bbm}
\usepackage[utf8]{inputenc}
\usepackage{graphicx}
\usepackage{color}
\usepackage[dvipsnames]{xcolor}
\numberwithin{equation}{section}
\allowdisplaybreaks[3]
\usepackage{braket}
\usepackage{subcaption}
\usepackage[linktocpage=true]{hyperref}
\hypersetup{
    colorlinks=true,
    linkcolor=blue,
    citecolor=blue,
    pdfpagemode=FullScreen,
    }

\edef\restoreparindent{\parindent=\the\parindent\relax}
\usepackage{parskip}
\restoreparindent

\usepackage{tikz}
\usetikzlibrary{arrows,arrows.meta,intersections, calc,positioning,decorations.pathreplacing,decorations.pathmorphing,shapes}
\usetikzlibrary{patterns}
\usetikzlibrary{decorations.markings}
\usetikzlibrary{knots}

\def\CF{{\cal F}}

\let\Re\relax
\DeclareMathOperator{\Re}{Re}
\let\Im\relax 
\DeclareMathOperator{\Im}{Im}
\DeclareMathOperator{\Res}{Res}

\begin{document}

\begin{titlepage}

\renewcommand{\thefootnote}{\fnsymbol{footnote}}
\begin{flushright}
\begin{tabular}{l}
YITP-23-64\\
\end{tabular}
\end{flushright}

\vfill
\begin{center}

\noindent{\Large \textbf{Complex saddles of Chern-Simons gravity}}

\medskip

\noindent{\Large \textbf{and dS$_3$/CFT$_2$ correspondence}}

\vspace{1.5cm}

\noindent{Heng-Yu Chen,$^{a}$\footnote{heng.yu.chen@phys.ntu.edu.tw} Yasuaki Hikida,$^b$\footnote{yhikida@yukawa.kyoto-u.ac.jp} Yusuke Taki$^b$\footnote{yusuke.taki@yukawa.kyoto-u.ac.jp} and Takahiro Uetoko$^c$\footnote{uetoko@kushiro-ct.ac.jp}}

\bigskip

\vskip .6 truecm

\centerline{\it $^a$Department of Physics, National Taiwan University, Taipei 10617, Taiwan }



\medskip

\centerline{\it $^b$Center for Gravitational Physics and Quantum Information, Yukawa Institute for Theoretical Physics, }
\centerline{\it Kyoto University, Kitashirakawa Oiwakecho, Sakyo-ku, Kyoto 606-8502, Japan}

\medskip

\centerline{\it $^c$Science in General Education, National Institute of Technology, Kushiro College,} 
\centerline{\it Otanoshike-Nishi 2-32-1, Kushiro, Hokkaido 084-0916, Japan}
\end{center}

\vfill
\vskip 0.5 truecm
\begin{abstract}

We examine the black hole solutions of dS$_3$ gravity by applying the explicit dS$_3$/CFT$_2$ correspondence.
The gravity theory is described by Chern-Simons theory with complex gauge group SL$(2,\mathbb{C})$, and the complexified theory is known to have too many saddle points. We determine the set of ``allowable geometry'' from dual CFT correlators. Concretely, we classify the possible complex solutions corresponding to dS$_3$ black holes from Liouville two-point functions. We extend the analysis to Liouville multi-point functions and among others we study geometry corresponding to two linked Wilson loops on $S^3$ by the monodromy matrix of Liouville four-point function. Some parts of the results were presented in a previous letter but here they are explained in more details and extended in various ways. In particular, we generalize the results to the case with higher-spin gravity by focusing the effects of higher-spin charges.

\end{abstract}
\vfill
\vskip 0.5 truecm

\setcounter{footnote}{0}
\renewcommand{\thefootnote}{\arabic{footnote}}
\end{titlepage}

\newpage

\hrule
\tableofcontents

\bigskip
\hrule
\bigskip

\section{Introduction}

According to the no-boundary proposal by Hartle and Hawking \cite{Hartle:1983ai}, the universe begins from nothing. More precisely speaking, the universe starts from Euclidean hemi-sphere and continues to Lorentzian de Sitter space-time. Similarly, the temperature of black hole can be read off from the periodicity of Euclidean time, where the smoothness condition is assigned at the horizon \cite{Hartle:1976tp}.  It is often argued whether these complex solutions {to Einstein equation} are purely mathematical objects or physically meaningful ones. If those unphysical ones were included when we integrate over complex metrics in the path integral of a gravity theory, then there would be too many saddle points, and the sum over all saddle points should not lead to a sensible answer. Recently, Witten investigated in \cite{Witten:2021nzp} a criterion of ``allowable complex geometry'' based on previous works \cite{Louko:1995jw,Kontsevich:2021dmb}. In this paper, we propose {an alternative} way to select the physical saddle points of quantum gravity by making use of a holography. Applying the recently proposed dS$_3$/CFT$_2$ correspondence in \cite{Hikida:2021ese,Hikida:2022ltr,Chen:2022ozy,Chen:2022xse}, we examine explicit examples of three-dimensional de Sitter (dS$_3$) gravity from dual two-dimensional conformal field theory (CFT$_2$).%
\footnote{See, e.g. \cite{Basile:2023ycy} for a related work on three-dimensional gravity with negative cosmological constant.}
We have already presented some parts of results in a previous letter \cite{Chen:2023prz}, and we will explain the derivation of the results in details and extend the analysis in this paper.

A canonical example of complex geometry is given by the no-boundary proposal for wave functional of universe mentioned above.
We may consider a complexified $(d+1)$-dimensional sphere $(S^{d+1})$ with the metric
\begin{align}
    ds^2 = \ell^2 \left[\left(\frac{d \theta (u)}{du} \right)^2 du^2 + \cos ^2 \theta (u) d \Omega_d^2 \right] \, ,
\end{align}
where $\ell$ is a length scale and $d \Omega_d^2$ is the metric of $S^d$.
Moreover, $\theta(u)$ is a complex function of a complex coordinate $u$. If $\theta (u) = u$ and $0 \leq u \leq \pi$, then the metric is that of Euclidean $S^{d+1}$. On the other hand, if  $\theta (u) = i u$ and $- \infty  \leq u \leq \infty$, then the universe is Lorentzian dS$_3$. 
Here we only complexify $u$-direction and not the $S^d$ part for simplicity.
We may assign that the universe starts from nothing, that is 
\begin{align}
\theta = \left(n + \frac12 \right) \pi, \quad n \in {\mathbb{Z}}
\label{thetau}
\end{align}
at the beginning, say, $u=0$ and ends as $\theta = i u$ at $u \to \infty$.
In this way, we consider a family of complex geometry labeled by an integer $n$. The criterion of $D$-dimensional allowable geometry in \cite{Kontsevich:2021dmb,Witten:2021nzp} is that the kinetic terms of $p$-form fields for all $p=0,1,\dots,D$ should have positive real parts. 
In our example, allowable geometry is instead only given by $n=-1,0$, which is nothing but the geometry considered by Hartle and Hawking in \cite{Hartle:1983ai}.

We would like to explicitly perform the path-integral of a quantum gravity and determine the set of saddle points we should take. It is a quite difficult task since we do not know how to formulate quantum gravity in general. In this paper, we study three-dimensional gravity theory with positive cosmological constant. The theory has a Chern-Simons description \cite{Witten:1988hc,Witten:1989ip,Witten:2010cx},%
\footnote{We summarize some properties of Chern-Simons gauge theory with complex gauge fields, in particular, the relation to gravity theory with positive/negative cosmological constant in appendix \ref{app:cs}.} moreover it also has an explicit holographic dual formulation as detailed in \cite{Hikida:2021ese,Hikida:2022ltr,Chen:2022ozy,Chen:2022xse}, thus we could study this theory in great depth. 
In dS$_3$, there are solutions to Einstein equation including the conical defect geometries examined in \cite{DESER1984405}. They are often called as dS$_3$ black hole solutions given by:
\begin{align}
ds^2 = \ell^2 \left[ - (r^2_+ - r^2) dt^2 + \frac{1}{r_+^2 - r^2} dr^2 + r^2 d \phi ^2 \right] \, . \label{dSBHmetric}
\end{align}
This geometry can be regarded as the dS$_3$ analog of BTZ black hole \cite{Banados:1992wn}. 
The parameter $r_+$ is related to the Newton constant $G_N$ and the black hole energy $E$ as
\begin{align}
r_+ = \sqrt{1 - 8 G_N E} \, . 
\end{align}
There is a horizon at $r = r_+$ and the Gibbons-Hawking entropy associated with the horizon is \cite{Bekenstein:1973ur,Hawking:1975vcx,Gibbons:1977mu,Gibbons:1976ue}
\begin{align}
S_\text{GH} = \frac{2 \pi \ell r_+}{4 G_N} =  \frac{ \pi \ell \sqrt{1 - 8 G_N E} }{ 2 G_N}\, . \label{BHentropy}
\end{align}
In terms of Chern-Simons gravity, we can construct a configuration of gauge fields corresponding to the dS$_3$ black hole geometry. Applying a large gauge transformation with winding number $n$ to the gauge fields, we can generate another configuration of gauge fields labeled by $n$. This $n$ turns out to be essentially the same as the one introduced in \eqref{thetau} for the complexified sphere. 
In the Chern-Simons gauge theory with complex gauge group, {large gauge transformations are known to generate new physically inequivalent configurations}, see, e.g., \cite{Witten:2010cx}. From the viewpoint of gravity theory, we do not have any criteria to determine which gauge configurations we should take a priori. We shall attack this problem for various black hole solutions by making use of the explicit dS$_3$/CFT$_2$ correspondence. We also examine its higher-spin generalizations and their properties focusing the effects of higher-spin charges.

We shall use the explicit dS$_3$/CFT$_2$ correspondence developed in \cite{Hikida:2021ese,Hikida:2022ltr,Chen:2022ozy,Chen:2022xse}, which may be regarded a lower dimensional version of dS$_4$/CFT$_3$ correspondence in \cite{Anninos:2011ui}. These are concrete examples of dS/CFT correspondence proposed in \cite{Witten:2001kn,Strominger:2001pn,Maldacena:2002vr}, see also \cite{Maldacena:1998ih,Park:1998qk,Park:1998yw}.
The duality of \cite{Anninos:2011ui} can be obtained as an ``analytic continuation'' of Klebanov-Polyakov duality involving higher-spin AdS$_4$ gravity \cite{Klebanov:2002ja}. Similarly, the duality used here is constructed by an ``analytic continuation'' of Gaberdiel-Gopakumar duality involving higher-spin AdS$_3$ gravity \cite{Gaberdiel:2010pz}. The gravity theory is given by Prokushkin-Vasiliev theory \cite{Prokushkin:1998bq} with gauge fields with higher-spin $s=2,3,\dots$ and two complex scalar fields with the mass parameterized as: 
\begin{align}
\ell^2_\text{AdS} m^2 = -1 + \lambda^2 \, . \label{AdSmass}
\end{align}
Here we denote the AdS radius by $\ell_\text{AdS}$ and set $0 \leq \lambda \leq 1$. The dual CFT is supposed to be W$_N$-minimal model described by a coset
\begin{align}
    \frac{\text{SU}(N)_k \otimes \text{SU}(N)_1}{\text{SU}(N)_{k+1}} \label{coset}
\end{align}
with the central charge
\begin{align}
    c = (N - 1) \left( 1 - \frac{N (N +1)}{(N+k) (N + k +1)} \right) \, .
\end{align}
We take the so-called 't Hooft limit, where $N,k$ is taken to be large with keeping
\begin{align}
    \lambda = \frac{N}{N +k} \label{thooft}
\end{align}
finite. In particular, the 't Hooft parameter $\lambda$ is identified with the mass parameter $\lambda$ appeared in \eqref{AdSmass}. A version of dS$_3$/CFT$_2$ correspondence can be constructed as follows \cite{Chen:2022ozy,Chen:2022xse}, see also \cite{Ouyang:2011fs}. For the higher-spin gravity, we just need to flip the sign of cosmological constant, which can be done by replacing $\ell_\text{AdS}$ by $i \ell$ with the dS radius $\ell$. In the CFT side, we perform an analytic continuation of parameters such that the central charge becomes $c = i c^{(g)}$ with $c^{(g)} \in \mathbb{R}$ while $\lambda$ is kept unchanged.

The (A)dS$_3$/CFT$_2$ correspondence introduced above involves an infinite tower of higher-spin fields, which makes analysis complicated. In three-dimension, a higher-spin gravity with gauge fields with truncated spin $s=2,3,\ldots,N$ can be constructed by Chern-Simons theory with gauge group SL$(N,\mathbb{C})$. We largely study the simplest case with $N=2$, which is supposed to be equivalent to the pure gravity theory \cite{Achucarro:1986uwr,Witten:1988hc}. The holography involving the Chern-Simons gravity with the finite dimensional group can be constructed with the help of triality relation of the higher-spin algebra \cite{Castro:2011iw,Gaberdiel:2012ku,Perlmutter:2012ds}. The dual CFT is given by the coset \eqref{coset} with a finite $N$ but a peculiar value of $k$ as
\begin{align}
    k = -1 - N + \frac{N (N^2 -1)}{c} + \mathcal{O}(c^{-2}) \, .
\end{align}
Here $c \in \mathbb{R}$ for the dual of AdS$_3$ and $c = i c^{(g)}$ with $c^{(g)} \in \mathbb{R}$ for the dual of dS$_3$, see \cite{Hikida:2021ese,Hikida:2022ltr}. For generic value of $k$ with finite $N$, it was shown in \cite{Creutzig:2021ykz} that the correlation functions of the coset model \eqref{coset}
are the same as those of $\mathfrak{sl}(N)$ Toda field theory (Liouville field theory for $N=2$).
We heavily use this version of (A)dS$_3$/CFT$_2$ correspondence since the Chern-Simons gravity with the finite dimensional gauge group is much more tractable than the Prokushkin-Vasiliev theory. Moreover, the dual CFT is quite well studied as in \cite{Dorn:1994xn,Zamolodchikov:1995aa,Witten:2010cx} for Liouville field theory and as in \cite{Fateev:2007ab} for $\mathfrak{sl}(N)$ Toda field theory.%
\footnote{It is not known how to couple matters to the Chern-Simons gravity, so the first version will be utilized when examining the propagation of matter fields on dS$_3$ black holes, see \cite{Castro:2023dxp,Castro:2023bvo} for recent related works.} 
The coset \eqref{coset} with a finite $N$ but a generic value of $k$ has the $W_N$-algebra symmetry with generators of spin $s=2,3,\dots,N$. There are degenerate representations of the algebra, which are labeled by two Young diagrams with no upper limit of the number of boxes, see, e.g. \cite{Prochazka:2015deb}. They are proposed to be dual to the bound states of scalar fields on a conical defect geometry \cite{Castro:2011iw,Gaberdiel:2012ku,Perlmutter:2012ds}. In terms of Liouville/Toda field theory, we examine correlation functions of so-called maximally degenerate operators. In this paper, we consider black hole solutions created due to the back reactions of heavy particles, which corresponds to the insertions of heavy operators in the dual CFT.

In dS/CFT correspondence, it is not so straightforward to compute bulk quantities from dual CFT in contrast with AdS/CFT correspondence. In order to illustrate this, we prepare the Hartle-Hawking wave functional $\Psi [\chi^{(0)}_j]$. It is obtained by the path integral over bulk fields $\chi_j$
\begin{align}
\Psi[\chi^{(0)}_j] = \int \mathcal{D} \chi_j e^{i S [\chi_j]} \, , 
\end{align}
where $S[\chi_j]$ is the action of dS gravity theory and the fields $\chi_j$ satisfy $\chi_j = \chi^{(0)}_j$ at the future infinity $t = t_\infty \to \infty$. 
We also denote a partition function of dual CFT by $Z_\text{CFT}[\chi^{(0)}_j]$, where $\chi^{(0)}_j$
 are sources for their dual operators $\mathcal{O}_j$. Then the proposal of \cite{Maldacena:2002vr} can be written as
\begin{align}
\Psi \left [\chi^{(0)}_j \right ] = Z_\text{CFT} \left [\chi^{(0)}_j \right ] \, . \label{GKPWdS}
\end{align}
With the wave functional, the gravity partition function at the semi-classical level can be evaluated as
\begin{align} \label{square}
Z_\text{dS}  = \int \prod_j  \mathcal{D} \chi_j^{(0)} \left | \Psi \left[ \chi_j^{(0)} \right] \right|^2  \, .
\end{align} 
We assume that there are several saddle points of bulk theory labeled by $n$. In the previous example, $n$ corresponds to the winding number of large gauge transformation generating new gauge configurations. It might be convenient to write it in terms of the Gibbons-Hawking entropy as
\begin{align}
    Z_\text{dS} \sim \sum_{n} \exp\left( S^{(n)}_\text{GH}\right) \, . \label{subGH}
\end{align}
Here $S^{(n)}_\text{GH}$ represents the contribution to the Gibbons-Hawking entropy from a saddle point labeled by $n$. 
In the classical limit, $G_N \to 0$, only the leading term dominates among the sum, and the other terms can be regarded as non-perturbative corrections.
In the previous example, it is given by (see, e.g. \cite{Witten:2021nzp})
\begin{align}
S_\text{GH}^{(n)}  =  \left( n + \frac12 \right)\frac{ \pi \ell \sqrt{1 - 8 G_N E} }{ G_N}
\end{align}
with $n=0,-1$. Therefore, $ S_\text{GH}^{(0)} $ is the leading contribution to the Gibbons-Hawking entropy $S_\text{GH}$ and $S_\text{GH}^{(-1)} $ is regarded as a non-perturbative correction.

In this paper we are mainly interested in black hole like objects, which may be created due to the back reactions of heavy objects. 
Let us assume that the geometry is created by the back reaction of heavy particle $\chi_j$. Then, the wave functional of the geometry is related to two-point function of dual operators $\mathcal{O}_j$ as
\begin{align}
 \Psi \sim  \left \langle \mathcal{O}_j \mathcal{O}_{j} \right \rangle \, . \label{W22pt}
\end{align}
From the geometry, we can specify the saddle points and compute the contribution from a saddle point to the Gibbons-Hawking entropy.
However, it is impossible to determine the set of allowable geometry among them unless the definite definition of quantum gravity is available. In the current case, the dual CFT is given by Liouville field theory or more generally Toda field theory, and its saddle points can be read off, e.g. as in \cite{Harlow:2011ny}. We can thus compute $S^{(n)}_\text{GH}$ from the dual CFT. For this purpose, it might be convenient to rewrite the wave functional as
\begin{align}
    \Psi \sim \sum_{n} \exp \left( S^{(n)}_{\rm GH} /2 + i \mathcal{I}^{(n)} \right) \, , \label{W2S}
\end{align}
where $i \mathcal{I}^{(n)}$ represents purely imaginary contributions.
We identify the allowable geometry of gravity theory by comparing the contributions to Gibbons-Hawking entropy from the both sides of duality. The above arguments can be generalized to higher-point function as well. We examine the three- and four-point functions in Liouville field theory and interpret the saddle point analysis in terms of SL$(2,\mathbb{C})$ Chern-Simons theory along the line of \cite{Harlow:2011ny}.
Moreover, we argue that the geometry corresponding to two linked Wilson loops on $S^3$ is dual to the monodromy matrix of four-point function and reproduce the results previously obtained in \cite{Hikida:2021ese,Hikida:2022ltr}. The results on geometries dual to Liouville two-point functions and those corresponding to two linked Wilson loops on $S^3$ were already presented in the previous letter \cite{Chen:2023prz}. In this paper, we explain details of the derivations and extend the analysis to the geometries dual to Liouville multi-point functions.

Since we deal with the pure gravity on dS$_3$ by Chern-Simons formulation, it is straightforward to extend the analysis to the higher-spin theory described by SL$(N,\mathbb{C})$ Chern-Simons theory. We examine the gravity theory by dual CFT$_2$, i.e. Toda field theory. We first construct higher-spin dS$_3$ black hole by analytically continuing the case of AdS$_3$ analyzed in \cite{Gutperle:2011kf}, see \cite{Ammon:2012wc} for a review.%
\footnote{See also \cite{Krishnan:2013zya} for a related work.}
We then study the saddle points of gravity theory corresponding to the solutions from Toda two-point functions as in the Liouville case. In the previous letter \cite{Chen:2023prz}, the higher-spin extension is only briefly mentioned, in particular only partial results on the geometries dual to Toda two-point functions were provided. In order to study the detailed properties of higher-spin dS$_3$ black hole, we extend the construction of the solutions to the Prokushkin-Vasiliev theory and probe the solutions by propagating a bulk scalar field, see \cite{Kraus:2011ds,Gaberdiel:2012yb,Kraus:2012uf,Gaberdiel:2013jca} for the AdS$_3$ case. Among others, we find a light-like singularity in the two-point function of bulk scalar field between the boundaries at the past and future infinities. The singularity should be the same as the one found in \cite{Strominger:2001pn} in the pure dS case.

This paper is organized as follows.
In the next section, we examine the simplest case with pure gravity described by SL$(2,\mathbb{C})$ Chern-Simons theory to a large extent. We describe the dS$_3$ black holes in terms of Chern-Simons gravity. In particular, we find non-trivial saddles of the complexified gravity, which are obtained by the large gauge transformations. We then determine the allowable set of saddles from two-point functions of dual Liouville field theory. In section \ref{sec:multi}, we extend the analysis to more complicated solutions dual to multi-point functions. 
We can insert monodromies along deficit lines in the Chern-Simons solutions, and the holonomies are read off from the Liouville correlation functions. Moreover, we also study geometry corresponding to two Wilson loops on $S^3$ constructed in \cite{Hikida:2021ese,Hikida:2022ltr}. We find that the entropy associated with the geometry can be obtained by the monodromy matrix of four-point function. In section \ref{sec:HSE}, we extend the analysis in section \ref{sec:dSBH} to the higher-spin gravity described by SL$(N,\mathbb{C})$ Chern-Simons gravity, whose dual description is given by Toda field theory.
In section \ref{sec:probe}, we construct higher-spin dS$_3$ black hole in the Prokushkin-Vasiliev theory and examine the behaviors of boundary-to-boundary two-point functions of bulk scalar field.
Section \ref{sec:conclusion} is devoted to conclusion and discussion. Several appendices follow, which explain technical details on the analysis in the main context.
In appendix \ref{app:cs}, we discuss subtleties associated Chern-Simons gauge theory based on complex gauge group and relation to the gravity theory with negative/positive cosmological constant.
In appendix \ref{app:Upsilon}, we summarize the properties of the Upsilon function, which is used to express three-point functions of Liouville field theory.
In appendices \ref{app:hsbh} and \ref{app:hslambda}, we explain the technical details of bulk analysis on higher-spin (A)dS$_3$ black holes. 
In appendix \ref{app:CFTcal}, we provide some CFT calculations as dual descriptions of higher-spin black holes.
In appendix \ref{app:Wilson}, we examine Wilson line operators in higher-spin dS$_3$ gravity and provide holographic computations of entanglement and thermal entropies.

\section{Three-dimensional dS black holes}\label{sec:dSBH}

In this section, we examine the black hole solutions of three-dimensional Einstein gravity both from the gravity theory and its dual CFT$_2$. 
We will first review the construction of black hole solutions 
of three-dimensional Einstein gravity with negative/positive cosmological constants. In particular, we find out semi-classical saddle points of path integral for the dS$_3$ gravity.
In subsection \ref{sec:Liouville}, we will analyze the CFT$_2$ dual to the dS$_3$ gravity, i.e. Liouville field theory. Finally we determine the set of allowable saddles of the dS$_3$ gravity from the two-point functions of dual Liouville field theory.

\subsection{Chern-Simons description of pure gravity}
\label{sec:gravity}

In this subsection, we review several useful results on the pure Einstein gravity on AdS$_3$ and dS$_3$ in the Chern-Simons formulation and black hole solutions to their equations of motion in order to prepare for the later sections, in particular, for higher-spin extensions.
The pure Einstein gravity in three space-time dimensions with negative cosmological constant can be described by $\text{SL}(2, \mathbb{R}) \times \text{SL}(2,\mathbb{R})$ Chern-Simons gauge theory \cite{Achucarro:1986uwr,Witten:1988hc}. See appendix \ref{app:cs} for some details. Its action is given by 
\begin{align}
 S = S_\text{CS} [A] - S_\text{CS} [\tilde A] \, , \quad
 S_\text{CS}[A]  = \frac{k}{4 \pi} \int \text{tr} \left( A \wedge d A + \frac{2}{3} A \wedge A \wedge A \right) \, . \label{CSaction}
\end{align}
Here the Chern-Simons level $k \, ( \in \mathbb{R} )$ is related to gravitational parameters as
\begin{align}
k = \frac{\ell_\text{AdS}}{4 G_N} \, .
\end{align}
The independent gauge fields $A,\tilde A$ are one-forms taking values in $\mathfrak{sl}(2)$ Lie algebra. 
The generators of $\mathfrak{sl}(2)$ Lie algebra are given by $L_n$ $(n=0,\pm 1)$ satisfying 
\begin{align}
[L_n , L_m] = (n - m)L_{n+m}\, .
\end{align}
We normalize the generators as $\text{tr} (L_0 L_0) = \frac{1}{2}$.
The solutions to the equations of motion can be put into the forms
\begin{align}
A = e^{- \rho L_0} a  e^{\rho L_0} + L_0 d \rho \, , \quad \tilde A = e^{\rho L_0} \tilde a  e^{-\rho L_0} -  L_0 d \rho  \label{gauge}
\end{align}
with
\begin{align}
    a = a_+ (x^+ ) d x^+ \, , \quad \tilde a = \tilde a_- (x^-) d x^- \, .
\end{align}
Here $a_+(x^+) , \tilde a_- (x^-)$ are arbitrary functions of $x^\pm = t \pm \phi$, where the periodicity $\phi \sim \phi + 2 \pi$ is assigned.
The bulk metric can be read off from the gauge fields as
\begin{align}
g_{\mu \nu} = \frac{\ell_\text{AdS}^2}{2} \text{tr} (A_\mu - \tilde A_\mu) (A_\nu - \tilde A_\nu) \, . \label{metricAdS}
\end{align}

The black hole solutions to the Einstein equations with negative cosmological constant are given by BTZ black holes with the metric
\cite{Banados:1992wn}
\begin{align}
\ell_\text{AdS}^{-2} ds^2 =   - (r^2 - r^2_+) dt^2 + \frac{1}{r^2 - r_+^2} dr^2 + r^2 d \phi ^2  \, ,\label{AdSBHmetric}
\end{align}
where the horizon is located at $r = r_+$ and the region outside the horizon is $r > r_+$.
Here and in the following, we only consider non-rotating black holes.
Let us consider a Euclidean time $t \to i t_E$ with $x^+ \to z = i t_E + \phi, x^- \to - \bar z = i t_E - \phi$. Then the absence of conical singularity at the horizon $r=r_+$ requires the periodicity 
$t_E \sim t_E + \beta^\text{AdS}$ with
\begin{align}
\tau^\text{AdS} = \frac{i \beta^\text{AdS}}{2 \pi} = \frac{i}{  r_+  }\, . \label{tauAdS}
\end{align}
The BTZ black hole can be realized by the configuration of gauge fields \eqref{gauge} with
\begin{align}
a _+ (x^+) = L_1 - \frac{2 \pi \mathcal{L}^\text{AdS}}{k}  L_{-1} \, , \quad
\tilde a _- (x^-) =  -  L_{-1} + \frac{2 \pi  \mathcal{L}^\text{AdS}}{k} L_{1} \, , \label{solution}
\end{align}
where we have set
\begin{align}
\mathcal{L}_\text{AdS} = \frac{\ell^\text{AdS} r^2_+}{32 \pi G_N } = \frac{k r_+^2}{8 \pi} .
\end{align}
We can read off the metric of the black hole from \eqref{metricAdS} as
\begin{align} 
\begin{aligned}
\ell_\text{AdS}^{-2} ds^2&= d \rho^2 - \left(e^\rho - \frac{2 \pi \mathcal{L} ^\text{AdS}}{k} e^{-\rho}\right)\left(e^\rho - \frac{2 \pi \mathcal{L}^\text{AdS} }{k} e^{- \rho}\right) dt^2  
\\& \qquad \qquad
+ \left(e^\rho + \frac{2 \pi \mathcal{L}^\text{AdS}}{k}  e^{ -\rho}\right)\left(e^\rho + \frac{2 \pi  \mathcal{L} ^\text{AdS}}{k} e^{- \rho}\right) d\phi^2 \, .
\end{aligned}
\end{align}
Performing a coordinate transformation
\begin{align}
    r = e^\rho + \frac{2 \pi \mathcal{L}^\text{AdS}}{k} e^{- \rho} \, ,  
\end{align}
the metric becomes that of BTZ black hole in \eqref{AdSBHmetric}.
In the pure gravity, the notion of horizon is gauge invariant, so there is no problem to define black hole. However, in higher-spin gravity theories, which we shall deal with, the notion of horizon is not gauge invariant generically, and the definition of black hole might be ambiguous.
Following \cite{Gutperle:2011kf,Ammon:2012wc}, we shall use Wilson loop to define black hole since it is a gauge invariant object. We thus introduce a holonomy matrix for $A$ along the thermal cycle 
\begin{align}
\mathcal{P} e^{ \oint A } = \mathcal{P} e^{i \oint d t_E A_+} = e^{- \rho L_0} e^\Omega e^{\rho L_0} \, . \label{AdShol}
\end{align}
In the current case, the eigenvalue of $\Omega$ can be evaluated as $(\pi i , - \pi i)$.

We then move to the case with positive cosmological constant.
The Einstein gravity can again be described by $\text{SL}(2,\mathbb{C}) \times \text{SL}(2,\mathbb{C})$ Chern-Simons theory with 
the action 
\begin{align} \label{CSactiondS}
S = S_\text{CS} [A] - S_\text{CS} [\bar A] \, , \quad S_\text{CS}[A]  = - \frac{\kappa}{4 \pi} \int \text{tr} \left( A \wedge d A + \frac{2}{3} A \wedge A \wedge A \right) \, .
\end{align}
See appendix \ref{app:cs} for some details.
The relation to gravity parameters is
\begin{align}
\kappa = \frac{\ell }{4 G_N} \, .
\end{align}
We may relate the Euclidean version of \eqref{CSaction} to \eqref{CSactiondS} by replacing the couplings $k \to i \kappa$, which corresponds to $\ell_\text{AdS} \to i \ell $.%
\footnote{In order to move from \eqref{CSaction} to the Euclidean version, we need to perform a Wick rotation $t \to it_E$, which provide a phase $i$. Combining the replacement $k \to i \kappa$, we have extra factor $i \cdot i = -1$ in the second equation of \eqref{CSactiondS}. }
The gauge fields $A,\bar A$ are one-forms taking values in $\mathfrak{sl}(2)$ Lie algebra. 
For generators, we assign the relation of complex conjugation as $(L_0)^* = - L_0$ and $(L_\pm)^* = L_{\mp}$.
As in \cite{Maldacena:2002vr}, we may perform furthermore the coordinate transformation
\begin{align} \label{tilderho}
 e^{- \rho} \to  - i e^{-\tilde\rho} = e^{- (\tilde\rho + \pi i /2 ) } \, .
\end{align}
Solutions to the equations of motion can be put into the form
\begin{align}
\begin{aligned}
&A = e^{- (\tilde\rho + \pi i /2) L_0} a  e^{(\tilde\rho + \pi i /2) L_0} + L_0 d \tilde\rho \, , \\
&\bar A = e^{(\tilde\rho + \pi i /2) L_0} \bar a  e^{- (\tilde\rho + \pi i /2) L_0} -  L_0 d \tilde\rho
\end{aligned}
\label{gaugedS}
\end{align}
with
\begin{align}
    a = a_+(x^+)  dx^+ \, , \quad \bar a = \bar a_-(x^-) dx^- \, .
\end{align}
Here $a_+(x^+) , \bar a_- (x^-)$ are arbitrary functions of $ x^\pm= i  t  \pm \phi$ (or $z = x^+ , \bar z = - x^-$) and the periodicity $\phi \sim \phi + 2 \pi$ is assigned.
The bulk metric can be read off from
\begin{align}
g_{\mu \nu} = - \frac{\ell^2}{2} \text{tr} (A_\mu - \bar A_\mu) (A_\nu - \bar A_\nu) \, . \label{metricdS}
\end{align}

Let us first consider the configuration of gauge fields of the form
\begin{align} 
\label{soldS}
&a _+ (x^+) = L_1 + \frac{2 \pi \mathcal{L}}{\kappa}  L_{-1} \, , \quad
\bar a _- (x^-) =  -  L_{-1} - \frac{2 \pi  \mathcal{L}}{\kappa} L_{1} \, ,
\end{align}
then the metric \eqref{metricdS} leads to
\begin{align}
\ell^{-2} ds^2 = - d \tilde\rho^2 +  \frac{8 \pi \mathcal{L} }{\kappa} \sinh ^2 \tilde\rho d t^2 
+ \frac{8 \pi \mathcal{L}}{\kappa} \cosh ^2 \tilde\rho d\phi^2 \, .
\end{align}
Here we have performed a shift $\tilde\rho \to \tilde\rho + \ln \sqrt{2 \pi  {\mathcal L}/\kappa} $.
If we regard $\tilde\rho$ as a {time-like coordinate}, then the geometry describes a cosmological universe.
Near the asymptotic future, the metric becomes
\begin{align}
\ell^{-2} ds^2 = - d \tilde\rho^2 +  \frac{8 \pi \mathcal{L} }{\kappa} e^{2\tilde\rho} (d t^2  + d\phi^2 )  \, ,
\end{align}
where the boundary geometry is with the flat metric
\begin{align}
d s^2 = d t^2 + d \phi^2 \, , \quad  - \infty < t < \infty \, , \quad \phi \sim \phi + 2\pi \, .
\end{align}
Notice that the infinities $t = \pm \infty$ are not included, thus the boundary topology is $\mathbb{R} \times S^1$, see, e.g. \cite{Balasubramanian:2001nb}.

The metric obtained above is outside the horizon of our static universe.
The static patch is obtained instead by an analytic continuation $\tilde\rho \to i \theta$ as 
\begin{align}
\ell^{-2} ds^2 = d \theta^2 - \frac{8 \pi {\mathcal L}}{\kappa} \sin ^2  \theta  d t^2  +  \frac{8 \pi {\mathcal L}}{\kappa} \cos ^2 \theta  d\phi^2 \, . \label{dSmetric}
\end{align}
A coordinate transformation
\begin{align}
    r = \sqrt{\frac{8 \pi \mathcal{L}}{\kappa}} \cos \theta \, , \quad  
\mathcal{L} = \frac{\ell r^2_+}{32 \pi G_N } = \frac{\kappa r_+^2}{8 \pi} \label{L}
\end{align}
indeed leads to the metric in \eqref{dSBHmetric}.
Considering a Euclidean time $i t \to t_E$, the absence of conical singularity at the horizon requires the periodicity 
$t_E \sim t_E + \beta$ with
\begin{align}
\tau = \frac{i \beta}{2 \pi} =  \frac{i \kappa}{2} \frac{1}{\sqrt{2 \pi \kappa \mathcal{L}}} \, .  \label{tau}
\end{align} 
As in the AdS$_3$ case, it is convenient to define
a holonomy matrix for $A$ along the time cycle by
\begin{align}
\mathcal{P} e^{\oint A} = \mathcal{P} e^{ \oint d t_E A_+}  = e^{- (i \theta  + \pi i/2) L_0} e^\Omega e^{(i \theta  + \pi i/2) L_0} \,  . \label{dShol}
\end{align}
The eigenvalues of $\Omega$ are computed as $( \pi i , - \pi i)$.

Let us examine more about the holonomy condition. The action of large gauge transformation changes the eigenvalues of $\Omega$ as $(2 \pi i(n +1/2), -2 \pi i(n +1/2)) $ with integer $n$.\footnote{We may require that the holonomy matrix \eqref{dShol} is trivial. The center of gauge group $\text{SL}(N) \times \text{SL}(N)$ is $\mathbb{Z}_2$ for even $N$, and the trivial holonomy means that the holonomy matrix is given by a center of the gauge group $\pm \mathbbm{1}$. Thus, the trivial holonomy condition allows $n \in \mathbb{Z} +1/2$, but such a case is not considered here as it is not generated by a large gauge transformation.} 
A configuration of gauge fields with the holonomy matrix is given by 
\begin{align}
a  =- \sqrt{\frac{2 \pi \mathcal{L}}{\kappa}} \sigma_1 ( d \phi + (2 n + 1) d t_E )\, , \quad 
\bar a =  - \sqrt{\frac{2 \pi \mathcal{L}}{\kappa}} \sigma_1 ( d \phi - (2 n +1) d t_E )
\end{align}
with
\begin{align}
\sigma_1 = \frac12
\begin{pmatrix}
0 & 1 \\
1 & 0 
\end{pmatrix} \, .
\end{align}
The metric from the configuration of gauge fields can be read off as
\begin{align}
\ell^{-2} ds^2 = d \theta^2 + \frac{8 \pi (2 n + 1)^2 {\mathcal L}}{\kappa} \sin ^2  \theta  d t^2_E  +  \frac{8 \pi {\mathcal L}}{\kappa} \cos ^2 \theta  d\phi^2 \, . \label{SGHn}
\end{align}
The classical action for the configuration is 
\begin{align}
 - S \, (\equiv S^{(n)}_\text{GH})= 4 \pi (2 n +1) \sqrt{2 \pi \kappa \mathcal{L}} =  (2 n + 1) \frac{ \pi \ell \sqrt{1 - 8 G_N E}}{2 G_N} \, ,
\end{align}
where $S^{(n)}_\text{GH}$ was introduced in \eqref{subGH}.
See \cite{Hikida:2021ese,Hikida:2022ltr}
for the case with conical defect geometry.
If we set $n = 0,-1$, then the entropy reproduces \eqref{BHentropy}.
With an integer level and a compact gauge group, a large gauge transformation is a symmetry of Chern-Simons theory. However, now the level is not integer valued nor the gauge group is complex (non-compact), thus a large gauge transformation generates a different configuration, see, e.g. \cite{Witten:2010cx,Harlow:2011ny}. In the Chern-Simons formulation of gravity theory, there are no criteria to choose the proper set of saddle points labeled by $n$. In the next subsection, we determine the proper set from its dual CFT description.

\subsection{Liouville description}
\label{sec:Liouville}

In \cite{Hikida:2021ese,Hikida:2022ltr}, the Gibbons-Hawking entropy \eqref{BHentropy}  of the dS$_3$ black hole \eqref{dSBHmetric} was evaluated from its dual CFT$_2$. At the leading order in $G_N$, the entropy is obtained from a particular limit of modular $S$-matrix of SU$(2)$ Wess-Zumino-Witten (WZW) model by applying the method of \cite{Witten:1988hf}. It was also argued that the same result can be obtained through Liouville field theory in the corresponding limit of large central charge. In fact, the entropy for pure dS$_3$ was reproduced also from the vacuum sphere amplitude of Liouville field theory in \cite{Hikida:2022ltr}. In this subsection, we extend the result to the case with insertions of two heavy operators, which is dual to the dS$_3$ black hole.
We then read off all the saddle points of complexified Liouville field theory in the corresponding limit.

The action of Liouville field theory is given by
\begin{align}
S_\text{L} = \frac{1}{4 \pi} \int d^2 \sigma \sqrt{\tilde g} \left[ \partial_a \phi \partial_{a'} \phi \tilde g^{aa'} + Q \tilde{\mathcal{R}} \phi + 4 \pi \mu e^{2 b \phi}\right] \, .
\end{align}
Here and in the following, we follow the notation of \cite{Harlow:2011ny}, see also \cite{Zamolodchikov:1995aa}. We use $\tilde g_{aa'}$ as a reference metric with $\tilde g = \det \tilde g_{aa'}$ and $\tilde{\mathcal{R}}$ as the curvature with respect to the reference metric.
The theory is invariant under the combination of the Weyl transformation $\tilde g_{aa'} \to \Omega (\sigma) \tilde g_{aa'}$ and the shift of field $\phi \to \phi - \frac{Q}{2} \ln \Omega(\sigma) $.
With the help of the symmetry, we may set the ``physical'' metric as $g_{aa'} = e^{\frac{2}{Q} \phi} \tilde g_{aa'}$.
The vertex operators are defined by
\begin{align}
V_\alpha = e^{2 \alpha \phi}
\end{align}
with conformal weights $h = \bar h = \alpha (Q - \alpha)$. The {Liouville} central charge $c$ is related to the background charge 
$Q = b + b^{-1}$ as
\begin{align}
c = 1 + 6 Q^2 = 1 + 6 (b + b^{-1})^2 \, . \label{Liouvillec}
\end{align}
We are interested in sphere amplitudes. We thus require that the Liouville field is regular everywhere on $S^2$. It is convenient to move to flat space with the reference metric $ds^2 = dz d \bar z$ with $z = \sigma_1 + i \sigma_2$ and $\bar z = \sigma_1 - i \sigma_2$ by making use of the symmetry of Liouville field theory. Then the regularity condition for the Liouville field on $S^2$ is mapped to the boundary condition
\begin{align}
\phi = - 2 Q \log |z| + \mathcal{O} (|z|^0) 
\end{align}
at the infinity $|z| \to \infty$. In order to remove the subtlety associated with the infinity $|z| \to \infty$, we consider a disc $D$ with radius $R \to \infty$ with the boundary $\partial D$. We then instead use the regularized action \cite{Zamolodchikov:1995aa}
\begin{align}
S_\text{L} = \frac{1}{4 \pi} \int_D d ^2 \sigma [\partial_a \phi \partial_a \phi + 4 \pi \mu e^{2 b \phi}] + \frac{Q}{\pi} \oint _{\partial D} \phi d \theta + 2 Q^2 \ln R \, ,
\end{align}
where $\theta$ represents the coordinate of the boundary $\partial D$.

We would like to study classical dS$_3$ gravity from the viewpoints of its dual CFT$_2$. The asymptotic symmetry near the future infinity is found to be Virasoro symmetry and its central charge is obtained as \cite{Strominger:2001pn} (see also \cite{Ouyang:2011fs})%
\footnote{The central charge $c$ is an exact quantity including full quantum corrections since it is fixed by commutation relations among Virasoro generators. The relation to gravitational parameters may receive quantum corrections, but they should not violate the condition $c^{(g)} \in \mathbb{R}$ as the physical gravitational parameters take real values.}
\begin{align}
    c \, (\equiv i c^{(g)}) = i\cdot 6 \kappa = i \frac{3 \ell}{2 G_N}  \label{cg}
\end{align}
as in the well-known case of AdS$_3$ by \cite{Brown:1986nw}.
We are interested in the leading order of $G_N$, which is dual to that of $1/c^{(g)}$. Solving \eqref{Liouvillec}, we find
\begin{align}
b^{-2} =  \frac{i c^{(g)}}{6} - \frac{13}{6}  + \mathcal{O} ((c^{(g)})^{-1})\, . \label{binc}
\end{align}
A unitary region of Liouville field theory is obtained by a real positive $b$, however the above equation implies that $b$ has to take a complex value.
The Liouville field theory with a complex $b$ has appeared before in the context of time-like Liouville theory describing rolling closed string tachyon, where we set $b=i$, see, e.g. \cite{Strominger:2003fn,Zamolodchikov:2005fy,Schomerus:2003vv}. 
In the current case, the central charge is large, so we need to set $b \sim 0$. Moreover, the order $\mathcal{O}((c^{(g)})^0)$ contribution implies that $\text{Re} \,  b^{-2} < 0$, which will be important later. 
At the leading order in $b \sim 0$,  the action may be written as
\begin{align}
b^2 S_\text{L} = \frac{1}{16 \pi} \int_D d^2 \sigma [\partial_a \phi_c \partial_a \phi_c + 16 \lambda e^{\phi_c}] + \frac{1}{2\pi} \oint_{\partial D} \phi_c d \theta + 2 \ln R + \mathcal{O} (b^2) 
\end{align}
with finite $\phi_c = 2 b \phi$ and $\lambda \equiv \pi \mu b^2$.
The boundary condition is
\begin{align}
\phi_c (z ) = - 4 \ln |z| + \mathcal{O}(|z|^0)
\end{align}
for $|z| \to \infty$.
Since $b$ is a complex number, the Liouville field is also assumed to take a complex value as emphasized in \cite{Harlow:2011ny}. In the study of rolling tachyon, we usually set $\mu$ as a real number, which makes $\lambda$ to be complex. Here we rather set $\lambda$ as a real number, which implies that $\mu$ is complex. We will explain the reason of this choice later.

As mentioned above, we would like to determine the saddle points of gravity path integral from Liouville two-point function. The operator should have a large conformal dimension if it is dual to a bulk field back reacting to create a black hole geometry. Such an operator is usually called as a heavy operator.
At $b \sim 0$, the two-point function can be evaluated as
\begin{align}
\left \langle V_{\alpha} (z_1) V_\alpha (z_2) \right \rangle \equiv 
\int \mathcal{D} \phi_c e^{- S_\text{L}} \exp \left( b^{-1} \alpha (\phi_c (z_1) + \phi_c (z_2))\right) \, . \label{classical2pt}
\end{align}
A heavy operator is defined with $\alpha = \eta /b$, where $\eta$ is kept finite for $b \to 0$. The conformal dimension $h \equiv i h^{(g)}$ of the heavy operator and the energy $E$ to create a black hole geometry is related as 
\begin{align}
    2 h \, (\equiv 2 \alpha (Q - \alpha)) = i \ell E \, , \quad 
    1 - 2 \eta= \sqrt{1-8G_NE} \, . \label{eta2E}
\end{align}
In particular, a black hole exists only if the condition
\begin{align}
    0 < \eta < \frac{1}{2} \label{Seiberg}
\end{align}
is satisfied. The upper bound is the same as so-called Seiberg bound \cite{Seiberg:1990eb}.

We regard the insertions of such heavy operators as a part of action. The equation of motion thus becomes:
\begin{align}
\partial \bar \partial \phi_c = 2 \lambda e^{\phi_c} -2 \pi \eta [\delta^{(2)} (\sigma - \sigma_1) + \delta^{(2)} (\sigma - \sigma_2) ] \, .
\end{align}
Notice that the equation is invariant under the constant shifts $\phi_c \to \phi_c + 2 \pi i n$ with integer $n$. Therefore, once $\phi_{c(0)}$ is a 
solution to the equation of motion, then the same is true for 
\begin{align} \label{Liouvillevac}
\phi_{c(n)} = \phi_{c(0)} + 2\pi i n \, .
\end{align}
Near $z \sim z_1$, the heavy operator behaves as
\begin{align}
\phi_c (z) \sim - 4 \eta |z - z_1| \, , 
\end{align}
which implies that the physical metric is
\begin{align}
ds^2 = \frac{1}{r^{4 \eta }} (dr^2 + r^2 d \theta^2) \label{conical}
\end{align}
near $z \sim z_1$. 
In the presence of heavy operators, we use the modified action \cite{Zamolodchikov:1995aa,Harlow:2011ny}
\begin{align}
\begin{aligned}
b^2 \tilde S_\text{L} &= \frac{1}{16 \pi} \int _{D- d_1 - d_2} d^2 \sigma [\partial^a \phi_c \partial_a \phi_c + 16 \lambda e^{\phi_c} ] 
+ \frac{1}{2 \pi} \oint_{\partial D} \phi_c d \theta + 2 \ln R \\
& \quad  - \sum_i \left[ \frac{\eta}{2 \pi} \oint_{\partial d_i} \phi_c d \theta_i + 2 \eta^2 \ln \epsilon \right] \, ,
\end{aligned}
\end{align}
where $d_i$ is a small disk with radius $\epsilon$ including $z_i$.
The modified action at a saddle point was evaluated in \cite{Harlow:2011ny} as
\begin{align}
\begin{aligned}
b^2 \tilde S_\text{L} &= 2 \pi i (n + 1/2)  (1- 2 \eta) + (2 \eta -1) \ln \lambda + 4 (\eta - \eta^2) \ln |z_{12}| \\
                       & \quad + 2 [ (1 - 2 \eta) \ln (1 - 2 \eta) - (1 - 2 \eta)] \, . 
\end{aligned}
\end{align}
 The semi-classical limit of two-point function \eqref{classical2pt} is thus given by the sum of $\exp (- \tilde S_\text{L})$ over some saddle points \eqref{Liouvillevac}.

Even in a non-gravitational theory, it is difficult to determine the set of all semi-classical saddle points of path integral. Fortunately, the exact expression of Liouville two-point function is known as \cite{Dorn:1994xn,Zamolodchikov:1995aa}
\begin{align} \label{exact2pt}
&\langle V_\alpha (z_1) V_\alpha (z_2) \rangle   = |z_{12}|^{- 4 \alpha (Q - \alpha) } \frac{2 \pi}{b^2} [\pi \mu \gamma (b^2)] ^{(Q - 2 \alpha)/b} \gamma \left ( \frac{2 \alpha }{b} -1 - \frac{1}{b^2} \right )
\gamma(2b\alpha -b^2 ) \delta (0) \, .
\end{align}
Here we defined
\begin{align} \label{smallgamma}
\gamma(x) = \frac{\Gamma(x)}{\Gamma(1-x)} \, .
\end{align}
The delta function in the right hand side of \eqref{exact2pt} comes from the fact that $\langle V_\alpha V_{\alpha '} \rangle \propto \delta (\alpha - \alpha ')$.
We then read off the set of semi-classical saddles from the exact expression. For $b \sim 0$, the two-point function reduces to
\begin{align}
\langle V_\alpha (z_1) V_\alpha (z_2) \rangle \sim \delta (0) |z_{12}|^{- 4 \eta (1 - \eta)/b^2 } \lambda^{(1 - 2 \eta) /b^2}
\left[ \frac{\gamma(b^2)}{b^2} \right]^{(1 - 2 \eta)/b^2} \gamma \left( \frac{2 \eta -1}{b^2}\right) \, .
\end{align}
At the region of $b \sim 0$, we can easily see%
\footnote{Throughout this paper, we promise that $\log z$ for $z\in\mathbb{C}$ takes the principal value, satisfying $-\pi<\Im \log z\le\pi$. By using this, $a^z$ with $a,z\in\mathbb{C}$ is defined as $a^z\equiv e^{z\log a}$. Note that some properties, e.g. $\log (ab)=\log a+\log b$, may be lost for complex case.}
\begin{align}
\left[ \frac{\gamma(b^2)}{b^2} \right]^{(1 - 2 \eta)/b^2} \sim \exp \left[ \frac{2 - 4\eta }{b^2}\left(\ln \frac{1}{b^2}-\pi i\right)\right] \, ,
\end{align}
where $-\pi i$ term comes from the difference of branches between $\ln\frac{1}{b^2}$ and $\ln\frac{1}{b^4}$.
Furthermore, we can find (see \cite{Harlow:2011ny})
\begin{align}
\gamma \left( \frac{2 \eta -1}{b^2}\right)  
\sim \left( e^{ -  \pi i (1 - 2 \eta) /b^2} - e^{ \pi i (1 - 2 \eta) /b^2} \right) \exp \left[ \frac{ 4 \eta - 2 }{b^2} \left(\ln (1 - 2 \eta) +\ln \frac{1}{b^2}-\pi i - 1\right)\right]
\end{align}
by recalling that now $\text{Re} \,  b^{-2} < 0$ as in \eqref{binc}.
Here we have used Stirling's approximation 
\begin{align}
\Gamma (x) = \exp (x \ln x - x  + \mathcal{O} (\ln x)) \, ,
\end{align}
for $\text{Re} \, x > 0$ and 
\begin{align}
\Gamma (x) =\frac{1}{e^{\pi i x} - e^{- \pi i x}} \exp (x \ln (- x) - x  + \mathcal{O} (\ln (- x)))\,,
\end{align}
for $\text{Re} \, x < 0$, and an identity $\ln(-1/b^2)=\ln(1/b^2)-\pi i $. 
In summary, the two-point function behaves near $b \sim 0 $ as
\begin{align}
\begin{aligned}\label{twopoint}
&\langle V_\alpha (z_1) V_\alpha (z_2) \rangle \sim \delta (0) |z_{12}|^{- 4 \eta (1 - \eta)/b^2 } \lambda^{(1 - 2 \eta) /b^2} \\ & \quad \times
\left( e^{ -  \pi i (1 - 2 \eta) /b^2} - e^{ \pi i (1 - 2 \eta) /b^2} \right)  \exp \left \{ - \frac{2}{b^2} \left[( 1 - 2 \eta )\ln (1 - 2 \eta) - (1 - 2 \eta )  \right] \right \}\, .
\end{aligned}
\end{align}
This is the same as the sum of $\exp (- \tilde S_\text{L})$ at the saddle points with $n =-1 , 0$.
For $b^{-2} \sim i c^{(g)}/6$ with $c^{(g)} \gg 1$,
the absolute value of two-point function behaves as
\begin{align}
\left|\langle V_\alpha (z_1) V_\alpha (z_2) \rangle \right| \sim \left | e^{ \frac{\pi c^{(g)}}{6} \sqrt{1 - 8 G_N E}} - e^{- \frac{\pi c^{(g)}}{6} \sqrt{1 - 8 G_N E}} \right | \, , \label{abs2pt}
\end{align}
where we have used \eqref{eta2E}. 
{Notice that the $|z_{12}|$ dependence is canceled due to the purely imaginary power of $-4\eta(1-\eta)/b^2$, see appendix \ref{app:cs} for the arguments. }
With \eqref{W22pt} and \eqref{W2S}, we can read off the contributions corresponding to those to the Gibbons-Hawking entropy. 
Compared with \eqref{SGHn}, we can see that the possible saddles are with $n = -1,0$. This reproduces the result obtained from the criteria of allowable complex geometry in \cite{Witten:2021nzp}.

Let us pause here to explain why we set the redefined parameter $\lambda \equiv \pi b^2 \mu$ to be real. 
For simplicity, let us focus on a vacuum amplitude.
In the CFT side, it was computed in \cite{Hikida:2022ltr} as
\begin{align}\label{CFTPF}
    Z_\text{CFT} \simeq C e^\frac{\pi c^{(g)} }{6} \lambda^{i \frac{c^{(g)}}{6}} \, , 
\end{align}
where $C$ is a constant independent of $c^{(g)}$. On the other hand, the Hartle-Hawking wave functional of universe should behave as \eqref{W2S},
\begin{align}
 \Psi \sim \exp \left( S_\text{GH} /2 + i\mathcal{I} \right)  \, .
\end{align}
We have already seen that the part $S_\text{GH}$ agrees among two dual descriptions. Therefore, we should identify as
\begin{align}
    \mathcal{I} = \frac{c^{(g)}}{6} \ln \lambda \, .
\end{align}
Since $\mathcal{I}$ is real, the above identification implies that $\lambda$ should be real as well.
In other words, if the Liouville theory at large central charge is dual to the geometry for Hartle-Hawking wave function, then $\lambda$ should be real.

\section{Geometries dual to multi-point functions}\label{sec:multi}

In the previous section, we have examined the two-point functions of heavy operators in Liouville field theory. From their behaviors at the semi-classical limit with $b \sim 0$, we have read off the saddle points of the path integral in the Chern-Simons gravity we should take. In this section, we extend the analysis to the cases with multi-point functions of heavy operators in Liouville field theory. Namely, we identify the wave functional of the Chern-Simons gravity with Liouville $n$-point functions as
\begin{align}
    \Psi = \langle V_{\alpha_1} (z_1) \dots   V_{\alpha_n} (z_n) \rangle \, .
    \label{multi}
\end{align}
Here we require that the Liouville momenta $\alpha_i$ of vertex operators scale as $\alpha_i = \eta_i/b$ with $\eta_i$ fixed for $b \to 0$. Generically, the geometries dual to $n$-point functions can be realized by $S^3$ with $n$ conical deficits connected inside the bulk.%
\footnote{The Hartle-Harking wave functional is realized by  a geometry starting from hemi-sphere and connecting to Lorentzian dS. After taking the square of the wave functional, only the part corresponding to the sphere remains as the part corresponding to the Lorentzian dS gives only a phase factor, see, e.g. \cite{Hikida:2022ltr}.}
As pointed out in \cite{Harlow:2011ny}, we can insert monodromies along the defect lines and the different monodromies lead to different saddle points of gravitational path integral. As in the case of Liouville two-point functions, we shall read off the saddle points of the gravitational path integral from the semi-classical analysis of Liouville multi-point functions. 
In the next subsection, we first examine the geometries dual to Liouville three-point functions. The thee-point coefficients are obtained by so-called DOZZ formula \cite{Dorn:1994xn,Zamolodchikov:1995aa}, and their semi-classical behaviors were already examined in \cite{Harlow:2011ny}. Applying their analysis, we read off the saddle points of gravitational path integral. In subsection \ref{Sec:LiouvilleHigher}, we similarly examine the Liouville four-point functions at the semi-classical limit and read off the saddle points of dual gravity theory, see \cite{Balasubramanian:2017fan} for a related work. The extensions to more higher-point functions of Liouville field theory are analyzed in appendix \ref{app:5pt}. 
In subsection \ref{Sec:Liouville 4pt}, we consider some specific geometries constructed in \cite{Hikida:2021ese,Hikida:2022ltr}, which are related to $S^3$ with two linked (unlinked) Wilson lines in Chern-Simons gauge theory. We determine the set of saddle points of gravitational path integral from Liouville four-point functions and compare with the previous results in \cite{Hikida:2021ese,Hikida:2022ltr}.

\subsection{Three-point functions}
\label{sec:Liouville3pt}

We would like to evaluate the semi-classical limits of three-point functions in Liouville field theory. The explicit form of three-point function is known in \cite{Dorn:1994xn,Zamolodchikov:1995aa}:
\begin{align}
    \langle V_{\alpha_1}(z_1,\bar{z}_1)V_{\alpha_2}(z_2,\bar{z}_2)V_{\alpha_3}(z_3,\bar{z}_3)\rangle=\frac{C(\alpha_1,\alpha_2,\alpha_3)}{|z_{12}|^{2(h_1+h_2-h_3)}|z_{13}|^{2(h_1+h_3-h_2)}|z_{23}|^{2(h_2+h_3-h_1)}} \, , 
\end{align}
\begin{align}
    \begin{aligned}\label{DOZZ}
C(\alpha_1,\alpha_2,\alpha_3)=&\left[\lambda\gamma(b^2)b^{-2b^2}\right]^{(Q-\sum_i\alpha_i)/b}\\
    &\times\frac{\Upsilon_b'(0)\Upsilon_b(2\alpha_1)\Upsilon_b(2\alpha_2)\Upsilon_b(2\alpha_3)}{\Upsilon_b(\sum_i\alpha_i-Q)\Upsilon_b(\alpha_1+\alpha_2-\alpha_3)\Upsilon_b(\alpha_2+\alpha_3-\alpha_1)\Upsilon_b(\alpha_3+\alpha_1-\alpha_2)} \, .
    \end{aligned}
\end{align}
See appendix \ref{app:Upsilon} for the definition and properties of the Upsilon function $\Upsilon_b(x)$.
Here $\Upsilon'_b(x)$ denotes the derivative with respect to $x$.
As mentioned above, we consider the case where all the external operators are heavy, i.e. $\alpha_i=\eta_i/b$ with $b\to 0$ while $\eta_i$ kept fixed. 
Here we will consider real $\eta_i$, related to the bulk energy $E_i$ as 
$
    2 \eta_i= 1 - \sqrt{1-8G_NE_i} 
$,
which satisfies the Seiberg bound 
$
0<\eta_i< 1/2
$.
As argued in \cite{Harlow:2011ny}, we can further divide them into two possible regions, called Region I and Region II in that paper:
\begin{align}
    \text{Region I:}&\quad \sum_i \eta_i>1 \, , \\
    \text{Region II:}&\quad\begin{cases}\displaystyle
    \sum_i \eta_i<1 \, , \\
    \displaystyle \eta_i+\eta_j-\eta_k>0 \, .
    \end{cases}\label{regionII}
\end{align}
The condition of Region I comes from requiring the convergence of path integral. On the other hand, the original path integral over real $\phi$ does not converge in Region II, which leads us to taking complex saddles into account. The second condition in \eqref{regionII} is also satisfied in Region I without any extra conditions. Further meanings of the conditions will be explained below. First we evaluate a three-point function for each region.

It is easily checked that the first term is universally behaved as 
\begin{align}
    \left[\lambda\gamma(b^2)b^{-2b^2}\right]^{(Q-\sum_i\alpha_i)/b}\simeq \exp\left[\frac{1-\sum_i\eta_i}{b^2}\log\lambda+\frac{1-\sum_i\eta_i}{b^2}\log \frac{1}{b^2} \right] \, .
\end{align}
The semi-classical limits of the other terms are evaluated by using the formula described in appendix \ref{app:Upsilon}. In particular, the behavior of $\Upsilon_b(\sum_i\alpha_i-Q)$ depends on which region $\eta_i$ belong to. 

\paragraph{Region I}

In this region, we can apply the asymptotic formula \eqref{upsasymp} to every $\Upsilon_b$ function in the DOZZ formula \eqref{DOZZ}. The result is 
\begin{align}\begin{aligned}
   & C(\alpha_1,\alpha_2,\alpha_3)
    \sim \lambda^{(1-\sum_i\eta_i)/b^2}\exp\bigg[\frac{1}{b^2}\bigg\{1-\sum_i\eta_i+F(2\eta_1)+F(2\eta_2)+F(2\eta_3)+F(0)\\
    &\quad -F(\sum_i\eta_i-1)-F(\eta_1+\eta_2-\eta_3)-F(\eta_2+\eta_3-\eta_1)-F(\eta_3+\eta_1-\eta_2)\bigg\}\bigg] \, ,
\end{aligned}\end{align}
where $F(\eta)$ is defined through integral in \eqref{Def:F(eta)}.
Since $\eta_i$ are all real-valued while $b^2$ and conformal weights are purely imaginary, the absolute value is
\begin{align}
    |\langle V_{\alpha_1}(z_1,\bar{z}_1)V_{\alpha_2}(z_2,\bar{z}_2)V_{\alpha_3}(z_3,\bar{z}_3) \rangle|^2\sim \mathcal{O}(1) \, .
\end{align}
This result may be related to the fact that there is no bulk dual to the three-point functions in this regime. In other words, we cannot construct the geometry of two-sphere with three conical deficits whose deficit angles are given by $\pi\eta_i$.

\paragraph{Region II}
Since $\sum_i\eta_i-1<0$, we have to use the recursion relation \eqref{defups} 
\begin{align}
    \Upsilon_b\left(\frac{\sum_i\eta_i-1}{b}\right)\sim \frac{b^{1-\frac{2(\sum_i\eta_i-1)}{b^2}}}{\gamma\left(\frac{\sum_i\eta_i-1}{b^2}\right)}\Upsilon_b\left(\frac{\sum_i\eta_i}{b}\right)\, .
\end{align}
By using this, \eqref{DOZZ} approximates
\begin{align}\begin{aligned}
    C(\alpha_1,\alpha_2,\alpha_3)&\sim\left(\lambda \frac{1}{b^4}\right)^{(1-\sum_i\eta_i)/b^2}\gamma\left(\frac{\sum_i\eta_i-1}{b^2}\right)\\
    &\times \frac{\Upsilon_b'(0)\Upsilon_b(2\alpha_1)\Upsilon_b(2\alpha_2)\Upsilon_b(2\alpha_3)}{\Upsilon_b(\sum_i\alpha_i)\Upsilon_b(\alpha_1+\alpha_2-\alpha_3)\Upsilon_b(\alpha_2+\alpha_3-\alpha_1)\Upsilon_b(\alpha_3+\alpha_1-\alpha_2)} \, .
\end{aligned}\end{align}
Furthermore, since $\Re\left[(\sum_i\eta_i-1)/b^2\right]>0$, we have
\begin{align}\begin{aligned}
    \gamma\left(\frac{\sum_i\eta_i-1}{b^2}\right)&\sim \left(e^{-\pi i\frac{1-\sum_i\eta_i}{b^2}}-e^{\pi i\frac{1-\sum_i\eta_i}{b^2}}\right)\\
    &\times\exp\left[-\frac{2}{b^2}\left(1-\sum_i\eta_i\right)\left(\log\left(1-\sum_i\eta_i\right)+\log \frac{1}{b^2}-\pi i-1\right)\right] \, .
\end{aligned}\end{align}
The $\pi i$ factor, which comes from $\log(-1/b^2)=\log(1/b^2)-\pi i$, is cancelled with 
\begin{align}
    \left(\frac{1}{b^4}\right)^{(1-\sum_i\eta_i)/b^2}=\exp\left(\frac{1-\sum_i\eta_i}{b^2}\log\frac{1}{b^4}\right)=\exp\left[\frac{2(1-\sum_i\eta_i)}{b^2}\left(\log\frac{1}{b^2}-\pi i\right)\right] \, .
\end{align}
Therefore, by using the asymptotic formula \eqref{upsasymp}, we find
\begin{align}\begin{aligned}
    C(\alpha_1,\alpha_2,\alpha_3)\sim &\left(e^{-\pi i\frac{1-\sum_i\eta_i}{b^2}}-e^{\pi i\frac{1-\sum_i\eta_i}{b^2}}\right)\lambda^{(1-\sum_i\eta_i)/b^2}\\
    &\times\exp\Bigg[\frac{1}{b^2}\Bigg\{F(2\eta_1)+F(2\eta_2)+F(2\eta_3)+F(0)-F\left(\sum_i\eta_i\right)\\
    &-F(\eta_1+\eta_2-\eta_3)-F(\eta_2+\eta_3-\eta_1)-F(\eta_3+\eta_1-\eta_2)\\
    &+2\left(1-\sum_i\eta_i\right)\log\left(1-\sum_i\eta_i\right)-2\left(1-\sum_i\eta_i\right)\Bigg\}\Bigg] \, .
\end{aligned}\end{align}
We can see that the two saddles $N=-1,0$ in eq. (4.31) of \cite{Harlow:2011ny} contribute to the DOZZ formula \eqref{DOZZ}. Thus the absolute value of the three-point function goes to
\begin{align}
    |\langle V_{\alpha_1}(z_1,\bar{z}_1)V_{\alpha_2}(z_2,\bar{z}_2)V_{\alpha_3}(z_3,\bar{z}_3) \rangle|^2\sim \exp\left[\frac{\pi c^{(g)}}{3}\left(1-\sum_i\eta_i\right)\right]\, .
\end{align}

Let us consider the bulk configurations dual to three-point functions. 
The three insertions are parametrized by $\eta_i$ with $i=1,2,3$ as above. We are interested in type II region: $0 \leq \eta_i \leq 1/2$, $\sum_i \eta_i < 1$ and $0 < \eta_i + \eta_j -\eta_k$. The operator insertion with $\eta_i$ is known to create a conical deficit with deficit angle $4 \pi \eta_i$. The way to create three conical deficits on $S^2$ is summarized in subsection 4.2 of \cite{Harlow:2011ny}. In particular, we can see that the area of the geometry is $(1- \sum_i \eta_i)$ times that of $S^2$. The dual geometry may be constructed with the metric of the form:
\begin{align}\label{dualgeo}
    ds^2 = d \theta^2 + \cos ^2 \theta d s^2_\text{con} \, , 
\end{align}
where $ds^2_\text{con}$ is the metric of $S^2$ with three conical deficits.%
\footnote{It would be nice if we can construct the metric $ds^2_\text{con}$ explicitly for any number of conical defects, see, e.g., \cite{Frolov:2001uf,10.1215/ijm/1255984954}.}
From the construction, we can see that the volume of the geometry is $(1 - \sum_i \eta_i)$ times that of $S^3$. This reproduces the result from Liouville three-point function.

\begin{figure}
  \centering
    \begin{tikzpicture}[thick]
        \begin{scope}
            \draw (0,0) circle (2);
            \draw (-2,0) arc (180:360:2 and 0.5);
            \draw[dotted] (2,0) arc (0:180:2 and 0.5);
            \draw[draw=none,fill=lightgray,opacity=0.5] (0,0) circle (2 and 0.5);
            \begin{scope}
                \draw[dashed,BrickRed] (0,2)--(0,-2);
                \draw[dashed,BrickRed] (0,2) arc (150:210:4 and 4);
                \draw[dashed,BrickRed] (0,2) arc (150:210:8 and 4);
                \draw[dashed,BrickRed] (0,2) arc (30:-30:4 and 4);
                \draw[dashed,BrickRed] (0,2) arc (30:-30:8 and 4);
            \end{scope}
            \begin{scope}
                \draw[fill=black,BrickRed] (0,0) circle (0.04);
                \draw[fill=black,BrickRed] (0.54,0) circle (0.04);
                \draw[fill=black,BrickRed] (1.07,0) circle (0.04);
                \draw[fill=black,BrickRed] (0,0) circle (0.04);
                \draw[fill=black,BrickRed] (-0.54,0) circle (0.04);
                \draw[fill=black,BrickRed] (-1.07,0) circle (0.04);
            \end{scope}
        \end{scope}
        \node at (1.5,0.8) {$S^3$};
    \end{tikzpicture}
 \caption{Utilizing the topological nature of Chern-Simons gravity, conical defect lines are set such as to start from (end on) the north (south) pole of $S^3$.}
  \label{fig:conical}
\end{figure}
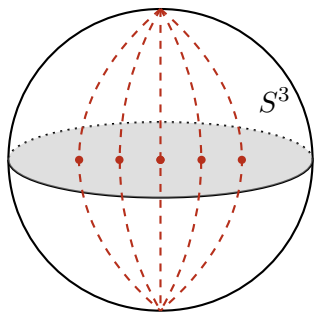

Let us discuss generic cases with $s$ insertions of vertex operators on $S^2$ in the Liouville field theory. The dual geometry in the Chern-Simons theory may be given by $S^3$ with $s$ defect lines.
The action of Chern-Simons theory is given by \eqref{CSactiondS}.
Precisely speaking, we need to add the boundary term 
\begin{align} \label{CSboundary}
\begin{aligned}
&S = S_\text{CS} [A] - S_\text{CS} [\tilde A] + \frac{ \kappa}{4 \pi} \int_{\partial \mathcal{M}} \text{Tr} (A \wedge \tilde A) \, ,
\end{aligned}
\end{align}
where $\partial \mathcal{M}$ represents the boundary of the base manifold $\mathcal{M}$.
Using the topological nature of Chern-Simons theory, we may put the starting (ending) points of the defect lines to the north (south) pole of $S^3$, see fig.~\ref{fig:conical}.
For solutions to the equations of motion, only non-trivial contributions come from the boundary (defect) of the manifold and the topologically non-trivial configuration. For one insertion of defect line, the value of action is shifted by
\begin{align}
       \frac{\pi}{3} c^{(g)} m_i \eta_i
\end{align}
with integer $m_i$.
Adding the topological contribution with the bulk winding number $n$, we have 
\begin{align} 
      \frac{\pi}{3} c^{(g)} (n - \sum_{i=1}^s m_i \eta_i) \, .  \label{bulkmono}
\end{align}
See section 6 of \cite{Harlow:2011ny} for details.
The results from Liouville three-point functions may be reproduced from setting
\begin{align}
n =  m_1 =  m_2 =  m_3 = \pm 1 \, .
\end{align}
The leading contribution comes from the case with $+1$.

\subsection{Four-point functions}
\label{Sec:LiouvilleHigher}
In this subsection, we consider the following generic scalar four-point function in Liouville field theory:
\begin{align}\label{4pt}
    G(z,\bar{z})\equiv \left\langle V_1(0)V_2(z,\bar{z})V_3(1)V_4(\infty)\right\rangle \, ,
\end{align}
where $V_i\ (i=1,2,3,4)$ have the momenta $\alpha_i=\eta_i/b$ and the conformal weights $h_i$. From the Seiberg bound, every $\eta_i$ satisfies $0<\eta_i<1/2$. If they satisfy $\Re \alpha_1+\Re\alpha_2>\Re( Q/2)$ and $\Re \alpha_3+\Re\alpha_4>\Re( Q/2)$, the four-point function can be decomposed as 
\begin{align}\label{cbdecom}
    G(z,\bar{z})=\frac{1}{2}\int_{-\infty}^\infty \frac{dP}{2\pi} C\left(\alpha_1,\alpha_2,\frac{Q}{2}-iP\right)C\left(\alpha_3,\alpha_4,\frac{Q}{2}+iP\right)\CF^{12}_{34}(h_P|z)\CF^{12}_{34}(h_P|\bar{z})
\end{align}
by using the three-point coefficient $C(\alpha_1,\alpha_2,\alpha_3)$ of the form \eqref{DOZZ} and the conformal block $\CF^{12}_{34}(h|z)$. 

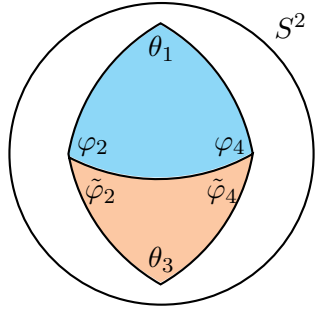
\begin{figure}
    \centering
    \begin{tikzpicture}[thick]
        \begin{scope}
            \draw[draw=none,fill=Orange!40](1.2,0) arc (300:242:2.5 and 2.5) arc (190:240:2.5 and 2.4) arc (300:350:2.5 and 2.4);
            \draw[draw=none,fill=Cyan!40,yshift=-0.02cm] (1.2,0) arc (10:60:2.5 and 2.5) arc (120:170:2.5 and 2.5) arc (242:300:2.5 and 2.5);
            \draw (0,0) circle (2);
            \draw (1.2,0) arc (10:60:2.5 and 2.5) arc (120:170:2.5 and 2.5) arc (190:240:2.5 and 2.5) arc (300:350:2.5 and 2.5) arc (300:242:2.5 and 2.5);
        \end{scope}
        \begin{scope}
            \node at (0,1.4) {$\theta_1$};
            \node at (0,-1.4) {$\theta_3$};
            \node at (-0.9,0.1) {$\varphi_2$};
            \node at (-0.8,-0.5) {$\tilde{\varphi}_2$};
            \node at (0.9,0.1) {$\varphi_4$};
            \node at (0.8,-0.5) {$\tilde{\varphi}_4$};
        \end{scope}
        \node at (1.7,1.7) {$S^2$};
    \end{tikzpicture}
    \caption{$S^2$ with four conical deficits can be embedded in $S^2$ as a spherical quadrangle with angles $\theta_i\ (i=1,2,3,4)$. This picture shows one way of triangulation by a geodesic that splits $\theta_2$ and $\theta_4$ into $\theta_2=\varphi_2+\tilde{\varphi}_2$ and $\theta_4=\varphi_4+\tilde{\varphi}_4$. We can also consider another way of triangulation by splitting $\theta_1$ and $\theta_3$.}
    \label{fig:spherical}
\end{figure}

We are interested in the insertions of four heavy operators $\alpha_i=\eta_i/b$ with real $\eta_i$ such that the four-point function \eqref{4pt} has a dual geometry with four conical deficits.
Let us discuss the condition on $\eta_i$ in order to exist such a dual geometry. 
The dual geometry is given in the form of \eqref{dualgeo}, where $ds_{\text{con}}^2$ is the metric of $S^2$ with four conical deficits in this case.
The sphere with conical deficits are embedded to a spherical quadrangle in a unit two-sphere.
Assuming there exists a spherical quadrangle with four angles $\theta_i=\pi(1-2\eta_i)\ (i=1,2,3,4)$. The spherical quadrangle can be split into two spherical triangles in the same way as the ordinary triangulation. There are multiple ways to triangulate. Here we triangulate it by a geodesic that splits $\theta_2$ and $\theta_4$, then we get two spherical triangles with angles $\theta_1,\varphi_2,\varphi_4$ and $\tilde{\varphi}_2,\tilde{\varphi}_4,\theta_3$, where $\varphi_2+\tilde{\varphi}_2=\theta_2$ and $\varphi_4+\tilde{\varphi}_4=\theta_4$, see fig.~\ref{fig:spherical}. Each spherical triangle has to satisfy the same condition as \eqref{regionII}, therefore 
\begin{align}
    \theta_1+\varphi_2+\varphi_4&>\pi \, ,&
    \theta_1+\varphi_2-\varphi_4&<\pi \, ,&
    \varphi_2+\varphi_4-\theta_1&<\pi \, ,& 
    \varphi_4+\theta_1-\varphi_2&<\pi \, , \\
    \theta_3+\tilde{\varphi}_2+\tilde{\varphi}_4&>\pi\, ,&
    \theta_3+\tilde{\varphi}_2-\tilde{\varphi}_4&<\pi \, ,&
    \tilde{\varphi}_2+\tilde{\varphi}_4-\theta_3&<\pi \, ,& 
    \tilde{\varphi}_4+\theta_3-\tilde{\varphi}_2&<\pi \, .
\end{align}
Combining the two inequalities in each column, and repeating the same analysis for all ways of triangulation, we then obtain necessary conditions
\begin{align}\begin{aligned}\label{4ptcond}
    \sum_i\eta_i&<1 \, ,\\
    0<\eta_i+\eta_j+\eta_k-\eta_l&<1 \, ,\quad (i\neq j\neq k\neq l) \, ,\\
    -1<\eta_i+\eta_j-\eta_k-\eta_l&<1 \, ,\quad (i\neq j\neq k\neq l) \, .
\end{aligned}\end{align}
In particular, due to the first condition of \eqref{4ptcond}, either $\eta_1+\eta_2>1/2$ or $\eta_3+\eta_4>1/2$ is broken. Hence $\alpha_i$'s do not satisfy the condition for the conformal block decomposition \eqref{cbdecom} to hold. Therefore we have to be careful for evaluating the four-point function. 

First we consider the case where the operators satisfy $\eta_1+\eta_2>1/2$ and $\eta_3+\eta_4<1/2$ for simplicity. 
The calculation for this case is similar to that done in section 5 of \cite{Balasubramanian:2017fan}, although the authors of the paper regard $\eta_3$ and $\eta_4$ as perturbatively small. 
When $\eta_3+\eta_4<1/2$, some poles of the DOZZ coefficient $C(\alpha_3,\alpha_4,Q/2+iP)$ cross the integral contour along the real axis in $P$-plane. In that case, the conformal block decomposition has contributions from the poles%
\footnote{In this paper, we focus on the contribution from the poles. We are currently examining whether the contribution from the integration over $P$ can be interpreted as semi-classical saddles or not. See also arguments at the end of section 5 of \cite{Balasubramanian:2017fan}.}
\begin{align}
\begin{aligned}
    G(z,\bar{z})=&\ i\sum_{\text{poles crossing $\mathbb{R}$}} C\left(\alpha_1,\alpha_2,\frac{Q}{2}-iP\right)\Res C\left(\alpha_3,\alpha_4,\frac{Q}{2}+iP\right)\CF^{12}_{34}(h_{P}|z)\CF^{12}_{34}(h_{P}|\bar{z})\\
    &+\frac{1}{2}\int_{-\infty}^\infty \frac{dP}{2\pi} C\left(\alpha_1,\alpha_2,\frac{Q}{2}-iP\right)C\left(\alpha_3,\alpha_4,\frac{Q}{2}+iP\right)\CF^{12}_{34}(h_P|z)\CF^{12}_{34}(h_P|\bar{z}) \, .
\end{aligned}
\end{align}

Let us analyze the pole structure of $C(\alpha_3,\alpha_4,Q/2+iP)$. The zero-point structure of $\Upsilon_b(x)$ was analyzed in \cite{Zamolodchikov:2005fy} (see also \cite{Ribault:2014hia}), where it was shown that the simple zeros of $\Upsilon_b(x)$ are at 
\begin{align}
    x\in-b\mathbb{N}-\frac{1}{b}\mathbb{N}\quad \text{or}\quad x\in Q+b\mathbb{N}+\frac{1}{b}\mathbb{N}
\end{align}
($\mathbb{N}$ includes $0$) and it has no poles. 
We can deduce that the factors in $C(\alpha_3,\alpha_4,Q/2+iP)$ that produce the poles are $\Upsilon_b(\alpha_3+\alpha_4+iP-Q/2)$ and $\Upsilon_b(\alpha_3+\alpha_4-iP-Q/2)$ in the denominator,
whose zero points are given by 
\begin{align}\label{pole1}
    \pm iP&=-\alpha_3-\alpha_4+\frac{Q}{2}-mb-\frac{n}{b} \, ,\\
    \pm iP&=-\alpha_3-\alpha_4+\frac{Q}{2}+(\tilde{m}+1)b+\frac{\tilde{n}+1}{b} \, ,
\end{align}
with non-negative integers $m,n,\tilde{m},\tilde{n}$. 
Since we are considering the range of $\Re\alpha_3+\Re\alpha_4<\Re (Q/2)$, taking the semi-classical limit $b\to0$, the poles that cross the real axis are \cite{Balasubramanian:2017fan}
\begin{align}
    \pm iP_m\equiv-\alpha_3-\alpha_4+\frac{Q}{2}-mb \, ,\quad m=0,1,\ldots \, , 
\end{align}
i.e. only \eqref{pole1} with $n=0$ contribute to the conformal block decomposition. Therefore we have
\begin{align}
    &G(z,\bar{z}) \nonumber \\
    &= i\sum_{m=0}^\infty C\left(\alpha_1,\alpha_2,Q-\alpha_3-\alpha_4-mb\right)\Res C\left(\alpha_3,\alpha_4,\alpha_3+\alpha_4+mb\right)\CF^{12}_{34}(h_{P_m}|z)\CF^{12}_{34}(h_{P_m}|\bar{z})\nonumber \\
    &\quad +\frac{1}{2}\int_{-\infty}^\infty \frac{dP}{2\pi} C\left(\alpha_1,\alpha_2,\frac{Q}{2}-iP\right)C\left(\alpha_3,\alpha_4,\frac{Q}{2}+iP\right)\CF^{12}_{34}(h_P|z)\CF^{12}_{34}(h_P|\bar{z}) \, .
\end{align}

Let us denote the term that comes from the residue at $\alpha_m=\alpha_3+\alpha_4+mb$ as 
\begin{align}
    a_m\equiv iC\left(\alpha_1,\alpha_2,Q-\alpha_3-\alpha_4-mb\right)\Res C\left(\alpha_3,\alpha_4,\alpha_3+\alpha_4+mb\right)\CF^{12}_{34}(h_{P_m}|z)\CF^{12}_{34}(h_{P_m}|\bar{z}) \, .
\end{align}
First we consider the semi-classical limit of the $m=0$ residue $a_0$. Since $\Upsilon_b(x)\simeq \Upsilon_b'(0)x$ for $x\ll1$,  we have
\begin{align}\begin{aligned}\label{res0}
    &iC(\alpha_1,\alpha_2,Q-\alpha_3-\alpha_4)\Res C\left(\alpha_3,\alpha_4,\alpha_3+\alpha_4\right)\simeq\left(\lambda\gamma(b^2)b^{-2b^2}\right)^{\frac{1}{b^2}(1-\sum_i\eta_i)}\\
    &\times\frac{\Upsilon'_b(0)\Upsilon_b(2\alpha_1)\Upsilon_b(2\alpha_2)\Upsilon_b(2\alpha_3+2\alpha_4)}{\Upsilon_b(\sum_i\alpha_i-Q)\Upsilon_b(\alpha_1+\alpha_2-\alpha_3-\alpha_4)\Upsilon_b(Q+\alpha_1-\alpha_2-\alpha_3-\alpha_4)\Upsilon_b(Q-\alpha_1+\alpha_2-\alpha_3-\alpha_4)} \, .
\end{aligned}\end{align}
From Seiberg bound \eqref{Seiberg}, the conditions for the existence of the dual geometry \eqref{4ptcond}, and the current assumption $\eta_1+\eta_2>1/2,\ \eta_3+\eta_4<1/2$, we can see that the arguments of all Upsilon functions but $\Upsilon_b(\sum_i\alpha_i-Q)$ are in the range $0<\Re x<\Re Q$, which is the condition that the asymptotic formula \eqref{upsasymp} can be used. Due to $\Re(\sum_i\alpha_i-Q)<0$ from the first condition of \eqref{4ptcond}, we have to use the defining recursion relation \eqref{defups}:
\begin{align}
    \Upsilon_b\left(\frac{\sum_i\eta_i-1}{b}\right)=\gamma\left(\frac{\sum_i\eta_i-1}{b^2}\right)^{-1}b^{1+\frac{2}{b^2}(1-\sum_i\eta_i)}\Upsilon_b\left(\frac{\sum_i\eta_i}{b}\right)\, ,
\end{align}
then the residue \eqref{res0} becomes 
\begin{align}\begin{aligned}
    &iC(\alpha_1,\alpha_2,Q-\alpha_3-\alpha_4)\Res C\left(\alpha_3,\alpha_4,\alpha_3+\alpha_4\right)\sim\left(\lambda b^{-4}\right)^{\frac{1}{b^2}(1-\sum_i\eta_i)}\gamma\left(\frac{\sum_i\eta_i-1}{b^2}\right)\\
    &\times \frac{\Upsilon'_b(0)\Upsilon_b(2\alpha_1)\Upsilon_b(2\alpha_2)\Upsilon_b(2\alpha_3+2\alpha_4)}{\Upsilon_b(\sum_i\alpha_i)\Upsilon_b(\alpha_1+\alpha_2-\alpha_3-\alpha_4)\Upsilon_b(Q+\alpha_1-\alpha_2-\alpha_3-\alpha_4)\Upsilon_b(Q-\alpha_1+\alpha_2-\alpha_3-\alpha_4)} \, .
\end{aligned}\end{align}
Due to $\Re b^2<0$, we have
\begin{align}\begin{aligned}
    \gamma\left(\frac{\sum_i\eta_i-1}{b^2}\right)\sim &\left(e^{-i\pi(1-\sum_i\eta_i)/b^2}-e^{i\pi(1-\sum_i\eta_i)/b^2}\right)\\
    &\times\exp\left[-\frac{2}{b^2}\left(1-\sum_i\eta_i\right)\left(\ln \left(1-\sum_i\eta_i\right)+\ln \frac{1}{b^2}-\pi i-1\right)\right] \, .
\end{aligned}\end{align}
The $\pi i$ factor is cancelled with 
\begin{align}
    \left(\frac{1}{b^4}\right)^{\frac{1}{b^2}(1-\sum_i\eta_i)}=\exp\left[\frac{1}{b^2}\left(1-\sum_i\eta_i\right)\log\left(\frac{1}{b^4}\right)\right]=\exp\left[\frac{2}{b^2}\left(1-\sum_i\eta_i\right)\log\left(\frac{1}{b^2}-\pi i\right)\right] \, .
\end{align}
By using the formula for the semi-classical limit of the $\Upsilon_b$ function \eqref{upsasymp}, we finally obtain
\begin{align}
    a_0&\sim \left(e^{-i\pi(1-\sum_i\eta_i)/b^2}-e^{i\pi(1-\sum_i\eta_i)/b^2}\right)\lambda^{(1-\sum_i\eta_i)/b^2} \nonumber \\
    &\times\exp\Bigg[\frac{1}{b^2}\Bigg\{F(2\eta_1)+F(2\eta_2)+F(2\eta_3+2\eta_4)+F(0)-F(\eta_1+\eta_2+\eta_3+\eta_4)\\
    &\quad-F(\eta_1+\eta_2-\eta_3-\eta_4)-F(1+\eta_1-\eta_2-\eta_3-\eta_4)-F(1-\eta_1+\eta_2-\eta_3-\eta_4) \nonumber \\
    &\quad-2\left(1-\sum_i\eta_i\right)\ln\left(1-\sum_i\eta_i\right)+2\left(1-\sum_i\eta_i\right)\Bigg\}\Bigg]\mathcal{F}^{12}_{34}(\alpha_3+\alpha_4|z)\mathcal{F}^{12}_{34}(\alpha_3+\alpha_4|\bar{z}) \, . \nonumber 
\end{align}
Next let us consider all terms $a_m$.
Note that the semi-classical conformal blocks behave as \cite{Zamolodchikov:426555} (see \cite{Besken:2019jyw} for a proof in that case where $c,h_i,h_P$ are all real)
\begin{align}\label{blockas}
    \CF^{12}_{34}(h_P|z)\sim \exp \left[ - \frac{c}{6} f \left (\frac{h_i}{c} , \frac{h_P}{c},z \right )  \right] \, .
\end{align}
Since $c$ is pure imaginary as in \eqref{cg} and the ratios $h_i/c, h_P/c$ are real at the leading order in $1/c$, we can regard the semi-classical conformal block as a pure phase.
In the semi-classical limit $b\to 0$, a ratio $a_{m+1}/a_m$ behaves as
\begin{align}
    \frac{a_{m+1}}{a_m}\sim \frac{A}{(m+1)b^2} \, , 
\end{align}
where $A$ is a real constant given by
\begin{align}
    A=-\frac{\gamma(2\eta_3+2\eta_4)^2\gamma(2\eta_3+2\eta_4-1)^2\gamma(\eta_1+\eta_2-\eta_3-\eta_4)}{\gamma(2\eta_3)\gamma(2\eta_4)\gamma(2\eta_3+2\eta_4-1)\gamma(-\eta_1+\eta_2+\eta_3+\eta_4)\gamma(\eta_1-\eta_2+\eta_3+\eta_4)\gamma(\sum_i\eta_i-1)}
\end{align}
by using the recursion relations \eqref{defups} and the asymptotic behavior of $\gamma(x)\sim x^{-2}$ for $x\ll1$. Also note that $\gamma(1-x)=\gamma(x)^{-1}$.
Each residue is then 
\begin{align}
    a_m&\simeq \frac{1}{m!}\left(\frac{A}{b^2}\right)^ma_0 \, .
\end{align}
Therefore the semi-classical limit of the sum of the discrete terms is 
\begin{align}
    \sum_{m=0}^\infty a_m\simeq \exp\left(\frac{A}{b^2}\right)a_0 \, .
\end{align}
Thus we finally obtain the semi-classical limit of the four-point function:
\begin{align}\begin{aligned}
    G(z,&\bar{z})\sim \left(e^{-i\pi(1-\sum_i\eta_i)/b^2}-e^{i\pi(1-\sum_i\eta_i)/b^2}\right)\lambda^{(1-\sum_i\eta_i)/b^2}e^{A/b^2}\\
    &\times\exp\Bigg[\frac{1}{b^2}\Bigg\{F(2\eta_1)+F(2\eta_2)+F(2\eta_3+2\eta_4)+F(0)-F(\eta_1+\eta_2+\eta_3+\eta_4)\\
    &\quad-F(\eta_1+\eta_2-\eta_3-\eta_4)-F(1+\eta_1-\eta_2-\eta_3-\eta_4)-F(1-\eta_1+\eta_2-\eta_3-\eta_4)\\
    &\quad-2\left(1-\sum_i\eta_i\right)\ln\left(1-\sum_i\eta_i\right)+2\left(1-\sum_i\eta_i\right)\Bigg\}\Bigg]\mathcal{F}^{12}_{34}(\alpha_3+\alpha_4|z)\mathcal{F}^{12}_{34}(\alpha_3+\alpha_4|\bar{z}) \, .
\end{aligned}\end{align}
We can see that two saddle points contribute. Since the conformal weights are purely imaginary in $b\sim0$, the conformal block \eqref{blockas} just gives a phase. Therefore in the semi-classical limit $b^{-2}\sim ic^{(g)}/6$, the absolute value of the four-point function is
\begin{align}\begin{aligned}
    |G(z,\bar{z})|&\sim \exp\left[\frac{\pi c^{(g)}}{6}\left(1-\sum_i\eta_i\right)\right]-\exp\left[\frac{-\pi c^{(g)}}{6}\left(1-\sum_i\eta_i\right)\right]\\
    &\sim\sinh\left[\frac{\pi c^{(g)}}{6}\left(1-\sum_i\eta_i\right)\right] \, .
\end{aligned}\end{align}

Even when $\eta_1+\eta_2<1/2$ and $\eta_3+\eta_4<1/2$, the calculation is essentially same to the above case. In this case both two DOZZ coefficients have poles, but the leading contribution comes from the ``minimal'' pole among $\alpha_1+\alpha_2$ and $\alpha_3+\alpha_4$. If $\eta_1+\eta_2>\eta_3+\eta_4$, then the minimal pole is at $\alpha_3+\alpha_4$ instead. Finally we have the same result as the above case.

\subsection{Higher-point functions}\label{app:5pt}

We can extend the above calculations to higher-point functions. For simplicity, we only consider five-point functions
\begin{align}
    \langle V_1(0)V_2(z_2,\bar{z}_2)V_3(z_3,\bar{z}_3)V_4(1)V_5(\infty)\rangle
\end{align}
for Liouville momenta $ \alpha_i=\eta_i/b$.
The extension to general correlation functions is straightforward. 

First of all, we would like to identify the regions of $\eta_i$ from the condition for the dual geometry to exist.
The dual geometry is expected to be $S^3$ with five conical defects with deficit angles $4\pi\eta_i$ with \eqref{dualgeo}, each section of which can be mapped to a spherical pentagon with five angles $\theta_i=\pi(1-2\eta_i)$ in $S^2$. A spherical pentagon can be split into a spherical triangle and a spherical quadrangle. Applying conditions \eqref{regionII} and \eqref{4ptcond} to each way of splitting, we finally obtain necessary conditions
\begin{align}\begin{aligned}\label{5ptregion}
    \sum_i\eta_i&<1\, ,\\
    0<\eta_i+\eta_j+\eta_k+\eta_l-\eta_m&<1\, ,\quad (i\neq j\neq k\neq l\neq m) \, ,\\
    -1<\eta_i+\eta_j+\eta_k-\eta_l-\eta_m&<1\, ,\quad (i\neq j\neq k\neq l\neq m) \, .
\end{aligned}\end{align}
Furthermore, we assume $\eta_1\ge\eta_2\ge\eta_3\ge\eta_4\ge\eta_5$ without loss of generality. 

Let us perform the conformal-block expansion in a channel 
\begin{align}
    \CF^{134}_{25}(h_{P_1},h_{P_2}|z_2,z_3)=\quad
    \begin{tikzpicture}[thick,baseline=-0.1]
        \begin{scope}
            \draw (-1,0)--(1,0);
            \draw (-1,0)--(-1.5,0.5);
            \draw (-1,0)--(-1.5,-0.5);
            \draw (0,0)--(0,0.5);
            \draw (1,0)--(1.5,0.5);
            \draw (1,0)--(1.5,-0.5);
        \end{scope}
        \begin{scope}
            \node at (-1.7,0.5) {$1$};
            \node at (-1.7,-0.5) {$2$};
            \node at (0,0.7) {$3$};
            \node at (1.7,0.5) {$4$};
            \node at (1.7,-0.5) {$5$};
            \node at (-0.5,-0.3) {$P_1$};
            \node at (0.5,-0.3) {$P_2$};
        \end{scope}
    \end{tikzpicture} \, .
\end{align}
If $\eta_i$ satisfied $\eta_1+\eta_2\ge1/2$ and $\eta_3+\eta_4\ge1/2$, the conformal block decomposition have the form 
\begin{align}\begin{aligned}
    &\langle V_1(0)V_2(z_2,\bar{z}_2)V_3(z_3,\bar{z}_3)V_4(1)V_5(\infty)\rangle\\
    &=\frac{1}{2}\int_{\mathbb{R}} \frac{dP_1}{2\pi}\ \frac{1}{2}\int_{\mathbb{R}}\frac{dP_2}{2\pi}\,C( \alpha_1, \alpha_2,\alpha_{P_1})C(Q-\alpha_{P_1},\alpha_3,\alpha_{P_2})C(Q-\alpha_{P_2},\alpha_4,\alpha_5)\\
     & \quad \times \CF^{134}_{25}(h_{P_1},h_{P_2}|z_2,z_3)\CF^{134}_{25}(h_{P_1},h_{P_2}|\bar{z}_2,\bar{z}_3)\, .
\end{aligned}\end{align}
However this does not hold for $\eta_i$'s in the region \eqref{5ptregion}, so we have to discuss the analytic continuation in $\eta_i$. Let us assume $\eta_1+\eta_2\ge1/2,\ \eta_3+\eta_4+\eta_5\le1/2$. First, the integral contour in $P_1$-plane picks up the poles 
\begin{align}
    \alpha_{P_2}= \alpha_4+\alpha_5+mb \, , \quad m=0,1,\ldots
\end{align} 
of $C(Q-\alpha_{P_2},\alpha_4,\alpha_5)$. As discussed in section \ref{Sec:LiouvilleHigher}, the contribution from summing over $m$ finally gives just a constant phase. Therefore we only focus on the contribution from the $m=0$ pole 
\begin{align}
    \alpha_{P_2}=\alpha_4+\alpha_5 \, .
\end{align}
The residue for this pole is 
\begin{align}
    \Res C(Q-\alpha_4-\alpha_5,\alpha_4,\alpha_5)=1 \, .
\end{align}
Next, the integral contour of $P_1$ picks up a pole 
\begin{align}
    \alpha_{P_1}=\alpha_3+\alpha_{P_2}=\alpha_3+\alpha_4+\alpha_5
\end{align}
of $C(\alpha_{P_1},\alpha_3,Q-\alpha_4-\alpha_5)$. The residue for this pole is again 
\begin{align}
    \Res C(\alpha_3+\alpha_4+\alpha_5,\alpha_3,Q-\alpha_4-\alpha_5)=1 \, .
\end{align}
Therefore the coefficients of the conformal block decomposition have the contributions from the first factor $C(\alpha_1,\alpha_2,\alpha_{P_1})$ at $\alpha_1=\alpha_3+\alpha_4+\alpha_5$:
\begin{align}\begin{aligned}
    &C(\alpha_1,\alpha_2,\alpha_3+\alpha_4+\alpha_5)=\left[\lambda\gamma(b^2)b^{-2b^2}\right]^{\frac{1}{b}(Q-\sum_i\alpha_i)}\\
    &\times\frac{\Upsilon'_b(0)\Upsilon_b(2\alpha_1)\Upsilon_b(2\alpha_2)\Upsilon_b(2\alpha_3+2\alpha_4+2\alpha_5)}{\Upsilon_b(\sum_i\alpha_i-Q)\Upsilon_b(\alpha_1+\alpha_2-\alpha_3-\alpha_4-\alpha_5)\Upsilon_b(\alpha_1-\alpha_2+\alpha_3+\alpha_4+\alpha_5)\Upsilon_b(-\alpha_1+\alpha_2+\alpha_3+\alpha_4+\alpha_5)} \, .
\end{aligned}\end{align}
From the conditions \eqref{5ptregion} and the assumption $\eta_1+\eta_2\ge1/2,\ \eta_3+\eta_4+\eta_5\le1/2$, we can see that only the factor $\Upsilon_b(\sum_i\alpha_i-Q)$ produces the $\gamma$ function in the same way as four-point functions. Thus finally we have 
\begin{align}\begin{aligned}
    &C(\alpha_1,\alpha_2,\alpha_3+\alpha_4+\alpha_5)\sim \left(e^{-i\pi(1-\sum_i\eta_i)/b^2}-e^{i\pi(1-\sum_i\eta_i)/b^2}\right)\lambda^{(1-\sum_i\eta_i)/b^2}\\
    &\times\exp\Bigg[\frac{1}{b^2}\Bigg\{F(2\eta_1)+F(2\eta_2)+F(2\eta_3+2\eta_4+2\eta_5)+F(0)-F(\eta_1+\eta_2+\eta_3+\eta_4+\eta_5)\\
    &\quad-F(\eta_1+\eta_2-\eta_3-\eta_4-\eta_5)-F(\eta_1-\eta_2+\eta_3+\eta_4+\eta_5)-F(-\eta_1+\eta_2+\eta_3+\eta_4+\eta_5) \\
    &\quad-2\left(1-\sum_i\eta_i\right)\ln\left(1-\sum_i\eta_i\right)+2\left(1-\sum_i\eta_i\right)\Bigg\}\Bigg]\, .
\end{aligned}\end{align}
In the semi-classical limit $b\to0$, it is natural to expect that the five-point conformal block behaves as%
\footnote{See, e.g. \cite{Cho:2017oxl} for a work on higher-point blocks.}
\begin{align}
    \CF^{134}_{25}(h_{P_1},h_{P_2}|z_2,z_3)\sim \exp \left[ - \frac{c}{6} \tilde f \left(\frac{h_i}{c},\frac{h_{P_1}}{c},\frac{h_{P_2}}{c};z_2;z_3 \right)\right] 
\end{align}
as in the case of four-point conformal block in \eqref{blockas}.
Therefore the conformal blocks contribute to just a phase factor. 
Thus finally we obtain the semi-classical approximation
\begin{align}
    |\langle V_1(0)V_2(z_2,\bar{z}_2)V_3(z_3,\bar{z}_3)V_4(1)V_5(\infty)\rangle|\sim e^{\frac{\pi c^{(g)}}{6}(1-\sum_i\eta_i)}-e^{-\frac{\pi c^{(g)}}{6}(1-\sum_i\eta_i)} \, .
\end{align}

In the same way, we can calculate general correlation functions with any number of heavy operator insertions satisfying certain conditions coming from the existence of the dual geometry. Expanding correlation functions by using the linear channel $s$-point conformal block 
\begin{align}
    \CF^{(\text{linear})}(\{h_i\}_{1\le i\le s},\{h_{P_i}\}_{1\le i\le s-3}|\{z_i\}_{2\le i\le s-2})=
    \begin{tikzpicture}[baseline=0.2cm]
        \centering
        \begin{scope}[thick]
            \draw (-2,0)--(2,0);
            \draw (-1.5,0)--(-1.5,0.8);
            \draw (-1,0)--(-1,0.8);
            \draw (-0.5,0)--(-0.5,0.8);
            \draw (1.5,0)--(1.5,0.8);
            \draw (1,0)--(1,0.8);
        \end{scope}
        \begin{scope}
            \node at (-2.2,0) {\small $1$};
            \node at (-1.5,1) {\small $2$};
            \node at (-1,1) {\small $3$};
            \node at (-0.5,1) {\small $4$};
            \node at (1.6,1) {\small $s-1$};
            \node at (2.2,0) {\small $s$};
        \end{scope}
        \begin{scope}
            \node at (-1.25,-0.3) {\small $P_1$};
            \node at (-0.75,-0.3) {\small $P_2$};
            \node at (1.4,-0.3) {\small $P_{s-3}$};
        \end{scope}
        \node at (0.3,0.5) {\large $\cdots$};
    \end{tikzpicture}\ ,
\end{align}
the correlation function would give the same semi-classical approximation as above calculations.

As mentioned at the end of subsection \ref{sec:Liouville3pt}, the saddle points of dual Chern-Simons gravity are classified by bulk winding number $n$ and monodromies around the defects labeled $m_i$ $(i=1,2,\dots,s)$ in case with Liouville $s$-point function.
Since the classical action for the saddle point with labels $n,m_i$ is evaluated as \eqref{bulkmono}, we conclude that the allowable saddles are given with
\begin{align}
n =  m_1 = \cdots = m_s = \pm 1 \, .
\end{align}
The leading contribution comes from the case with $+1$.

\subsection{Four-point functions dual to two bulk Wilson lines}
\label{Sec:Liouville 4pt}

In \cite{Hikida:2021ese,Hikida:2022ltr}, a dS/CFT correspondence was proposed between classical pure dS$_3$ gravity and SU(2) WZW model at the critical level $k \to -2$ or at the large central charge. Applying the CS/WZW correspondence \cite{Witten:1988hf}, the WZW model can be described by SU(2) Chern-Simons theory. In this way, we have a curious relation between dS$_3$ gravity and Chern-Simons theory. We should emphasize here that the relation is not the same as the Chern-Simons description of dS$_3$ gravity. For the Chern-Simons formulation of classical gravity, we take the level to infinity $k \to i \infty$, though the Chern-Simons description of WZW model uses the critical level $k \to -2$. Therefore, the two Chern-Simons descriptions are quite different. It is natural to suspect that they should be related as the triality relation of higher spin symmetry in the dual CFT \cite{Gaberdiel:2012ku}. In any cases, we can construct a geometry corresponding to Chern-Simons theory at the critical level possible with Wilson loop operators. In particular, the gravity solution was identified in \cite{Hikida:2021ese,Hikida:2022ltr} corresponding to the configuration of two linked (unlinked) Wilson loops on S$^3$ in the Chern-Simons theory at the critical level. According to \cite{Witten:1988hf}, the configuration corresponds to a four-point function of SU(2) WZW model at the large central charge. Indeed, the partition function was computed from the both sides of the holographic duality and agreement was obtained.

In \cite{Hikida:2021ese,Hikida:2022ltr}, it was assumed that the expressions for the SU(2) WZW model at an integer level $k$ can be analytically continued to those at a complex $k$. In order to justify this, it was first argued that SU(2) WZW model and the coset
\begin{align}
    \frac{\text{SU}(2)_k \times \text{SU}(2)_1}{\text{SU}(2)_{k+1}} \label{su2coset}
\end{align}
becomes the same at the critical level $k \to -2$ as the SU(2)$_k$ part of the coset dominates at the limit. Moreover, it was shown in \cite{Creutzig:2021ykz} that correlation functions of the coset \eqref{su2coset} and Liouville field theory are the same. Therefore, it should be possible to describe the geometry corresponding to the two linked (unlinked) Wilson loops on $S^3$ in the Chern-Simons theory by a four-point function in Liouville field theory at the semi-classical limit $b \to 0$. Indeed, we show in this subsection that the results obtained from the native analytic continuation of $S$-matrix analysis lead to the correct answers even including sub-leading terms at the semi-classical limit.

We start by relating the CFT computations by four-point functions and those by modular $S$-matrices. In order to illustrate this, it is convenient to work with a rational CFT like SU$(N)_k$ WZW model. We will later consider a non-rational CFT, i.e. Liouville field theory. We examine the following four-point function
\begin{align} \label{Cij}
C_{ij} (z , \bar z ) = \frac{\left \langle \mathcal{O}^\dagger_i (\infty) \mathcal{O}^\dagger_j (1) \mathcal{O}_i (z) \mathcal{O}_j (0) \right \rangle}{\left \langle \mathcal{O}^\dagger_i\mathcal{O}_i \right \rangle \left \langle \mathcal{O}^\dagger_j\mathcal{O}_j \right \rangle} \, ,
\end{align}
where $\mathcal{O}_j$ is a CFT operator and $\mathcal{O}_j^\dagger$ is its conjugate. We define the function $C_{ij}(z, \bar z)$ with a suitable normalization. Inserting a complete set $\mathbbm{1} = \sum_p \ket{p} \bra{p}$, we can decompose the four-point function by the sum of conformal blocks as
\begin{align}
C_{ij} (z , \bar z ) = \sum_{p} \mathcal{F}^{ii}_{jj} (p|z) \bar{\mathcal{F}}^{ii}_{jj} (p|\bar z) \, . 
\end{align}
We are interested in the region with large central charge. Then, we can argue that the identity block with $p=0$ dominates, see \cite{Hartman:2013mia}.
When considering $z \sim 0$, then the function behaves as 
\begin{align}
C_{ij} (z , \bar z ) \sim 1 \, . \label{C1}
\end{align}

\begin{figure}
  \centering
  \begin{tikzpicture}[thick,scale=0.8]
        \begin{scope}[scale=0.7,xshift=-9cm]
            \begin{knot}
                \strand[BrickRed,ultra thick] (-2.5,-0.5) parabola bend(-1,-3) (0.5,0);
                \strand[BrickRed, ultra thick] (-1,0) parabola bend(0.5,-2.5) (2.5,0.5);
            \end{knot}
            \begin{scope}
                \draw (-4,-1)--(2,-1)--(4,1)--(-2,1)--cycle;
                \fill[lightgray,opacity=0.5] (-4,-1)--(2,-1)--(4,1)--(-2,1)--cycle;
                \fill (-2.5,-0.5) circle (0.05) node[left] {0};
                \fill (0.5,0) circle (0.05) node[left] {1};
                \fill (2.5,0.5) circle (0.05) node[right] {$\infty$};
                \fill (-1,0) circle (0.05) node[left] {$z$};
            \end{scope}
            \draw[->,OliveGreen] (-1,0) arc (-180:120:1 and 0.4);
            \node at (-3,0.8) {$S^2$};
        \end{scope}
        \begin{scope}[scale=0.7]
            \begin{knot}[flip crossing=2]
                \strand[BrickRed,ultra thick] (-2.5,-0.5) parabola bend(-1,-3) (0.5,0);
                \strand[BrickRed, ultra thick] (-1,0) ..controls (1,-1) and (1,-1.3) ..(-0.6,-1.8)..controls (-1,-2) and (-1,-2.5) ..(0,-2.5) arc (-90:-30:2.9 and 6) ;
            \end{knot}
            \begin{scope}
                \draw (-4,-1)--(2,-1)--(4,1)--(-2,1)--cycle;
                \fill[lightgray,opacity=0.5] (-4,-1)--(2,-1)--(4,1)--(-2,1)--cycle;
                \fill (-2.5,-0.5) circle (0.05) node[left] {0};
                \fill (0.5,0) circle (0.05) node[left] {1};
                \fill (2.5,0.5) circle (0.05) node[right] {$\infty$};
                \fill (-1,0) circle (0.05) node[left] {$z$};
            \end{scope}
            \node at (-3,0.8) {$S^2$};
        \end{scope}
        \begin{scope}[scale=0.7,xshift=9cm]
            \begin{knot}[flip crossing=2]
                \strand[BrickRed,ultra thick] (-2.5,-0.5) parabola bend(-1,-3) (0.5,0);
                \strand[BrickRed, ultra thick] (-1,0) ..controls (1,-1) and (1,-1.3) ..(-0.6,-1.8)..controls (-1,-2) and (-1,-2.5) ..(0,-2.5) arc (-90:-30:2.9 and 6) ;
            \end{knot}
            \begin{scope}
                \draw (-4,-1)--(2,-1)--(4,1)--(-2,1)--cycle;
                \fill[lightgray,opacity=0.5] (-4,-1)--(2,-1)--(4,1)--(-2,1)--cycle;
                \fill (-2.5,-0.5) circle (0.05) node[left] {0};
                \fill (0.5,0) circle (0.05) node[left] {1};
                \fill (2.5,0.5) circle (0.05) node[right] {$\infty$};
                \fill (-1,0) circle (0.05) node[left] {$z$};
            \end{scope}
            \begin{knot}
                \strand[BrickRed,ultra thick] (-2.5,-0.5) parabola bend(-1,2) (0.5,0);
                \strand[BrickRed, ultra thick] (-1,0) parabola bend(0.5,2.5) (2.5,0.5);
            \end{knot}
            \node at (-3,0.8) {$S^2$};
        \end{scope}
        \draw[->] (-3.7,0)--(-2.7,0);
        \draw[->] (2.7,0)--(3.7,0);
    \end{tikzpicture}
\caption{From the left to middle figures, we move the coordinate $z$ around $1$ anti-clockwise. From the middle to right figures, we glue the holomorphic and anti-holomorphic parts of four-point function. In the Chern-Simons description, there appears two Wilson loops and they are wrapped around each other.}
  \label{fig:monodromy}
\end{figure}
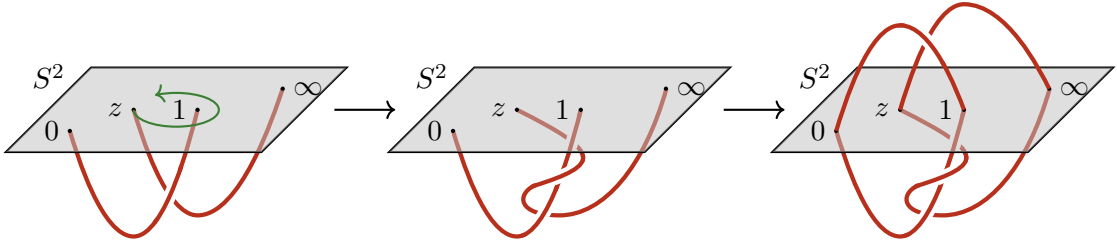

In the Chern-Simons formulation of WZW model, the four-point function \eqref{Cij} describes two unlinked Wilson loops. In order to obtain a configuration with two linked Wilson loops, we perform a move of the holomorphic coordinate $z$ as in fig.~\ref{fig:monodromy}. Namely, we start from $z \sim 0$, go around $z =1$ anti-clockwise, then come back to $z \sim 0$.%
\footnote{This procedure was used to compute out-of-time-order correlators in 2d CFTs \cite{Roberts:2014ifa}.}
This move yields a monodromy matrix as
\begin{align}
\mathcal{F}^{ii}_{jj} (p|z) \to \sum_q \mathcal{M}_{pq} \mathcal{F}^{ii}_{jj} (q|z) \, .
\end{align}
Combining the anti-holomorphic part, where we do not perform the move of $\bar z$, we have
\begin{align}
C_{ij} (z , \bar z) \sim \mathcal{M}_{00} \mathcal{F}^{ii}_{jj} (0|z) \bar{\mathcal{F}}^{ii}_{jj} (0|\bar z)
\end{align}
for $z \sim 0$. The monodromy matrix is \cite{Moore:1988uz,Moore:1988ss}
\begin{align}
\mathcal{M}_{00} = \frac{S^*_{ij} S_{00} S_{00}}{S_{00} S_{0i} S_{0j}} \, . \label{M00}
\end{align}
As in fig.~\ref{fig:monodromy}, we can identify the four-point function as the partition function of Chern-Simons theory with two linked Wilson loops for the Chern-Simons theory, see \cite{Caputa:2016tgt}.
We thus find
\begin{align}
\left |\left \langle \mathcal{O}^\dagger_i (\infty) \mathcal{O}^\dagger_j (1) \mathcal{O}_i (z) \mathcal{O}_j (0) \right \rangle \right| \sim \left|\frac{S_{ij}}{S_{00}}\right| \, ,
\end{align}
which also implies that 
\begin{align} \label{2ptS0j}
\left |\left \langle \mathcal{O}^\dagger_j\mathcal{O}_j \right \rangle \right | \sim \left|\frac{S_{0j}}{S_{00}}\right|
\end{align}
with $i=0$. Here the correlator with the insertion of the identity operator $\mathcal{O}_0 \equiv \mathbbm{1}$ is normalized as one,
\begin{align}
\left \langle \, \mathbbm{1} \, \right \rangle = 1 \, ,
\end{align}
as usual. In \cite{Hikida:2021ese,Hikida:2022ltr}, unnormalized sphere correlators are used by adopting the convention of \cite{Witten:1988hf}.
Specifically, the normalization of partition function,
\begin{align}
\left \langle \, \mathbbm{1} \, \right \rangle = S_{00} \, ,
\end{align}
was used. Taking into account the normalization, we have
\begin{align}
\left |\left \langle \mathcal{O}^\dagger_j\mathcal{O}_j \right \rangle \right | \sim |S_{0j}| \, ,  \label{2pt0j}
\end{align}
and
\begin{align}
\left |\left \langle \mathcal{O}^\dagger_i (\infty) \mathcal{O}^\dagger_j (1) \mathcal{O}_i (z) \mathcal{O}_j (0) \right \rangle \right| \sim |S_{ij}| \, . \label{4ptij}
\end{align}
These results reproduce those in \cite{Hikida:2021ese,Hikida:2022ltr} from the modular $S$-matrix of WZW model.
In case with two unlined Wilson loops, we should have \eqref{C1}.
Using \eqref{2pt0j}, we find
\begin{align}
\left |\left \langle \mathcal{O}^\dagger_i (\infty) \mathcal{O}^\dagger_j (1) \mathcal{O}_i (z) \mathcal{O}_j (0) \right \rangle \right| \sim \left|\frac{S_{0i} S_{0j}}{S_{00}}\right|\, .
\end{align}
Here we should keep track of non-trivial normalization $S_{00}$. This also reproduces a finding in \cite{Hikida:2021ese,Hikida:2022ltr}.

Up to now we have assumed that the dual CFT is a rational one. In the following, we show that the same is true also for the case with Liouville field theory, which is a non-rational CFT. We start from the two-point function. In subsection \ref{sec:Liouville}, we have shown that the Liouville two-point function behaves as \eqref{abs2pt}. The modular $S$-matrix of SU(2) WZW model with level $k$ is given by
\begin{align}
    S_j^{~l} = \sqrt{\frac{2}{k+2}}\sin \left[ \frac{\pi}{k+2} (2 j+1) (2 l+1) \right] \, .
\end{align}
Plugging $k \sim -2 + 6i/c^{(g)}$ and $2 j + 1 \sim \sqrt{1 - 8 G_N E_j}$ into the above expression as in \cite{Hikida:2021ese,Hikida:2022ltr}, we find
\begin{align} \label{Liouville2pt}
|S_{0j}| \sim \left   | e^{\frac{\pi}{6} c^{(g)} \sqrt{1 - 8 G_N E_j} } -  e^{- \frac{\pi}{6} c^{(g)} \sqrt{1 - 8 G_N E_j} } \right | \, .
\end{align}
This reproduces the sub-leading non-perturbative corrections.
We obtain the same conclusion even applying the modular $S$-matrix of Liouville field theory between the identity state $p=0$ and non-degenerate state given by \cite{Zamolodchikov:2001ah}
\begin{align}
S_{0j} = -2\sqrt{2} \sin 2 \pi b (\alpha_j - Q/2) \sin {2\pi}{b^{-1}} (\alpha_j - Q/2) \, . \label{S0jLiou}
\end{align}

We then consider the four-point function corresponding to two linked Wilson loops in the Chern-Simons theory as in fig.~\ref{fig:monodromy}. The modular $S$-matrix of Liouville theory among non-degenerate operators are given by \cite{Zamolodchikov:2001ah}
\begin{align}
 S_{ij} = \sqrt{2} \cos 4 \pi (\alpha_i - Q/2) (\alpha_j - Q/2) \, . \label{SijLiou}
\end{align}
Therefore, \eqref{4ptij} leads to
\begin{align} \label{4ptresult}
  \left| \left \langle \mathcal{O}^\dagger_i(\infty) \mathcal{O}^\dagger_j (1) \mathcal{O}_i (z) \mathcal{O}_j (0) \right \rangle  \right| \sim  \left   | c_1 e^{\frac{\pi}{6} c^{(g)} \sqrt{1 - 8 G_N E_i}\sqrt{1 - 8 G_N E_j} } + c_2 e^{- \frac{\pi}{6} c^{(g)} \sqrt{1 - 8 G_N E_i}\sqrt{1 - 8 G_N E_j} } \right | 
 \end{align}
with real coefficients $c_1,c_2$ both for SU(2) WZW model and Liouville field theory for the large central charge. In the following, we derive \eqref{4ptresult} directly from four-point functions of Liouville field theory, or more precisely speaking, four-point conformal blocks.
We would like to read off the monodromy matrix of four-point conformal blocks around $z =1$. For this, we use the asymptotic behavior of conformal block $ \mathcal{F}^{ii}_{jj} (p|z) $ near $z \sim 1$, which was obtained in  \cite{Fitzpatrick:2016mjq} as
\begin{align} \label{cbsaddle}
  \mathcal{F}^{ii}_{jj} (p|z) \sim \sum_{n,\delta_\kappa = \pm}c_n^{\delta_\kappa}
 (1-z)^{\frac{c}{6} \kappa_n^{\delta_\kappa }}
 \end{align}
with 
\begin{align} \label{kappa}
 \kappa_n^{\delta_\kappa} = n (1 - n) - \frac12 + \left( \frac12 - n \right) ((1 - 2 \eta_i) + \delta_\kappa (1 - 2 \eta_j)) - \delta_\kappa \frac{(1 - 2 \eta_i) (1 - 2 \eta_j)}{2} \, .
\end{align}
We take $\delta_\kappa = \pm$ and $n \in \mathbb{Z}$.
The monodromy can be read off as
\begin{align}
    \mathcal{F}^{ii}_{jj} (p|z) \to \exp \left(\frac{\pi i c}{3} \kappa_n^{\delta_\kappa} \right)\mathcal{F}^{ii}_{jj} (p|z) \, .
\end{align}
The phase does not depend on the intermediate state $p$, but it is anyway set as $p=0$.
Performing the procedure in fig.~\ref{fig:monodromy}, the absolute value of four-point function becomes
\begin{align}
  \left| \left \langle \mathcal{O}^\dagger_i(\infty) \mathcal{O}^\dagger_j (1) \mathcal{O}_i (z) \mathcal{O}_j (0) \right \rangle  \right| \sim  \sum_{\delta_i , \delta_j ,\delta_\kappa = \pm , n }\left| \left \langle \mathcal{O}^\dagger_i\mathcal{O}_i \right \rangle_{\delta_i} \left \langle
  \mathcal{O}^\dagger_j\mathcal{O}_j \right \rangle_{\delta_j} \exp \left( - \frac{\pi c^{(g)} \kappa_{n}^{\delta_\kappa}}{3}  \right) \right| 
  \label{four2cb}
 \end{align}
 up to some coefficients. We already knew that the two-point functions behave as
 \begin{align}
 \left \langle
  \mathcal{O}^\dagger_j\mathcal{O}_j \right \rangle = \sum_{\delta_j = \pm} c_{\delta_j}\left \langle
  \mathcal{O}^\dagger_j\mathcal{O}_j \right \rangle_{\delta_j} \, , \quad \left \langle
  \mathcal{O}^\dagger_j\mathcal{O}_j \right \rangle_{\delta_j}
  \sim \exp \delta_j  \frac{\pi c^{(g)}(1 - 2 \eta_j)}{6}
  \label{kappadelta}
\end{align} 
with real coefficients $c_\pm$. We are now interested in the identity block, which is defined by dividing the two-point functions as in \eqref{Cij}. Therefore, it is natural to think that the terms linear in $(1 - 2 \eta_i)$ and $(1 - 2 \eta_j )$ come from the two-point functions. This means that we should choose $n = (\delta_i+1)/2 = (\delta_j \delta_\kappa + 1)/2$. This choice indeed leads to \eqref{4ptresult}.

\section{Higher-spin extension}
\label{sec:HSE}

With the Chern-Simons description of pure gravity theory, it is straightforward to extend the previous analysis to the case with higher-spin theory described by SL$(N,\mathbb{C})$ Chern-Simons gauge theory. As in the case of pure gravity, we will first review the construction of AdS higher-spin black holes, then perform the relevant analytic continuation to obtain their dS higher-spin counterparts and classify the possible saddles.
We then determine the allowed set of complex saddles from its dual CFT$_2$, i.e. Toda field theory in subsection \ref{sec:Toda}.

\subsection{Chern-Simons higher-spin gravity}
\label{sec:csgra}

In this subsection, we examine dS$_3$ black hole solutions of a higher-spin gauge theory described by $\text{SL}(N, \mathbb{C})$ Chern-Simons theory. For simplicity, we will first focus on the simplest but non-trivial case with $N=3$. We again start from the case with negative cosmological constant by reviewing \cite{Gutperle:2011kf,Ammon:2012wc}. We then move to the case with positive cosmological constant, see \cite{Krishnan:2013zya} for a related work.

\subsubsection{Higher-spin \texorpdfstring{AdS$_3$}{AdS3} black holes}
\label{sec:hssl3bh}

The higher-spin theory with spin $s=2,3$ gauge fields can be described by the action \eqref{CSaction} but now the one-forms $A , \bar A$ take values in $\mathfrak{sl}(3)$. We may introduce the $\mathfrak{sl}(3)$ generators $L_i$ $(i=\pm1 ,0)$ and $W_m$ $(m=\pm2 ,\pm1,0)$ satisfying
\begin{align}
\begin{aligned}
&[L_i , L_j ] = (i -j) L_{i+j } \, , \quad [L_i , W_m] = (2 i - m) W_{i+m} \, ,  \\
&[W_m , W_n] = - \frac13 (m - n) (2 m^2 + 2 n^2 - mn - 8) L_{m+n} \, .
\end{aligned}
\end{align}
We consider the configuration of gauge fields of the form \eqref{gauge} as
\begin{align} \label{gaugeE}
 A = e^{- \rho L_0} a e^{\rho L_0} d \rho \, , \quad 
 \bar A = e^{\rho L_0} \bar a e^{- \rho L_0} - L_0 d \rho 
\end{align}
with
\begin{align}\label{gaugeE2}
 a = a_z dz + a_{\bar z} d\bar z \, , \quad \bar a = \bar a_z dz + \bar a_{\bar z}
 d \bar z \, .
\end{align}
The configuration with Lorentzian configuration is given by replacing $z \to x^+ = t + \phi$ and $\bar z \to - x^- = - t + \phi$ as before.
It was claimed in \cite{Gutperle:2011kf} (see \cite{Ammon:2013hba} for a review) that the gauge configuration,
\begin{align}
\begin{aligned}
a 
&
= \left(L_1 - \frac{2 \pi{ \mathcal L}^\text{AdS}}{k}  L_{-1}- \frac{\pi { \mathcal W}^\text{AdS}}{2 k}  W_{-2}\right) dz  
\\& \quad 
- \mu^\text{AdS} \left(W_2 - \frac{4 \pi { \mathcal L}^\text{AdS}}{k} W_0 + \frac{4 \pi^2 ({  \mathcal L}^{\text{AdS}})^2}{k^2} W_{-2} + \frac{4 \pi \mathcal{W}^\text{AdS}}{k} L_{-1}\right) d \bar z \, ,
\end{aligned} \label{connection}
\end{align}
and
\begin{align}
\begin{aligned}
\bar a &=  \left(L_{-1 }- \frac{2 \pi \bar{\mathcal L}^\text{AdS}}{k}  L_{1} - \frac{\pi \bar{\mathcal W}^\text{AdS}}{2 k}  W_{2}\right) d \bar z \\
& \quad - \bar \mu^\text{AdS} \left(W_{-2} - \frac{4 \pi \bar{ \mathcal L}^\text{AdS}}{k} W_0 + \frac{4 \pi^2 ({\bar{\mathcal L}}^\text{AdS})^2}{k^2} W_{2} + \frac{4 \pi \bar{ \mathcal W}^\text{AdS}}{k} L_{1}\right) d  z \, , 
\end{aligned}
\end{align}
represents a higher-spin black hole geometry.
In the following, we consider the non-rotating case as
\begin{align}
\bar {\mathcal L}^\text{AdS} = {\mathcal L}^\text{AdS} \, , \quad \bar {\mathcal W}^\text{AdS} = - {\mathcal W}^\text{AdS} \, , \quad \bar \mu^\text{AdS} = - \mu^\text{AdS} \, .
\end{align}
The configuration reduces to the case of BTZ black hole in \eqref{solution} if we set $\mathcal{W}^\text{AdS} = \mu^\text{AdS} = 0$.

In the presence of higher-spin gauge symmetry, the definition of black hole is not obvious as mentioned above. In \cite{Gutperle:2011kf}, the authors have declared the conditions of higher-spin black hole as follows; 
(i) The Euclidean geometry is smooth and the spin-three field is non-singular at the horizon. 
(ii) In the limit $\mu^\text{AdS} \to 0$, the solution becomes smoothly the BTZ black hole.
(iii) The charge assignment $\mathcal{L}^\text{AdS} \equiv \mathcal{L}^\text{AdS} (\tau^\text{AdS} , \alpha^\text{AdS})$ and $\mathcal{W}^\text{AdS} \equiv \mathcal{W}^\text{AdS} (\tau^\text{AdS} , \alpha^\text{AdS})$, where $\tau^\text{AdS}$ is inverse temperature as introduced earlier in the AdS-BTZ black hole, while we have also introduced a new parameter here
$\alpha^\text{AdS}\,( \equiv \bar \tau^\text{AdS} \mu^\text{AdS})$, 
such that $\mathcal{L}^\text{AdS}$ and $\mathcal{W}^\text{AdS}$ satisfy the integrability condition:
\begin{align}
\frac{\partial \mathcal{L}^\text{AdS}}{\partial \alpha^\text{AdS}} = \frac{\partial \mathcal{W}^\text{AdS}}{\partial \tau^\text{AdS}} \, . \label{integrability}
\end{align} 
Here we clarify the meaning of the condition (iii) as the others do not seem to need explanations.
Let us consider the partition function
\begin{align}
\begin{aligned}
    Z (\tau^\text{AdS} , \alpha^\text{AdS} , \bar \tau^\text{AdS} , \bar \alpha^\text{AdS})
    & = \text{tr}\, e^{4 \pi^2 i [\tau^\text{AdS} \hat{\mathcal L}^\text{AdS} + \alpha^\text{AdS} \hat{\mathcal W}^\text{AdS} - \bar \tau^\text{AdS} \hat{\bar{\mathcal L}}^\text{AdS} - \bar \alpha^\text{AdS} \hat{\bar {\mathcal W}}^\text{AdS}]} \\
   &  = \text{tr}_\text{CFT}\, q^{L_0 - \frac{c}{24}} u^{W_{0}}  \bar q^{\bar L_0 - \frac{c}{24}} \bar u^{\bar W_{0}}\, .
   \end{aligned}
\end{align}
Here we treat the hatted quantities as operators, i.e.
\begin{align}
2 \pi \hat{\mathcal{L}}^\text{AdS} = L_0 \, , \quad 2 \pi \hat{\bar{\mathcal{L}}}^\text{AdS} = \bar L_0 \, , \quad  2 \pi \hat{\mathcal{W}}^\text{AdS} = W_0 \, , \quad 2 \pi \hat{\bar{\mathcal{W}}}^\text{AdS} = \bar W_0 \, .
\end{align}
Further, we set
\begin{align}
q = e^{2 \pi i \tau^\text{AdS}} \, , \quad \bar q = e^{2 \pi i \bar \tau^\text{AdS}} \, , \quad u = e^{2 \pi i \mu \bar \tau^\text{AdS}} = e^{2 \pi i \alpha^\text{AdS}} \, , \quad \bar u = e^{2 \pi i \bar \mu \tau^\text{AdS}} = e^{2 \pi i \bar \alpha^\text{AdS}} \, .
\end{align}
In order to obtain the thermodynamic meaning of ${\mathcal L}^\text{AdS}$ and ${\mathcal W}^\text{AdS}$, we should have
\begin{align} \label{intZ}
    \mathcal{L}^\text{AdS} = \langle \hat{\mathcal L}^\text{AdS} \rangle = - \frac{i}{4 \pi^2} \frac{\partial \ln Z}{ \partial \tau^\text{AdS}} \, , \quad 
    \mathcal{W}^\text{AdS} = \langle \hat{\mathcal W}^\text{AdS} \rangle = - \frac{i}{4 \pi^2} \frac{\partial \ln Z}{ \partial \alpha^\text{AdS}} \, .
\end{align}
The integrability of these equations leads to the condition (iii).

As in \eqref{AdShol}, it is convenient to introduce the holonomy matrix
\begin{align} \label{hsholonomyAdS}
\mathcal{P} e^{ \int a  } = e ^ \Omega \, , \quad \Omega =  i \beta^\text{AdS} (a_{z} - a_{\bar z}) 
\end{align}
as it is invariant under the higher-spin gauge transformations.
In \cite{Gutperle:2011kf}, it is claimed that the conditions are satisfied by requiring that the eigenvalues of $\Omega$ are the same as the BTZ case. When the BTZ black hole is realized as a classical solution of SL(3) Chern-Simons theory, the eigenvalues of $\Omega$ are given by $(2 \pi i , 0, - 2\pi i)$.
Equivalently, the conditions
\begin{align}
\text{tr} ( \Omega ^2) = - 8 \pi^2 \, , \quad \text{tr} (\Omega ^3) = 0  \label{hol}
\end{align}
are required.
For the gauge configuration \eqref{connection}, the conditions \eqref{hol} become
\begin{align}
\begin{aligned} \label{hol2}
0 &= - 2048 \pi^2 (\mu^\text{AdS})^3 ({\mathcal L }^\text{AdS})^3 + 576 \pi k \mu^\text{AdS} (\mathcal{L}^\text{AdS})^2 - 864 \pi k (\mu^\text{AdS})^2 \mathcal{W}^\text{AdS} \mathcal{L}^\text{AdS} \\
& \quad + 864 \pi k (\mu^\text{AdS})^3 ({\mathcal W}^\text{AdS})^2 - 27 k^2 \mathcal{W}^\text{AdS} \, , \\
&0 = 256 \pi^2 (\mu^\text{AdS})^2 (\mathcal{L}^\text{AdS})^2 + 24 \pi k \mathcal{L}^\text{AdS} - 72 \pi  \mu ^\text{AdS}\mathcal{W}^\text{AdS} + \frac{3 k^2}{(\tau^\text{AdS})^2} \, .
\end{aligned}
\end{align}
We can check that solutions to these equations satisfy the condition (iii), see \cite{Gutperle:2011kf,Ammon:2012wc}. Moreover, defining
\begin{align}
\zeta^\text{AdS} = \sqrt{\frac{k}{32 \pi ({\mathcal L}^\text{AdS})^3}} \mathcal{W}^\text{AdS} \, , \quad \gamma^\text{AdS} = \sqrt{\frac{2 \pi \mathcal{L}^\text{AdS}}{k}} \mu^\text{AdS} \, , 
\end{align}
the solutions to the conditions \eqref{hol2} are obtained as
\begin{align}
\begin{aligned}
&\zeta^\text{AdS} = \frac{1 + 16 (\gamma^\text{AdS})^2 - (1 - \frac{16}{3} (\gamma^\text{AdS})^2) \sqrt{1 + \frac{128}{3} (\gamma^\text{AdS})^2}}{128 (\gamma^\text{AdS})^3} \, , \\
&\beta^\text{AdS} = \frac{\sqrt{\frac{\pi k}{2 {\mathcal L}^\text{AdS}}}}{\sqrt{1 + \frac{16}{3} (\gamma^\text{AdS})^2 - 12 \gamma^\text{AdS} \zeta^\text{AdS}}} \, .
\end{aligned}
\end{align}
If we take $\gamma^\text{AdS}$ to zero, then $\zeta^\text{AdS}$ also goes to zero. This means that the solution satisfies the condition (ii).

It is actually hard to see the last condition (i).
Fortunately, a good gauge transformation was found in \cite{Ammon:2011nk}, and it was shown that the condition (i) is satisfied as well.
The authors considered the gauge transformation of the form
\begin{align} \label{gaugetrans}
A \to g(\rho)^{-1} A(\rho) g(\rho) + g(\rho)^{-1} d g(\rho) \, , \quad \bar A \to g(\rho) \bar A(\rho) g(\rho) - d g(\rho)  g(\rho)^{-1} \, ,
\end{align}
where $g(\rho)$ takes a value in SL$(3,\mathbb{R})$.
The metric can be put into the form of
\begin{align} \label{metric}
\begin{aligned}
g_{rr} &=\frac{(C^\text{AdS} -2 )(C^\text{AdS} - 3) }{(C^\text{AdS} -2 - \cosh^2 r)^2} \, ,  \\
g_{tt} &= - \left( \frac{8 \pi \mathcal{L}^\text{AdS}}{k} \right)\left( \frac{C^\text{AdS} - 3}{(C^\text{AdS})^2} \right) \frac{(a_t + b_t \cosh^2 r) \sinh^2 r}{(C^\text{AdS} -2 - \cosh^2 r)^2} \, , \\
g_{\phi \phi} &= \left( \frac{8 \pi \mathcal{L}^\text{AdS}}{k} \right)\left( \frac{C^\text{AdS} - 3}{(C^\text{AdS})^2} \right) \frac{(a_\phi + b_\phi \cosh^2 r) \sinh^2 r}{(C^\text{AdS} -2 - \cosh^2 r)^2} \\
& \quad + 
\left(\frac{8 \pi \mathcal{L}^\text{AdS}}{k} \right) (1 + \frac{16}{3} (\gamma^\text{AdS})^2 + 12 \gamma^\text{AdS} \zeta^\text{AdS}) \, , 
\end{aligned}
\end{align}
where $r=\rho-\rho_+$ with horizon $\exp (2\rho_+)= 2\pi \mathcal{L}^\text{AdS} / k$ and 
\begin{align}
\zeta^\text{AdS} = \frac{C^\text{AdS}-1}{(C^\text{AdS})^{3/2}} \, .
\end{align}
Moreover, $a_t$ and $b_t$ are functions of $\gamma^\text{AdS}$ and $C^\text{AdS}$ as in \eqref{ab}.
These coordinates describe the region outside the black hole horizon. Inside the horizon, we should replace $r$ by $i \theta$. 
In terms of $C^\text{AdS}$, the entropy can be expressed in a quite simple way as
\begin{align}
S = 2 \pi \sqrt{2 \pi k \mathcal{L}^\text{AdS}} \sqrt{1 - \frac{3}{4 C^\text{AdS}}} \, .
\end{align}
We obtain the expression by solving the thermodynamic relations
\begin{align}
\tau^\text{AdS} = \frac{i}{4 \pi^2} \frac{\partial S}{\partial \mathcal{L}^\text{AdS}} \, , \quad
\alpha^\text{AdS} = \frac{i}{4 \pi^2} \frac{\partial S}{\partial \mathcal{W}^\text{AdS}}
\end{align}
and the equations \eqref{hol2}.
This concludes our review of higher-spin AdS black hole. We will now generalize the analysis to the higher-spin dS cases in the next subsection.

\subsubsection{Higher-spin \texorpdfstring{dS$_3$}{dS3} black holes}

We now extend the analysis above to the dS$_3$ case as in subsection \ref{sec:gravity}. 
We consider the Chern-Simons action \eqref{CSactiondS} but with the Chern-Simons level $\kappa \in \mathbb{R}$.
The solutions to the equations of motion can be put into the forms \eqref{gaugedS}. Thus, in the inflation patch near the future infinity, we may use the gauge configuration \eqref{gauge} but $a$ is replaced by $e^{-(\pi i/2)L_0} a e^{(\pi i/2)L_0}$, which is evaluated as
\begin{align}
\begin{aligned}
e^{-(\pi i/2)L_0} a e^{(\pi i/2)L_0}&= i \left(L_1 - \frac{2 \pi { \mathcal L}}{\kappa}  L_{-1}- \frac{\pi { \mathcal W}}{2 \kappa}  W_{-2}\right) dz \\
&\quad- i \mu \left(W_2 - \frac{4 \pi { \mathcal L}}{\kappa} W_0 + \frac{4 \pi^2 {  \mathcal L}^2}{\kappa} W_{-2} + \frac{4 \pi  \mathcal{W}}{\kappa} L_{-1}\right) d \bar z \, . \label{dSconfig}
\end{aligned}
\end{align}
We define $\bar a$ in a similar manner for non-rotating solution. The rule of replacement we use is
\begin{align} \label{rule}
 k \to  i \kappa \, , \quad \mathcal{L}^\text{AdS} \to -i \mathcal{L} \, , \quad \mathcal{W}^\text{AdS} \to  \mathcal{W} \, , \quad \mu^\text{AdS} \to - i \mu \, , \quad \tau^\text{AdS} \to  i \tau \, .
\end{align}

As in the case of AdS$_3$, we can transform the gauge field such that the metric is of the form
\begin{align}
 ds^2 = g_{\tilde\rho\tilde\rho} (\tilde\rho) d\tilde\rho^2 + g_{tt} (\tilde\rho) dt^2 + g_{\phi \phi} (\tilde\rho) d\phi^2
\end{align}
by applying the gauge transformation similar to \eqref{gaugetrans}.
Moreover, by changing $\tilde\rho \to i \theta$, the metric becomes
\begin{align}
 ds^2 =  \tilde g_{\theta \theta } (\theta) d \theta ^2 + \tilde g_{t t } (\theta) dt ^2 + \tilde g_{\phi \phi} (\theta ) d\phi^2 \, . 
\end{align}
We may perform analytic continuations as $i t \to t_E$ with $t_E \sim t_E + \beta$. We then define the holonomy matrix for $a$ along the thermal cycle
\begin{align}
\mathcal{P} e^{ \oint a } = e^ \Omega \, , \quad \Omega =  \beta (a_z + a_{\bar z})  \, . \label{holdS0}
\end{align}
The eigenvalues of $\Omega$ are gauge invariant, 
in particular, they do not change under the gauge transformation of the form \eqref{gaugetrans}.
We require them to be the same as those for the dS black hole without any higher-spin charges, i.e. $(2 \pi i , 0, - 2\pi i )$ or equivalently
\begin{align}
\text{tr} (  \Omega ^2) = - 8 \pi^2 \, , \quad \text{tr} (\Omega ^3) = 0 \, , \label{holdS}
\end{align}
which become
\begin{align}
\begin{aligned} \label{hol2dS}
0 &= - 2048 \pi^2 \mu^3 {\mathcal L }^3 + 576 \pi \kappa \mu \mathcal{L}^2 - 864 \pi \kappa \mu^2 \mathcal{W} \mathcal{L}  + 864 \pi \kappa \mu^3 {\mathcal W}^2 - 27 \kappa^2 \mathcal{W} \, , \\
&0 = 256 \pi^2 \mu^2 \mathcal{L}^2 + 24 \pi \kappa \mathcal{L} - 72 \pi \kappa \mu \mathcal{W} - \frac{3 \kappa^2}{\tau^2} \, .
\end{aligned}
\end{align}
We may further define $\zeta$ and $\gamma$ by
\begin{align}
\zeta  =  \sqrt{\frac{\kappa}{32 \pi {\mathcal L}^3}} \mathcal{W} \, , \quad \gamma = \sqrt{\frac{2 \pi \mathcal{L}}{\kappa}} \mu \, , 
\end{align}
and $C$ is defined through the following relation: 
\begin{align}
\zeta = \frac{C-1}{C^{3/2}} \, .
\label{CdS}
\end{align}
The entropy associated with the higher-spin black hole is
\begin{align}
S_\text{GH} = 2 \pi \sqrt{2 \pi \kappa \mathcal{L}} \sqrt{1 - \frac{3}{4 C}} \, . \label{hsentropy}
\end{align}

In above, we have assigned the condition that the eigenvalues of the holonomy matrix defined by \eqref{holdS0} are the same as those of dS black hole without any higher-spin charges  \eqref{hol2dS}. However, it might be possible to consider the gauge configuration obtained by a large gauge transformation as in the case with $N=2$. We examine this issue for generic $N$. We can see that the eigenvalues of $\Omega$ for the gauge configuration corresponding to dS black hole without any higher-spin charges are $2 \pi i (\rho_1,\rho_2, \cdots , \rho_N)$, where
\begin{align}
    \rho_j = \frac{N+1}{2} - j \quad (j=1,2,\dots,N) \, . \label{rhoj}
\end{align}
As we will define below, these are the components of Weyl vector of SU$(N)$ in the orthogonal basis. If we require only that the holonomy matrix $\exp \Omega$ is the same as the one for dS black hole without any higher-spin charges, then the eigenvalues of $\Omega$ can take
\begin{align}
 2 \pi i(\Lambda_1 , \Lambda_2 , \ldots , \Lambda_N) \, , \quad  \Lambda_j = m_j + \rho_j \, , \quad m_j \in \mathbb{Z} 
\end{align}
satisfying $\sum_j m_j = 0$. 
As in the case of $N=2$, the saddle points of SL$(N)$ Chern-Simons gravity may be labeled by $(n-1)$ integers $m_i$ or $\Lambda_i = m_i + \rho_i$.
Defining $[e_{ij}]_{kl} = \delta_{i,k} \delta_{j,l}$, the corresponding gauge configuration may be given in a diagonal form as
\begin{align}
 a = - i \sum_{j=1}^N e_{jj} ( (\rho - \eta)_j d \phi +  \Lambda_j d t_E) \, , \quad 
\bar  a =  i \sum_{j=1}^N e_{jj} ( (\rho - \eta)_j d \phi - \Lambda_j d t_E) \, . \label{congauge}
\end{align}
The action corresponding to the configuration can be evaluated as in \cite{Hikida:2022ltr}
\begin{align}
  - S \, (\equiv S_\text{GH}^{(\Lambda)})= \frac{\pi}{3} c^{(g)} \frac{(\rho - \eta ,\Lambda)}{(\rho , \rho)}  \, ,  \quad \Lambda = \sum_j \Lambda_j \epsilon_j \, , \label{HSCS}
\end{align}
which takes a different value depending on the label $\Lambda$ of the saddle points. In the next subsection, we read off the possible saddles from $\mathfrak{sl}(N)$ Toda field theory by comparing the classical action at each saddle point as in the case with Liouville field theory.

\subsection{Toda description}
\label{sec:Toda}

In this subsection, we extend the Liouville theory analysis in subsection \ref{sec:Liouville} to that by $\mathfrak{sl}(N)$ Toda field theory.
We first introduce notations for $\mathfrak{sl}(N)$ Lie algebra. Let us denote the orthonormal basis of $\mathbb{R}^N$ by $\epsilon_j$ $(j=1,2,\ldots,N)$ satisfying $(\epsilon_i, \epsilon_j ) = \delta_{i,j}$. Then, the simple roots are given by 
\begin{align}
e_j = \epsilon_j - \epsilon_{j+1} \quad (j=1,2,\ldots,N-1) \, ,
\end{align}
which satisfy $(e_i ,e_j) = K_{ij}$ with $K_{ij}$ being the Cartan matrix for $\mathfrak{sl}(N)$.
The fundamental weights $\omega_j$ $(j=1,2,\ldots,N-1)$ satisfy $(\omega_i , e_j) = \delta_{i,j}$ and given by
\begin{align}
\omega_j = \sum_{l=1}^j \epsilon_l - \frac{j}{N} \sum_{l=1}^N \epsilon_l \, .
\end{align}
The Weyl vector $\rho$ is the half of the sum over all positive root or equivalently the sum over fundamental weights as
\begin{align}
 \rho = \sum_{j=1}^{N-1} \omega_j = \sum_{j=1}^N \rho_j \epsilon_j
\end{align}
with $\rho_j$ defined in \eqref{rhoj}.

We consider the Toda field $\phi = \sum_{j=1}^{N-1} \phi ^j e_j$ or $\phi_j = (e_j ,\phi)$ with $\phi_j = \sum_{i=1}^{N-1} K_{ji}\phi^i$.
The action is given by
\begin{align}
S_\text{T} = \frac{1}{4 \pi} \int d^2 \sigma \sqrt{\tilde g } \left[ \frac{1}{2} (\partial_a \phi , \partial_{a'} \phi)  \tilde g^{aa'} + (Q,\phi) \tilde{\mathcal{R}} + 4 \pi \mu \sum _{j=1}^{N-1} e^{b (e_j , \phi)} \right] \, .
\end{align}
The background charge here is 
\begin{align}
Q = \left( b + b^{-1} \right) \rho \, , \label{Q}
\end{align}
and the central charge is 
\begin{align}
c = N -1 + 12 (Q,Q)  = (N -1) (1 + N(N+1) (b + b^{-1})^2) \, .
\end{align}
We consider sphere amplitudes as in the Liouville case.
The theory is invariant under the combination of Weyl transformation $\tilde g_{aa'} \to \Omega (\sigma) \tilde g_{aa'}$ and $\phi \to \phi - Q \ln \Omega (\sigma)$. Making use of it, we set $\tilde g_{aa'} = \delta_{aa'}$ except at the infinity $|z| \to \infty$ and the boundary condition 
\begin{align}
\phi = - Q \ln |z| + \mathcal{O}(1)
\end{align}
at $|z| \to \infty$.

We consider the vertex operators of the form
\begin{align}
V_\alpha = e^{(\alpha , \phi)} \, ,
\end{align}
whose conformal weights are
\begin{align}
 h_\alpha = \bar h_\alpha = \frac{(\alpha , 2 Q - \alpha)}{2} \, .
\end{align}
The correlation functions are defined in the path integral formulation as
\begin{align}
 \langle V_{\alpha_1} (z_1) \cdots V_{\alpha_n} (z_n) \rangle 
= \int \mathcal{D} \phi e^{- S_\text{T}} V_{\alpha_1} (z_1) \cdots V_{\alpha_n} (z_n) \, .
\end{align}
We use the normalization of two-point function as
\begin{align}
\langle V_\alpha (z_1) V_{2Q -\alpha} (z_2) \rangle = \frac{\delta(0)}{|z_{12}|^{4 h_\alpha}} \, .
\end{align}
Performing a reflection relation, we obtain (see, e.g., \cite{Fateev:2007ab})
\begin{align}
\langle V_\alpha (z_1) V_{\alpha^*} (z_2) \rangle = \frac{\delta(0)R^{-1}(\alpha)}{|z_{12}|^{4 h_\alpha}} \, ,
\end{align}
where the conjugate parameter is defined as
\begin{align}
(\alpha , e_j) = (\alpha^* , e_{N-j}) \, .
\end{align}
The coefficient is
\begin{align}
R^{-1}(\alpha)  = \frac{A(\alpha)}{A(2 Q-\alpha)} \label{Toda2pt}
\end{align}
with
\begin{align}
A(\alpha) = (\pi \mu \gamma(b^2))^{(Q - \alpha,\rho)/b} \prod_{e > 0} \Gamma (1 - b (Q - \alpha ,e)) \Gamma( - b^{-1} (Q - \alpha ,e)) \, , \label{A}
\end{align}
where the product is over the positive roots. In the following, we set
\begin{align}
\text{Re} \, \alpha < \text{Re} \, Q
\end{align}
by performing a certain reflection to the vertex operator.

As in the Liouville case, we set $b \sim 0$, $\text{Re} \, b^{-2} < 0$, and $\lambda \equiv \pi \mu b^2$ real.
In the semi-classical limit with finite $N$,
the central charge behaves as \cite{Hikida:2022ltr}
\begin{align}
c = \frac{N(N^2-1)}{b^2} + N-1 + \mathcal{O} (b)\, ,
\end{align}
which implies
\begin{align}
b^{-2} = \frac{i c^{(g)}}{N (N^2 -1)} - \frac{1}{N(N+1)} + \mathcal{O} ((c^{(g)})^{-1}) \, .
\end{align}
We consider a two-point function of heavy operators with $\alpha = \eta/b$. The equations of motion reduce to
\begin{align}
 \partial^a \partial_{a}\phi^j = 4 \pi \mu b^2 e^{\phi^j} - 4 \pi (\eta ,e^j) [\delta^{(2)} (\sigma - \sigma_1) + \delta^{(2)} (\sigma -\sigma_2) ] \, .
\end{align}
As in the case of Liouville description, suppose there exists $\phi_{c(0)}$ as a classical solution to the equations of motion, then we have multiple solutions
\begin{align}
 \phi_{c(n)}= \phi_{c(0)} + 2\pi i n
\end{align}
with $(e_j ,n) \in \mathbb{Z}$ for all $j$. In other words, there are different solutions labeled by $N-1$ integers $(e_j , n)$.

We can read off the semi-classical saddles of Toda field theory from the exact expression \eqref{Toda2pt} with \eqref{A} by taking the limit of $b \to 0$ with $\text{Re} \, b^{-2} < 0$.
Near $b \sim 0$, the two-point function can be written as
\begin{align}
R^{-1} (\alpha) \sim  \lambda^{2 (\rho -\eta , \rho)/b^2} \left[ \frac{\gamma (b^2)}{b^2}\right]^{2 (\rho -\eta , \rho)/b^2} 
\prod_{e > 0} \frac{\Gamma(b^{-2} (\eta - \rho , e))}{\Gamma( - b^{-2} (\eta - \rho , e))} \, .
\end{align}
Using the Stirling's formula, we obtain:
\begin{align}
\left[ \frac{\gamma (b^2)}{b^2}\right]^{2 (\rho -\eta , \rho)/b^2}  \sim \exp \left[ - \frac{8 (\rho - \eta , \rho) \ln b }{b^2}\right]
\end{align}
and
\begin{align}
\begin{aligned}
\frac{\Gamma(b^{-2} (\eta - \rho , e))}{\Gamma( - b^{-2} (\eta - \rho , e))} 
\sim \left( e^{- \pi i (\rho - \eta ,e)/b^2} - e^{\pi i (\rho - \eta ,e)/b^2} \right) \exp  \left[ \frac{2 (\eta - \rho ,e)}{b^2}( \ln (\rho - \eta ,e) - 2 \ln b -1 ) \right] \, ,
\end{aligned}
\end{align}
the asymptotic behavior of two-point function is found to be
\begin{align}
\begin{aligned}
R^{-1} (\alpha)
\sim\lambda^{2 (\rho -\eta , \rho)/b^2} \prod_{e > 0} \left( e^{- \pi i (\rho - \eta ,e)/b^2} - e^{ \pi i (\rho - \eta ,e)/b^2} \right) 
\exp\left[ \sum_{e > 0}\frac{2 (\eta - \rho ,e)}{b^2}( \ln (\rho - \eta ,e) -1 ) \right]\, .
\end{aligned}
\end{align}
Taking the absolute value of two-point function of heavy operators, we find
\begin{align} \label{2ptheavy}
\left| \langle V_\alpha (z_1) V_{\alpha^*} (z_2) \rangle \right| \sim \prod_{e > 0} \left | e^{\frac{\pi}{6} c^{(g)} \frac{(\rho - \eta ,e)}{(\rho , \rho)}} - e^{- \frac{\pi}{6} c^{(g)} \frac{(\rho - \eta ,e)}{(\rho , \rho)}} \right| \, .
\end{align}
Here we consider the Weyl group of SL$(N)$ denoted by $W$, such that an element $w \in W$ exchanges the indices of orthogonal basis $\epsilon_j$ $(j=1,2,\dots,N)$. Then, the above expression can be rewritten as (see \cite{Drukker:2010jp} for the modular S-matrix of Toda field theory)
\begin{align} 
\left| \langle V_\alpha (z_1) V_{\alpha^*} (z_2) \rangle \right| \sim 
 \left | \sum_{w \in W} \epsilon (w) e^{\frac{\pi}{6} c^{(g)} \frac{( \rho -\eta , w(\rho)) }{(\rho , \rho)}  } \right | \sim  
|S_{0\alpha}| \, ,
\end{align}
where $\epsilon (w)$ is a sign related to $w$. 
This is consistent with \eqref{2pt0j} even after including the sub-leading non-perturbative corrections.
Compared with \eqref{HSCS}, we find that the allowable set of saddle points is given by
\begin{align}
\Lambda = w (\rho) \, . \label{wrho}
\end{align}
The final answer is quite natural since the gauge configurations labeled by \eqref{wrho} can be mapped to each other by acting elements of Weyl group of SU$(N)$. Namely, the gauge configurations related by this way should represents the same higher-spin dS$_3$ black hole.

\section{Probing higher-spin \texorpdfstring{dS$_3$}{dS3} black holes}
\label{sec:probe}

In the previous section, we have constructed higher-spin dS$_3$ black holes and computed their partition functions (or Gibbons-Hawking entropy) both from the higher-spin theory and its dual Toda field theory. In particular, we have identified all geometries realized by path integral saddles. In this section, we examine in more details  the properties of higher-spin dS$_3$ black holes realized as the dominant contributing saddle. 
For the purpose, it is convenient to study the boundary-to-boundary two-point function of bulk scalar field on the black hole geometry. 
{ If we view the dS$_3$ black hole arises due to the back reaction of a pair of heavy operators insertion on the boundary, we can alternatively view this computation as a special case of so-called heavy-heavy-light-light limit \cite{Hijano:2015rla}.}
However, it is known to be difficult to couple matter fields to higher-spin theory described by Chern-Simons gauge theory with finite dimensional group like SL$(N,\mathbb{C})$. We can avoid such a difficulty by working with the Prokushkin-Vasiliev theory on dS$_3$ \cite{Prokushkin:1998bq} instead of the Chern-Simons gravity with the finite dimensional group. In the next subsection, we introduce the Prokushkin-Vasiliev theory and compute the partition function of its black hole solution. In subsection \ref{sec:CFT2pt}, we will compute the boundary-to-boundary two-point function of bulk scalar field and examine its properties. In this section, we mainly explain the analysis by the dual CFT description, see appendix \ref{app:hslambda} for details on the analysis from the bulk viewpoints.

\subsection{Partition functions}
\label{sec:probepf}

The Prokushkin-Vasiliev theory contains an infinite tower of higher-spin gauge fields with $s=2,3,\ldots$ and two complex scalar fields with mass 
\begin{align}
\ell^2 m^2 = 1  - \lambda^2 \, . \label{dSmass}
\end{align}
Note that we need to replace $\ell_\text{AdS} \to i \ell$ as in \eqref{AdSmass} in order to move from the case with negative cosmological constant to that with positive cosmological constant. The higher-spin gauge fields can be described by Chern-Simons gauge theory based on the infinite dimensional higher-spin algebra $\mathfrak{hs}[\lambda]$. The generators of the algebra can be expressed as
\begin{align}
    V_n^s \, , \quad s=2,3,\ldots \, , \quad n = - s+ 1 , - s+2 , \dots , s-1 \, .
\end{align}
Here $V_{0}^{2}$ and $V_{\pm 1}^{2}$ form an $\mathfrak{sl}(2)$ sub-algebra, and the commutation relation with the other remaining generators are
\begin{align}
[V^2_m,V^s_n]=(-n+m(s-1))V^s_{m+n} \, . \label{comm}
\end{align}
See \cite{Pope:1989sr,Gaberdiel:2011wb} for generic commutation relations. A feature of $\mathfrak{hs}[\lambda]$ is that it can be truncated to $\mathfrak{sl}(N)$ at $\lambda = \pm N$ by dividing ideal formed.
We would like to construct a black hole solution as in the case with SL$(3,\mathbb{C})$ Chern-Simons gravity analyzed in subsection \ref{sec:csgra}. Using $\mathfrak{sl}(2) \in \mathfrak{hs}[\lambda]$, the dS$_3$ black hole in subsection \ref{sec:gravity} can be embedded into the Prokushkin-Vasiliev theory. As in subsection \ref{sec:csgra}, we introduce a non-trivial spin-3 charge to the black hole. In the current infinite dimensional case, there are gauge fields with spin $s >3$. They have also induced charges, which can be evaluated by solving equations of motion.

We would like to compute the partition function of higher-spin black holes. In the AdS$_3$ case, the higher-spin black hole as a solution to the Prokushkin-Vasiliev theory was constructed in \cite{Kraus:2011ds} as reviewed in appendix \ref{app:hslambdabh}. The gravity partition function of the black hole solution with the higher-spin charges was also computed in \cite{Kraus:2011ds} as the following perturbative expansion:
\begin{align} \label{AdSPF}
&\ln Z_\text{AdSBH} ( \tau^\text{AdS} , \alpha^\text{AdS}) \\
&\quad = \frac{i \pi c}{12 \tau^\text{AdS}}
\left[ 1 - \frac{4}{3} \frac{(\alpha^\text{AdS})^2}{( \tau^\text{AdS})^4} + \frac{400}{27} \frac{\lambda^2 - 7}{\lambda^2 - 4} \frac{(\alpha^\text{AdS})^4}{(\tau^\text{AdS})^8} - \frac{1600}{27} \frac{5 \lambda ^4 - 85 \lambda ^2 + 377}{(\lambda^2 -4)^2} \frac{(\alpha^\text{AdS})^6}{(\tau^\text{AdS})^{12}}+ \cdots \right] \, , \nonumber
\end{align}
where $\tau^\text{AdS}$ is the parameter related to mass of black hole given by \eqref{tauAdS}.
The leading factor $\exp \frac{i \pi c}{12 \tau^\text{AdS}}$ represents (the holomorphic part of) the partition function without introducing higher spin charges. The parameter $\alpha^\text{AdS} = \bar{\tau}^\text{AdS}\mu^\text{AdS} $ play the role of chemical potential for the spin-3 charge as in subsection \ref{sec:hssl3bh}, see appendix \ref{app:hslambdabh}. The partition function is evaluated perturbatively in the dimensionless parameter $\alpha^\text{AdS}/(\tau^\text{AdS})^2$.
It was reproduced from the CFT computation in \cite{Gaberdiel:2012yb} by
\begin{align}
Z_\text{CFT} = \text{Tr} \left( e^{- \beta^\text{AdS} H + 2 \pi i \alpha^\text{AdS} W_0} \right)
\end{align}
up to the order in \eqref{AdSPF}.
In the gravity side, the partition function can be mapped to the one for dS black hole by changing the parameters as
\begin{align}
 c \to  i c^{(g)} \, , \quad \tau^\text{AdS} \to  i \tau \, , \quad \alpha^\text{AdS} =\bar{\tau}^\text{AdS} \mu^\text{AdS}  \to   \alpha \, ,
\end{align}
as in \eqref{rule}, see also appendix \ref{app:hslambdabh}.
Thus we have
\begin{align} \label{ZdSBH}
\ln Z_\text{dSBH} (\tau , \alpha)
= \frac{i \pi c^{(g)}}{12 \tau}
\left[ 1 - \frac{4}{3} \frac{\alpha^2}{ \tau^4} + \frac{400}{27} \frac{\lambda^2 - 7}{\lambda^2 - 4} \frac{\alpha^4}{\tau^8} - \frac{1600}{27} \frac{5 \lambda ^4 - 85 \lambda ^2 + 377}{(\lambda^2 -4)^2} \frac{\alpha^6}{\tau^{12}}+ \cdots \right] 
\end{align}
as in \eqref{ZdSBHf}.
In order to obtain  the gravity partition function, we have to multiply by a factor two, since we need to consider the square of wave functional of universe as in \eqref{square}. 
We also need to multiply the anti-holomorphic part as well. Putting (see \eqref{L} and \eqref{tau})
\begin{align}
  \tau = \frac{i}{ \sqrt{1 - 8G_NE} } \, ,
\end{align}
the leading order in $\alpha$ becomes $\frac{\pi}{6} c^{(g)} \sqrt{1 - 8 G_N E}$ by combining the anti-holomorphic part.
It reproduces $S_\text{GH}/2$, where $S_\text{GH}$ is \eqref{BHentropy} with \eqref{cg}. In the rest of this subsection, we will derive this from its dual CFT.

We would like to extend the CFT analysis in \cite{Gaberdiel:2012yb} for AdS$_3$ to that for dS$_3$. As argued above, the map from AdS$_3$ to dS$_3$ is not so difficult in the gravity side. What we have to do is just inserting $-1$ and/or $i$ in a proper way. However, it does not look straightforward in the dual CFT side. In the AdS/CFT correspondence, the dual CFT is supposed to live on the spatial boundary of AdS$_3$ black hole. The boundary of Euclidean BTZ black hole is given by a torus, thus the dual CFT is living on a torus. In the dS/CFT correspondence, the dual CFT is supposed to live on the future or past infinity. The geometry of the boundary is given by a cylinder as we saw in subsection \ref{sec:gravity}. Therefore, it seems impossible to directly map the result for AdS$_3$ to that for dS$_3$ as in the gravity analysis. However, we are only interested in thermodynamic quantities, which are obtained in the high temperature limit. We shall show that the difference between torus and cylinder amplitudes disappears at the high temperature limit.

In subsection \ref{sec:Liouville}, we have reproduced the gravity partition function from the two-point function of heavy operators in Liouville field theory. The heavy operators create conical defects on $S^2$ as explained around \eqref{conical}, thus the same quantity can be computed as the partition function on $S^2$ with two conical defects as shown in appendix \ref{app:cd}. The discussion was done for the pure gravity (or Liouville field theory), but it can be extended for the Prokushkin-Vasiliev theory (or the 't Hooft limit of Toda field theory) as argued in appendix \ref{app:thooft}. We compute the CFT partition function in the expansion of $\alpha$, thus the insertions of spin-3 current can be treated perturbatively.
Therefore, we can compute the partition function on $S^2$ with two conical defects and also with the deformation of spin-3 currents.
As mentioned before, the Toda field theory has the symmetry under the combination of the Weyl transformation and the shift of fields. Utilizing it, we can transform $S^2$ with two conical defects to a cylinder with coordinates $(\sigma_0 ,\sigma_1)$ satisfying
\begin{align}
 \sigma_0 \sim \sigma_0 +  \beta \, , \quad 
0  \leq \sigma_1 \leq L \, .
\end{align}
Here we have introduced an infra-red cut off $L$, which takes a very large value. Applying again the Weyl transformation and the shift of fields, we may set
\begin{align}
 \sigma_0 \sim \sigma_0 + 2 \pi \tilde \beta \, , \quad \tilde \beta =  \frac{\beta}{L} \, , \quad 
0  \leq \sigma_1 \leq 2 \pi \, .
\end{align}
We need to redefine another parameter $\alpha$.
Since the parameter has conformal
dimension $-2$, we have to rescale $\tilde \alpha = (2 \pi /L)^2 \alpha$.

Now the computation reduces to that of CFT partition function on the cylinder as
\begin{align}
Z_\text{CFT} = \text{Tr}_r \left( e^{- \tilde \beta H + 2 \pi i \tilde \alpha W_0} \right) \, ,
\end{align}
where $r$ denote a representation.
We expand the partition function in the parameter $\tilde \alpha$, thus we compute the cylinder amplitude of the form at order $\tilde \alpha^n$
\begin{align}
\text{Tr} _r \left( W_0^n e^{- \tilde \beta H } \right) = \frac{1}{(2 \pi i )^n} \oint \frac{dz_1}{z_1} \cdots \oint \frac{dz_n}{z_n}
F_r ((W,z_1) , \cdots , (W , z_n) ; \tilde \tau)
\end{align}
with $z = \sigma_0 + i \sigma_1$ and $\tilde \tau = i \beta /(2 \pi L)$.
Here we use the notation in (2.12) of \cite{Gaberdiel:2012yb} as
\begin{align}
F_r((a^1 ,z_1) , \cdots (a^n , z_n); \tilde \tau) = z_1^{h_1} \cdots z_n^{h_n} \text{Tr}_r (V(a^1 , z_1) \cdots V(a^n , z_n) \tilde q^{L_0 - \frac{c}{24}}) 
\end{align}
with $h_j$ as the conformal weight of $a^j$ and $\tilde q=\exp (2 \pi i \tilde \tau)$.
Moreover, we set
\begin{align}
V(a ,z) = \sum_{m \in \mathbb{Z} } a_m z^{- m - h} \, .
\end{align}
Performing the $S$-transformation, we obtain
\begin{align}
(2 \pi i)^n \text{Tr} _r \left( W_0^n e^{- \beta H } \right) \sim
\tau^{2n} \int_{1}^q \frac{d \tilde z_1}{\tilde z_1} \cdots  \int_{1}^q \frac{d \tilde z_n}{\tilde z_n} \langle 0 | V(W,\tilde z_1) \cdots V(W , \tilde z_n)  \hat q^{L_0 - \frac{c}{24}}| 0 \rangle \, .
\end{align}
Since $L$ is a large number acting as an IR regulator, $ \hat \tau =  -1/{\tilde \tau} = i L / \beta$ is now very large. Therefore, the dominant contribution comes from the amplitude where the in and out states are given by the identity one as above.

With very large $\hat \tau$, we can also use the fact that the dominant contribution comes from the primary state $|0 \rangle$ among the descendant states represented by $|\{k_n\} \rangle$ as
\begin{align}
\begin{aligned}
 \langle 0 | V(W,\tilde z_1) \cdots V(W , \tilde z_n) \hat q^{L_0 - \frac{c}{24}}| 0 \rangle 
&\simeq \sum_{\{k_n\}} \langle \{k_n\}  | V(W,\tilde z_1) \cdots V(W , \tilde z_n) \hat q^{L_0 - \frac{c}{24}}| \{k_n\} \rangle  \\
&\equiv \text{Tr}_0 ( V (W,\tilde z_1)  \cdots  V (W,\tilde z_n) \hat q^{L_0 - \frac{c}{24}} ) \, .
\end{aligned}
\end{align}
Here the sum is over all descendants labeled by $\{k_n\}$. In this way, we can relate the cylinder amplitude to the torus one at the large $\tau$ limit.
Borrowing the result from \cite{Gaberdiel:2012yb}, the CFT partition function can be obtained as
\begin{align}
&\frac{2 \pi}{L} \ln Z_\text{CFT} ( \tilde \tau , \tilde \alpha) \\
&= \frac{ i \pi c^{(g)}}{6 \tilde \tau }
\left[ 1 - \frac{4}{3} \frac{\tilde \alpha^2}{\tilde \tau^4} + \frac{400}{27} \frac{\lambda^2 - 7}{\lambda^2 - 4} \frac{\tilde \alpha^4}{\tilde \tau^8} - \frac{1600}{27} \frac{5 \lambda ^4 - 85 \lambda ^2 + 377}{(\lambda^2 -4)^2} \frac{\tilde \alpha^6}{\tilde  \tau^{12}}+ \cdots \right] \, . \nonumber
\end{align}
The pre-factor is set such that the partition function without higher-spin charge should be invariant under the Weyl transformation and the shift of fields. The above expression reproduces the gravity computation \eqref{ZdSBH}.

\subsection{Two-point functions of boundary operators}
\label{sec:CFT2pt}

In this subsection, we examine the two-point functions of scalar operators in the dual CFTs, which correspond to the boundary-to-boundary two-point functions of bulk scalar fields. We first focus on the case without any higher-spin charges and then move to the case with higher-spin charges.

We start from the BTZ black hole, whose metric may be expressed as in  \eqref{AdSBHmetric}. For simplicity we set $\ell_\text{AdS} = 1$ here.
There are two disconnected boundaries at $r= \pm \infty$.
If the CFT operators are inserted in the same boundary, then the two-point function is \cite{Keski-Vakkuri:1998gmz}
\begin{align}
    \langle \bar{\mathcal{O}}_1 (t, \phi) \mathcal{O}_1 (0, 0) \rangle 
    = \sum_{n = - \infty}^\infty [- \cosh (r_+ t) + \cosh r_+(\phi + 2 \pi n)]^{- 2 h} \, , \label{O1O1}
\end{align}
where $h$ denotes the conformal dimension of $\mathcal{O}$ and the subscript in $\mathcal{O}_{1,2}$ denotes the two boundaries.
The two-point function exhibits a light-like singularity at $t = \phi + 2 \pi n$. If two CFT operators are inserted in different boundaries, then the two-point function is \cite{Maldacena:2001kr}
\begin{align}
    \langle \bar{\mathcal{O}}_1 (t, \phi) \mathcal{O}_2 (0, 0) \rangle 
    = \sum_{n = - \infty}^\infty [\cosh (r_+ t) + \cosh r_+(\phi + 2 \pi n)]^{- 2 h} \, . \label{O1O2}
\end{align}
There are no singularities in the two-point function. This is related to the fact that the two boundaries are separated by the horizon in the bulk. The two-point function may explore the insider of horizon, but it does not show any divergences associated with black hole singularity as explained in \cite{Maldacena:2001kr,Kraus:2002iv}.

Let us move to the dS$_3$ black hole with the metric \eqref{dSBHmetric}. The metric can be used for the region with $r_+ < r$, but the role of time and space is exchanges as $r \leftrightarrow t$. There are boundaries at the past and future infinities $(|r| \to \infty)$. We again consider the two-point functions of CFT operator dual to a bulk scalar operator.%
\footnote{It is interesting to see the relation to a recent work \cite{Cotler:2023xku}.}
If CFT operators are inserted only in the future (or past) infinity, then the two-point function is
\begin{align} \label{dS2pt}
    \langle \bar{\mathcal{O}}_1 (t, \phi) \mathcal{O}_1 (0, 0) \rangle 
    = \sum_{n = - \infty}^\infty [- \cosh (r_+ t) + \cos r_+(\phi + 2 \pi n)]^{- 2 h} \, ,
\end{align}
see \eqref{bdS2pt}.
Since the CFT is defined on a Euclidean space, there is only a singularity at the equal point $t = \phi + 2 \pi n = 0$.
If two CFT operators are inserted in different boundaries, then the two-point function is
\begin{align}\label{dS2pts}
    \langle \bar{\mathcal{O}}_1 (t, \phi) \mathcal{O}_2 (0, 0) \rangle 
    = \sum_{n = - \infty}^\infty [\cosh (r_+ t) + \cos r_+(\phi + 2 \pi n)]^{- 2 h} \, ,
\end{align}
see \eqref{bdS2pts}.
There is a singularity at $t=0$ and $\phi + 2 \pi n = \pi/r_+$.
This should be the same light-like singularity, which was observed for pure de Sitter case in \cite{Strominger:2001pn}.
The two-point function does not have any divergences associated with black hole (conical deficit) singularities. This may be explained along the line of \cite{Kraus:2002iv}.

Up to now, we have considered a scalar propagation with generic mass on BTZ or dS$_3$ black holes. In the following, we focus on the case with the Vasiliev theory. The theory includes two complex scalar fields, whose dual operators $\mathcal{O}^\pm$ have the conformal dimensions $\Delta_\pm = 2 h_\pm = 1 \pm \lambda$. We first consider the case of BTZ black hole. 
It is convenient to rewrite the two-point function of CFT operator in \eqref{O1O1} and \eqref{O1O2} as
\begin{align}
 &\langle {\bar{\mathcal{O}}}^\pm_1 (z , \bar z) \mathcal{O}^\pm_1 (0,0) \rangle ^{(0)} 
= \frac{(\hat \tau \hat {\bar \tau} )^{2h_\pm}}{\left( 4 \sin \frac{\hat \tau z  }{2} \sin \frac{\hat {\bar \tau} \bar z }{2} \right)^{2h_\pm}} \, , \label{O1O1v2} \\
& \langle {\bar{\mathcal{O}}}^\pm_1 (z , \bar z) \mathcal{O}^\pm_2  (0,0) \rangle ^{(0)} 
= \frac{(\hat \tau \hat {\bar \tau} )^{2h_\pm}}{\left( 4 \cos \frac{\hat \tau z  }{2} \cos \frac{\hat {\bar \tau} \bar z }{2} \right)^{2h_\pm}} \, .\label{O1O2v2} 
\end{align}
Here we set $z = \phi + i t_E, \bar z = \phi - it_E$ and neglect the sum over $n\neq 0$ terms since it is not important in the current analysis.
Furthermore, we set 
\begin{align}
\hat \tau = - \frac{1}{\tau} \, , \quad \hat {\bar \tau} = - \frac{1}{\bar \tau } 
\end{align}
with $\tau = \bar \tau$ as the moduli parameter of boundary torus. In the Lorentzian section with $z = \phi + t$ and $\bar z = \phi -t$, we see the light-like singularity at $\phi \pm t= 0$ in \eqref{O1O1v2} but no singularity in \eqref{O1O2v2}.
In order to move to dS$_3$ black hole, we just need to replace the parameters as
\begin{align}
 \hat \tau \to i \frac{2 \pi}{L} \hat \tau \, , \quad  \hat{\bar \tau} \to i \frac{2 \pi}{L} \hat{\bar \tau} \, , \quad z \to  \frac{L}{2 \pi} z \, , \quad \bar z \to  \frac{L}{2 \pi} \bar z \, . \label{replace}
\end{align}
Note here that we set $z = \phi + i t, \bar z = \phi - it$. Thus we find a singularity at $\phi = t= 0$ in \eqref{O1O1v2} and a light-like singularity at $t=0, \phi = \pi /|\hat \tau|$ in \eqref{O1O2v2} as mentioned above.

We would like to deform the background by inserting spin-3 charge with the deformation parameter $\alpha$ as in the previous subsection. In perturbative expansion with respect to $\alpha$, the two-point function after the deformation was obtained in \cite{Kraus:2012uf,Gaberdiel:2013jca}.
If CFT operators are inserted in the same boundary, then the two-point function is
\begin{align}
\frac{ \langle \bar{\mathcal{O}}^\pm_1 (z , \bar z) \mathcal{O}^\pm_1 (0,0) \rangle ^{(\alpha)} }{ \langle \bar{\mathcal{O}}^\pm_1 (z , \bar z) \mathcal{O} ^\pm_1 (0 , 0) \rangle ^{(0)} } = 1 + \frac{\alpha w_\pm}{\tau^2} 
\frac{-3 \sin (\hat \tau z) + (\hat \tau z - \hat {\bar \tau} \bar z) (2 + \cos (\hat \tau z))}{2 \sin ^2 \frac{\hat \tau z}{2}} + \mathcal{O} (\alpha^2) \label{def2pt}
\end{align}
with $w_\pm = (1 \pm \lambda) (2 \pm \lambda)/6$.
Thus the deformation only change the singularity structure from $z^{-2h}$ to $z^{-2h-2}$.
If two CFT operators inserted in different boundaries, then the two-point function is
\begin{align}
\frac{ \langle \bar{\mathcal{O}^\pm_1} (z , \bar z) \mathcal{O}^\pm_2 (0,0) \rangle ^{(\alpha)} }{ \langle \bar{\mathcal{O}}^\pm_1 (z , \bar z) \mathcal{O}^\pm_2 (0 , 0) \rangle ^{(0)} } = 1 + \frac{\alpha w_\pm}{\tau^2} 
\frac{ - \sin (\hat \tau z) + (\hat \tau z - \hat {\bar \tau} \bar z) (2 - \cos (\hat \tau z))}{2 \cos ^2 \frac{\hat \tau z}{2}} + \mathcal{O} (\alpha^2) \, , \label{def2pt2}
\end{align}
which does not produce any new singularity.
As discussed above, in order to move to the case with positive cosmological constant, we just need to replace the parameters 
$
\alpha \to  ( 2 \pi /L  )^2\alpha 
$
together with the parameter changes in \eqref{replace}. 
Therefore, we conclude that the deformation changes the singularity structure from $z^{-2h}$ to $z^{-2h-2}$ for the both types of two-point functions \eqref{O1O1v2} and \eqref{O1O2v2}.

\section{Conclusion and discussion}
\label{sec:conclusion}

In this paper, we examined solutions of Chern-Simons (higher-spin) gravity corresponding to dS$_3$ (higher-spin) black holes both from the bulk theory and dual CFT. Some parts of results were presented in a previous letter \cite{Chen:2023prz}, and here the details of their derivations were explained and the analysis was extended in several directions.

We first focused on the simplest case with pure gravity. The gravity theory is described by SL$(2,\mathbb{C})$ Chern-Simons gauge theory and we found a family of solutions obtained by large gauge transformations labeled by winding number $n$. Its CFT dual description is given by Liouville field theory as proposed in \cite{Hikida:2021ese,Hikida:2022ltr,Chen:2022ozy,Chen:2022xse}.
We examined the saddle points of Liouville two-point function by following \cite{Harlow:2011ny} and found that the saddle points of Chern-Simons gravity are realized by $n=0,-1$. We also examined Chern-Simons solutions dual to Liouville multi-point functions. Generic solutions are expected to correspond to Wilson line networks on $S^3$ as in fig.~\ref{fig:conical}. The solutions are labeled by monodromies along the deficit lines. From the detailed analysis of Liouville multi-point functions, we determined the monodromies realized in the gravity theory and found that two saddles contribute to the semi-classical limit of correlators. 
There are special geometries corresponding to two Wilson loops on $S^3$ as analyzed in \cite{Hikida:2021ese,Hikida:2022ltr}. We found that their entropy can be described by the monodromy matrix of four-point functions. From the analysis of Wilson loops in Chern-Simons gauge theory as in \cite{Witten:1988hf}, it is natural to guess that all geometries can be described by combining these approaches in the dual CFT. In any case, it is an important open problem to classify all the possible saddles of Chern-Simons gravity and describe all of them in terms of dual CFT.

We also extended the analysis to the higher-spin gravity on dS$_3$ described by SL$(N,\mathbb{C})$ Chern-Simons gauge theory. As in the pure gravity case, we classified the solutions by large gauge transformations or equivalently monodromy matrix defined in \eqref{dShol}. We then examined two-point functions in the dual CFT, i.e. $\mathfrak{sl}(N)$ Toda field theory as in the Liouville case.
We found out that the allowed saddles of Chern-Simons higher-spin gravity are given by the solutions labeled by \eqref{wrho}. We also examined the properties of higher-spin dS$_3$ black holes in Prokushkin-Vasiliev theory from the propagation of bulk scalar field. In particular, we found a light-like singularity in the two-point function between two boundaries at past and future infinities.
It should be important to analyze gravity solutions dual to correlation functions of $s$ heavy operators in $\mathfrak{sl}(N)$ Toda field theory as was done in section \ref{sec:multi} for Liouville field theory. We can evaluate the action of the Chern-Simons theory for the configuration in fig.\,\ref{fig:conical}. Near the defect lines, we may put the gauge field configuration as in \eqref{congauge}. For one insertion of defect line, the value of action is shifted by
\begin{align}
    \frac{\pi}{3} c^{(g)} \frac{(\rho - \eta^{(i)} , \Lambda^{(i)})}{(\rho ,\rho)} \, .
\end{align}
Adding the topological contribution with bulk winding numbers, the total contribution is
\begin{align}
      \frac{\pi}{3} c^{(g)} \frac{\left[(\rho , \Lambda^{(0)}) - \sum_i (\eta^{(i)} , \Lambda^{(i)})\right]}{(\rho ,\rho)} \, . 
\end{align}
It is an important problem to determine the set of possible $\Lambda$ from dual Toda field theory. 

There are several open problems to be pursued. Among them, we would like to consider the followings in near future. Firstly, we have examined only the semi-classical contributions to the Gibbons-Hawking entropy, however there are also perturbative corrections in $1/c^{(g)}$. These are expected to be asymptotic series, and it is important to see the relation to other saddles realized in Chern-Simons gravity.%
\footnote{See \cite{Benjamin:2023uib} for an interesting work for the AdS$_3$ case.}
In the dual CFT, we can obtain exact expressions, so it should be possible to analyze for all orders in $1/c^{(g)}$.
Furthermore, we would like to extend the analysis to more generic cases. It should be possible to include non-zero rotations and/or Maxwell fields. It is also interesting to consider higher dimensional holography, e.g. dS$_4$/CFT$_3$ by \cite{Anninos:2011ui}. It is also nice if we can find the relation to superstring theory, see the series of previous works \cite{Gaberdiel:2013vva,Gaberdiel:2014cha} and 
\cite{Creutzig:2011fe,Creutzig:2013tja,Creutzig:2014ula} for the attempts to relate stringy and higher-spin holographic dualities on AdS$_3$.

\subsection*{Acknowledgements}

We are grateful to Katsushi Ito, Tatsuma Nishioka, Hongfei Shu, Evgeny Skvortsov, Shigeki Sugimoto and Tadashi Takayanagi for useful discussions. We also thank the organizers of the workshop ``Higher Spin Gravity and its Application'' at APCTP for the hospitality.
The work is partially supported by JSPS Grant-in-Aid for Transformative Research Areas (A) ``Extreme Universe'' No.\,21H05187
and JSPS Grant-in-Aid for Scientific Research (B) No.\,23H01170.
The work of H.\,Y.\,C. is supported in part by Ministry of Science and Technology (MOST) through the grant 1
11-2628-M-002-004-MY3.
This work of Y.\,H. is supported by JSPS Grant-in-Aid for Scientific Research (B) No.\,19H01896 and JSPS Grant-in-Aid for Scientific Research (A) No.\,21H04469.
Y.\,T. is supported by Grant-in-Aid for JSPS Fellows No.\,22J21950, No.\,22KJ1971.
The work of T.\,U. is supported by JSPS Grant-in-Aid for Early-Career Scientists No.\,22K14042.

\appendix

\section{Chern-Simons gravity on \texorpdfstring{(A)dS$_3$}{(A)dS3}}
\label{app:cs}

In this appendix, we explain in some details about the Chern-Simons descriptions of gravity theory on AdS$_3$ and dS$_3$ by following \cite{Witten:1988hc,Witten:1989ip,Witten:2010cx}.
Related works may be found in, e.g., \cite{Castro:2020smu,Anninos:2021ihe}.
We consider complex Chern-Simons gauge theory with the action
\begin{align}
S = \frac{t}{2} S_\text{CS}[A] + \frac{\tilde t}{2} S_\text{CS}[\tilde{ A}] \, , \quad S_\text{CS}[A] = \frac{1}{4 \pi} \int  \text{tr} \left(A \wedge d A + \frac{2}{3} A \wedge A \wedge A \right) \, . \label{CCS}
\end{align}
We set the gauge group as $G = \text{SL}(2,\mathbb{C})$ (but it is easy to generalize to the case with $G = \text{SL}(N,\mathbb{C})$).
Let us first set $\tilde{A} = \bar{A}$ with $\bar{A}$ as a complex conjugate of $A$. Moreover, we set
\begin{align}
t = n + i s , \quad \tilde t = n - i s 
\end{align}
with $n \in \mathbb{Z}$ and $s \in \mathbb{R}$. Then the action can be written as
\begin{align}
 S = - s \, \text{Im}\,  S_\text{CS} + n \, \text{Re}\,  S_\text{CS} \, .
\end{align}
The quantum theory is given by a path integral of $\exp (i S)$ over $A, \tilde{A}$.
When the real slice of $\text{SL}(2,\mathbb{C})$ is $\text{SU}(2)$, then $\text{Re}\, S_\text{CS}$ is gauge invariant modulo $2\pi$ if the trace, $\text{tr}$, is properly normalized. Thus the gauge invariance of $\exp (i S)$ requires $n \in \mathbb{Z}$. However, there is no such constraint for $\text{Im} \, S_\text{CS}$, thus we can use any real $s$ (or any complex $s$ after analytic continuation).
It is convenient to use $A = {\mathcal A} + i {\mathcal B}$, where ${\mathcal A},{\mathcal B}$ take in some real forms of $\mathfrak{sl}(2,\mathbb{C})$. 
The action is rewritten as
\begin{align}
\begin{aligned}
S &= \frac{n}{4 \pi} \int \text{tr} \left({\mathcal A} \wedge d {\mathcal A} - {\mathcal B} \wedge d {\mathcal B} + \frac{2}{3} {\mathcal A} \wedge {\mathcal A} \wedge {\mathcal A} - 2 {\mathcal A} \wedge {\mathcal B} \wedge {\mathcal B} \right) \\
& \quad - \frac{s}{2\pi} \int \text{tr} \left( {\mathcal A} \wedge d {\mathcal B} + 2 {\mathcal A} \wedge {\mathcal A} \wedge {\mathcal B} - \frac{2}{3} {\mathcal B} \wedge {\mathcal B} \wedge {\mathcal B} \right) \, .
\end{aligned}
\end{align}

Let us consider the cases with positive and negative cosmological constants \cite{Witten:1989ip}. 
We first consider $(2+1)$-dimensional gravity in a Lorentzian space-time with positive cosmological constant. In this case, we set $s \in \mathbb{R}$ and $\tilde A$ as the complex conjugate of $A$ without taking its transposition. This implies that our real slice is ${\mathcal A},{\mathcal B} \in \mathfrak{sl}(2,\mathbb{R})$. ${\mathcal A}$ and ${\mathcal B}$ are identified with spin connection $\omega$ and vielbein $e$, respectively. The case with non-zero $n$ may lead to {additional} gravitational Chern-Simons theory \cite{Witten:1988hc}, but we set $n=0$ for our purpose.
A vacuum solution is given by dS$_3$. 
We then consider three-dimensional gravity in an Euclidean space-time with negative cosmological constant. In this case, we set $s = - 2 i k$ with real $k$ and  $\tilde A$ as the complex conjugate of $A$ with taking its transposition. This implies that our real slice is $\mathcal{A}, i \mathcal{B} \in \mathfrak{su}(2)$. A vacuum solution is given by AdS$_3$.

In previous works \cite{Hikida:2021ese,Hikida:2022ltr,Chen:2022ozy,Chen:2022xse}, we construct Chern-Simons gravity in a Lorentzian space-time with positive cosmological constant by performing an analytic continuation as $k \to i s/2$ with $s$ real. However, as indicated above, this is not enough. For AdS$_3$, we assign $\tilde A = A^\dagger$ by combining the simple complex conjugate and the transposition. However, for dS$_3$, we set $\tilde A = A^*$ by the simple complex conjugate together with $k \to i s /2 $. Therefore, the simple change of variable from $k \to  i s /2 $ is not enough. For instance, in the case of dS$_3$, $A$ and $\tilde A$ and related by a simple complex conjugation, the central charge of dual energy momentum tensor would be $c =3is$ and $\bar c = - 3is$ for holomorphic and anti-holomorphic sector, respectively, with large $s$. However, for AdS$_3$, the central charge of dual energy momentum tensor is $c = \bar c = 6k$ both for holomorphic and anti-holomorphic sector with large $k$. This implies that naive analytic continuation of $k \to i s /2 $ leads to $c = \bar c = 3 is$. Therefore, in order to map the energy-momentum tensor in the anti-holomorphic part of CFT and the Chern-Simons gauge field $\tilde A$, we need to perform a field redefinition of one of them, see, e.g. the end of section 2.4 of \cite{Ouyang:2011fs}.

Here we would like to mention the cases where $A$ and $\tilde{A}$ are independent with each other. We may write $A = {\mathcal A} + {\mathcal B}$ and $\tilde{A} = {\mathcal A} - {\mathcal B}$. Then the action \eqref{CCS} becomes
\begin{align}
\begin{aligned}
S &= \frac{n}{4 \pi} \int \text{tr} \left({\mathcal A} \wedge d {\mathcal A} + {\mathcal B} \wedge d {\mathcal B} + \frac{2}{3} {\mathcal A} \wedge {\mathcal A}\wedge {\mathcal A} + 2 {\mathcal A} \wedge {\mathcal B} \wedge {\mathcal B} \right) \\
& \quad +  \frac{i s}{2\pi} \int \text{tr} \left( {\mathcal A} \wedge d {\mathcal B} + 2 {\mathcal A} \wedge {\mathcal A} \wedge {\mathcal B} + \frac{2}{3} {\mathcal B} \wedge {\mathcal B} \wedge {\mathcal B} \right) \, .
\end{aligned}
\end{align}
Setting $n = 0$, the action can be identified with Einstein-Hilbert action in the first order formulation, where ${\mathcal A},{\mathcal B}$ are spin-connection $\omega$, vielbein $e$, respectively. Let us first set ${\mathcal A},{\mathcal B} \in \mathfrak{su}(2)$, then the gravity theory describes three-dimensional Euclidean space with positive cosmological constant. A vacuum solution is given by Euclidean dS$_3$, i.e. $S^3$. In order to describe the Hartle-Hawking universe, then we use $\tilde{A} = A^*$ for $T \geq 0$ with global time $T$ and $A, \tilde{A} \in \mathfrak{su}(2)$ independent with each other for $T < 0$. Let us next set ${\mathcal A} ,{\mathcal B} \in \mathfrak{sl}(2,\mathbb{R})$ with $s \to - 2 i k$ $(k \in \mathbb{R})$. In this case, the gravity theory describes (2 +1)-dimensional Lorentzian space-time with a negative cosmological constant. A vacuum solution is given by Lorentzian AdS$_3$.

\section{Semi-classical limit of \texorpdfstring{$\Upsilon_b$}{Upsilon} function}
\label{app:Upsilon}
Here we summarize the properties of the Upsilon function $\Upsilon_b(x)$, especially on the asymptotic behavior in the semi-classical limit. The basic properties of the Upsilon function are summarized for real $b$ in, e.g. appendix A of \cite{Harlow:2011ny}. Since our interests include an imaginary central charge, we extend to the case with complex $b$. See also \cite{Ribault:2014hia} for more detailed discussions on analytic continuation of the Upsilon function. 

The Upsilon function $\Upsilon_b(x)$ is defined as the unique solution to the recursion relations
\begin{align}\label{defups}
    \begin{aligned}
    \Upsilon_b(x+b)&=\gamma(bx)b^{1-2bx}\Upsilon_b(x) \, ,\\
    \Upsilon_b\left(x+\frac{1}{b}\right)&=\gamma\left(\frac{x}{b}\right)b^{\frac{2x}{b}-1}\Upsilon_b(x) \, ,
    \end{aligned}
\end{align}
where $\gamma(x)$ is defined by \eqref{smallgamma}.
When $\Re b>0$, there exists an integral form of the $\Upsilon_b$ function:
\begin{align}\label{upsint}
    \log\Upsilon_b(x)=\int_0^\infty\frac{dt}{t}\left[\left(\frac{Q}{2}-x\right)^2e^{-t}-\frac{\sinh^2\left(\left(\frac{Q}{2}-x\right)\frac{t}{2}\right)}{\sinh\frac{tb}{2}\sinh\frac{t}{2b}}\right],\qquad 0<\Re x<\Re Q \, , 
\end{align}
where $Q=b+1/b$.
By using the defining recursion relations \eqref{defups}, we can extend the range of $x$ to the whole complex plane. 

For applications to CFT, it is useful to introduce the central charge $c=1+6Q^2\in\mathbb{C}$. Note that we can express it as $c=13+6(b^2+b^{-2})$. For any $c\notin(-\infty,1]$, we can choose the branch of $b$-plane such that it satisfies $\Re b>0$. Therefore we can define $\Upsilon_b(x)$ in the way described above for any CFTs that has the central charge $c\notin (-\infty,1]$. On the other hand, for $c\in(-\infty,1]$, $b$ should be purely imaginary. Therefore the integral form \eqref{upsint} cannot be defined. In order to consider CFTs that have $c\in (-\infty,1]$, we have to use another function $\hat{\Upsilon}_b(x)$ defined by the modified recursion relations \cite{Zamolodchikov:2005fy}
\begin{align}
    \begin{aligned}
        \hat{\Upsilon}_b(x+b)&=\gamma(bx)(ib)^{1-2bx}\hat{\Upsilon}_b(x) \, ,\\
        \hat{\Upsilon}_b\left(x+\frac{1}{b}\right)&=\gamma\left(\frac{x}{b}\right)(ib)^{\frac{2x}{b}-1}\hat{\Upsilon}_b(x) \, .
    \end{aligned}
\end{align}
For example, to discuss so-called time-like Liouville theory \cite{Strominger:2003fn} we have to use $\hat{\Upsilon}_b(x)$ since it has negative central charge. The function $\hat{\Upsilon}_b(x)$ can be defined for $\Im b<0$, equally $c\notin [25,\infty)$. Therefore we can use either $\Upsilon_b(x)$ or $\hat{\Upsilon}_b(x)$ for CFTs with $c\notin (-\infty,1]\cup[25,\infty)$. Because the theory of our interest has imaginary $c$, henceforth we will discuss only the unhatted Upsilon function $\Upsilon_b(x)$ by promising we always take a branch $\Re b>0$.

Let us discuss the semi-classical limits $b\to0$ of $\Upsilon_b(x)$ assuming that $x$ scales as $x=\eta/b$. For simplicity, we restrict ourselves to real $\eta$,\footnote{When both $b$ and $\eta$ are imaginary, the condition $0<\Re \eta<1$ does not imply $0<\Re (\eta/b)<\Re Q$. We then would need some additional condition for $\Im \eta$.} which is the situation we are interested in. We first consider $x=\frac{\eta}{b}+\frac{b}{2}$, keeping $\eta$ fixed under $b\to0$. When $0< \eta<1$, we find 
\begin{align}
\begin{aligned}
    b^2\log\Upsilon_b\left(\frac{\eta}{b}+\frac{b}{2}\right) 
      &=-\left(\eta-\frac{1}{2}\right)^2\log b
     \\  & \quad +\int_0^\infty\frac{dt}{t}\left[\left(\eta-\frac{1}{2}\right)^2e^{-t}-\frac{2}{t}\frac{\sinh^2\left((\eta-1/2)\frac{t}{2}\right)}{\sinh\frac{t}{2}}\right]+\mathcal{O}(b^4) \, .
     \end{aligned}
\end{align}
If we define:
\begin{align}
    F(\eta)\equiv\int_0^\infty\frac{dt}{t}\left[\left(\eta-\frac{1}{2}\right)^2e^{-t}-\frac{2}{t}\frac{\sinh^2\left((\eta-1/2)\frac{t}{2}\right)}{\sinh\frac{t}{2}}\right]\label{Def:F(eta)}
\end{align}
and use the integral representation \eqref{upsint} together with an identity 
\begin{align}
    \log x=\int_0^\infty\frac{dt}{t}\left(e^{-t}-e^{-xt}\right) \, ,\quad \Re x>0 \, .
\end{align}
We find the asymptotic formula for $\Upsilon_b$:
\begin{align}\label{upsasymp}
    \Upsilon_b\left(\frac{\eta}{b}\right)=\exp\left(\frac{1}{b^2}\left[-(\eta-1/2)^2\log b+F(\eta)+\mathcal{O}(b\log b)\right]\right) \, .
\end{align}
Here we again emphasize that this formula is applicable for any $b$ with $\Re b>0$, equally $c\notin (-\infty,1]$, but only for $0<\eta<1$. 
We can obtain the asymptotic formula for other ranges of $\eta$ by applying the recursion relations \eqref{defups}. For example, let us consider $-1< \eta<0$, and use inversely the second equation in \eqref{defups},
\begin{align}
    \Upsilon_b\left(\frac{\eta}{b}+\frac{b}{2}\right)=\frac{b^{1-\frac{2\eta}{b^2}}}{\gamma\left(\frac{\eta}{b^2}\right)}\Upsilon_b\left(\frac{\eta+1}{b}+\frac{b}{2}\right) \, ,
\end{align}
we obtain:
\begin{align}\label{upsasymp2}
    \Upsilon_b\left(\frac{\eta}{b}\right)=\frac{b^{1-\frac{2\eta}{b^2}}}{\gamma\left(\frac{\eta}{b^2}\right)}\exp\left(\frac{1}{b^2}\left[-(\eta+1/2)^2\log b+F(\eta+1)+\mathcal{O}(b\log b)\right]\right)\, .
\end{align}


\section{SL(3) Chern-Simons gravity}
\label{app:hsbh}

In section \ref{sec:csgra}, we examined black hole solutions with higher-spin charges of SL$(3)$ Chern-Simons gravity with negative/positive cosmological constant. In this appendix, we summarize the technical details of the bulk analysis for the higher-spin black holes.

\subsection{Higher-spin \texorpdfstring{AdS$_3$}{AdS3} black holes}

We start by examining black hole solutions with higher-spin charges of SL$(3)$ Chern-Simons gravity with negative cosmological constant. In \cite{Gutperle:2011kf}, the configuration of gauge fields corresponding to the black hole solutions was obtained as in \eqref{connection} in the form of \eqref{gauge}. From the gauge field configuration, the metric can be read off from \eqref{metricAdS} as
\begin{align}
\begin{aligned}
	\ell_{\text{AdS}}^{-2}ds^2&=d\rho^2-\left\{\left(2\mu^{\text{AdS}} e^{2\rho}+\frac{\pi{\cal W}^{\text{AdS}}}{k}e^{-2\rho}-\frac{8\pi^2\mu^{\text{AdS}}\left({\cal L}^{\text{AdS}}\right)^2 }{k^2}e^{-2\rho}\right)^2\right.\\
	&\qquad\qquad\left.+\left(e^{\rho}-\frac{2\pi {\cal L}^{\text{AdS}}}{k}e^{-\rho}+\frac{4\pi \mu^{\text{AdS}}{\cal W}^{\text{AdS}}}{k}e^{-\rho}\right)^2\right\}dt^2\\
	&\quad+\left\{ \left(2\mu^{\text{AdS}} e^{2\rho}+\frac{\pi{\cal W}^{\text{AdS}}}{k}e^{-2\rho}+\frac{8\pi^2\mu^{\text{AdS}}\left({\cal L}^{\text{AdS}}\right)^2 }{k^2}e^{-2\rho}\right)^2\right.\\
	&\qquad\left.+\left(e^{\rho}+\frac{2\pi {\cal L}^{\text{AdS}}}{k}e^{-\rho}+\frac{4\pi \mu^{\text{AdS}}{\cal W}^{\text{AdS}}}{k}e^{-\rho}\right)^2+\frac{64\pi^2\left(\mu^{\text{AdS}}\right)^2\left({\cal L}^{\text{AdS}}\right)^2 }{3k^2}\right\}d\phi^2\,.
\end{aligned} 
\label{eq:metric3}
\end{align}
In \cite{Gutperle:2011kf}, they proposed the conditions of black hole as (i), (ii), and (iii) given above \eqref{integrability}.
The authors of \cite{Ammon:2011nk} have found a good gauge transformation and shown that the condition (i) is satisfied. We review this transformation below.

We consider the connection with the gauge transformation \eqref{gaugetrans}.
Then the metric and spin-3 field may take the form
\begin{align}
\begin{aligned}
	ds^2 &= g_{rr}(r)dr^2+g_{tt}(r)dt^2+g_{\phi\phi}(r)d\phi^2\,,\\
	\varphi_{\alpha\beta\gamma}dx^{\alpha}dx^{\beta}dx^{\gamma} &= \varphi_{\phi rr}(r)d\phi dr^2+\varphi_{\phi tt}(r)d\phi dt^2+\varphi_{\phi \phi\phi}(r)d\phi^3\,,
\end{aligned}
\end{align}
where $r=\rho-\rho_+$ and $\rho=\rho_+$ are the event horizons, we also set $\exp (2\rho_+)= 2\pi {\cal L}^{\text{AdS}} /k$. 
In order to satisfy the condition (i), we demand 
\begin{align}
	\beta = 2\pi\sqrt{\frac{2g_{rr}(0)}{-g_{tt}''(0)}} = 2\pi\sqrt{\frac{2\varphi_{\phi rr}(0)}{-\varphi_{\phi tt}''(0)}}\,,
\end{align}
and the metric components enjoy the following symmetries:
\begin{align}
\begin{aligned}
	&g_{rr}(-r)=g_{rr}(r)\,,\quad g_{tt}(-r)=g_{tt}(r)\,,\quad g_{\phi\phi}(-r)=g_{\phi\phi}(r)\,,\\
	&\varphi_{\phi rr}(-r)=\varphi_{\phi rr}(r)\,,\quad \varphi_{\phi tt}(-r)=\varphi_{\phi tt}(r)\,,\quad \varphi_{\phi\phi\phi}(-r)=\varphi_{\phi\phi\phi}(r)\,.
\end{aligned}
\end{align}
Then the transformed metric becomes \eqref{metric}
with
\begin{align}
\begin{aligned}
	a_t&=(C-1)^2\left( 4\gamma^{\text{AdS}}-\sqrt{C} \right)^2\, , \quad 
	a_\phi=(C-1)^2\left( 4\gamma^{\text{AdS}}+\sqrt{C} \right)^2\,,\\
	b_t&=16(\gamma^{\text{AdS}})^2(C-2)(C^2-2C+2)-8\gamma^{\text{AdS}}\sqrt{C}(2C^2-6C+5)+C(3C-4)\,,\\
	b_\phi&=16(\gamma^{\text{AdS}})^2(C-2)(C^2-2C+2)+8\gamma^{\text{AdS}}\sqrt{C}(2C^2-6C+5)+C(3C-4)\,.
\end{aligned} \label{abdS}
\end{align}

\subsection{Higher-spin \texorpdfstring{dS$_3$}{dS3} black holes}

We then move to the black hole solutions with higher-spin charges of SL$(3)$ Chern-Simons gravity with positive cosmological constant. We consider the configuration of gauge fields in the form of \eqref{gaugedS}.
We here use a non-rotating solution
\begin{align}
\begin{aligned}
	&a = \left(L_1 + \frac{2 \pi {\cal L}}{\kappa} L_{-1} +i \frac{\pi {\cal W}}{2\kappa}  W_{-2}\right) dz +i \mu \left(W_2 + \frac{4 \pi {\cal L}}{\kappa} W_0 + \frac{4 \pi^2 {\cal L}^2}{\kappa^2} W_{-2} -i \frac{4 \pi {\cal W}}{\kappa} L_{-1}\right) d{\bar z}\,,  \\
	&\bar{a} = \left(L_{-1 }+ \frac{2 \pi {\cal L}}{\kappa}  L_{1} -i \frac{\pi{\cal W}}{2 \kappa}  W_{2}\right) d{\bar z} -i \mu \left(W_{-2} + \frac{4 \pi {\cal L}}{\kappa} W_0 + \frac{4 \pi^2 {\cal L}^2}{\kappa^2} W_{2} + i\frac{4 \pi {\cal W}}{\kappa} L_{1}\right) dz\,,
\end{aligned} 
\end{align}
with $z = i t + \phi,\bar z=-(it-\phi)$, then we have \eqref{dSconfig}
\begin{align}
\begin{aligned}
e^{-(\pi i/2)L_0} a e^{(\pi i/2)L_0}&= i \left(L_1 - \frac{2 \pi { \mathcal L}}{\kappa}  L_{-1}- \frac{\pi { \mathcal W}}{2 \kappa}  W_{-2}\right) dz \\
&\quad
- i \mu \left(W_2 - \frac{4 \pi { \mathcal L}}{\kappa} W_0 + \frac{4 \pi^2 {  \mathcal L}^2}{\kappa} W_{-2} + \frac{4 \pi  \mathcal{W}}{\kappa} L_{-1}\right) d \bar z \, .
\end{aligned}
\end{align}
The metric after changing $\tilde\rho\to i\theta$ is 
\begin{align}
\begin{aligned}
	&\ell^{-2}ds^2=d\theta^2\\
	&+\left\{\left(2\mu e^{2i\theta}+\frac{\pi{\cal W}}{\kappa}e^{-2i\theta}-\frac{8\pi^2\mu{\cal L}^2 }{\kappa^2}e^{-2i\theta}\right)^2+\left(e^{i\theta}-\frac{2\pi {\cal L}}{\kappa}e^{-i\theta}+\frac{4\pi \mu{\cal W}}{\kappa}e^{-i\theta}\right)^2\right\}dt^2\\
	&+\left\{ \left(2\mu e^{2i\theta}+\frac{\pi{\cal W}}{\kappa}e^{-2i\theta}+\frac{8\pi^2\mu{\cal L}^2 }{\kappa^2}e^{-2i\theta}\right)^2
	+\left(e^{i\theta}+\frac{2\pi {\cal L}}{\kappa}e^{-i\theta}+\frac{4\pi \mu{\cal W}}{\kappa}e^{-i\theta}\right)^2+\frac{64\pi^2\mu^2{\cal L}^2 }{3\kappa^2}\right\}d\phi^2\,.
\end{aligned} 
\label{eq:metric3d}
\end{align}

Let us consider the condition (i) for dS$_3$ black hole case. We again consider the connection with the gauge transformation $g(\tilde \theta)\in \text{SL}(3,\mathbb{R})$,
where $\tilde\theta=\theta-\theta_+$ and $\exp ( 2i\theta_+)= 2\pi {\cal L} /{\kappa}$. Then the metric and spin-3 field may take the form
\begin{align}
\begin{aligned}
	ds^2 &= g_{\tilde\theta\tilde\theta}(\tilde\theta)d\tilde\theta^2+g_{tt}(\tilde\theta)dt^2+g_{\phi\phi}(\tilde\theta)d\phi^2\,,\\
	\varphi_{\alpha\beta\gamma}dx^{\alpha}dx^{\beta}dx^{\gamma} &= \varphi_{\phi \tilde\theta\tilde\theta}(\tilde\theta)d\phi dr^2+\varphi_{\phi tt}(\tilde\theta)d\phi dt^2+\varphi_{\phi \phi\phi}(\tilde\theta)d\phi^3\,.
\end{aligned}
\end{align}
In order to satisfy the condition (i), we also demand 
\begin{align}
	\beta = 2\pi\sqrt{\frac{2g_{\tilde\theta\tilde\theta}(0)}{-g_{tt}''(0)}}= 2\pi\sqrt{\frac{2\varphi_{\phi \tilde\theta\tilde\theta}(0)}{-\varphi_{\phi tt}''(0)}}\,,
\end{align}
and 
\begin{align}
\begin{aligned}
	&g_{\tilde\theta\tilde\theta}(-\tilde\theta)=g_{\tilde\theta\tilde\theta}(\tilde\theta)\,,\quad g_{tt}(-\tilde\theta)=g_{tt}(\tilde\theta)\,,\quad g_{\phi\phi}(-\tilde\theta)=g_{\phi\phi}(\tilde\theta)\,,\\
	&\varphi_{\phi \tilde\theta\tilde\theta}(-\tilde\theta)=\varphi_{\phi \tilde\theta\tilde\theta}(\tilde\theta)\,,\quad \varphi_{\phi tt}(-\tilde\theta)=\varphi_{\phi tt}(\tilde\theta)\,,\quad \varphi_{\phi\phi\phi}(-\tilde\theta)=\varphi_{\phi\phi\phi}(\tilde\theta)\,.
\end{aligned}
\end{align}
Then the transformed metric becomes
\begin{align}
\begin{aligned}
	g_{\tilde\theta\tilde\theta}&=\frac{(C-2)(C-3)}{(C-2-\cos^2(\tilde\theta))^2}\,,\\
	g_{tt}&=-\left(\frac{8\pi{\cal L}}{\kappa}\right)\left(\frac{C-3}{C^2}\right)\frac{(a_t+b_t\cos^2(\tilde\theta))\sin^2(\tilde\theta)}{(C-2-\cos^2(\tilde\theta))^2}\,,\\
	g_{\phi\phi}&=-\left(\frac{8\pi{\cal L}}{\kappa}\right)\left(\frac{C-3}{C^2}\right)\frac{(a_\phi+b_\phi\cos^2(\tilde\theta))\sin^2(\tilde\theta)}{(C-2-\cos^2(\tilde\theta))^2}+\left(\frac{8\pi{\cal L}}{\kappa}\right)\left(1+\frac{16}{3}\gamma^2 +12\gamma\zeta\right)
\end{aligned}
\end{align}
with
\begin{align}
\begin{aligned}
	a_t&=(C-1)^2\left( 4\gamma-\sqrt{C} \right)^2\,, \quad
	a_\phi=(C-1)^2\left( 4\gamma+\sqrt{C} \right)^2\,,\\
	b_t&=16\gamma^2(C-2)(C^2-2C+2)-8\gamma\sqrt{C}(2C^2-6C+5)+C(3C-4)\,,\\
	b_\phi&=16\gamma^2(C-2)(C^2-2C+2)+8\gamma\sqrt{C}(2C^2-6C+5)+C(3C-4)\,.
\end{aligned} \label{ab}
\end{align}


\section{Prokushkin-Vasiliev theory}
\label{app:hslambda}

In section \ref{sec:probe}, we have examined black hole solutions with higher-spin charges in the Prokushkin-Vasiliev theory \cite{Prokushkin:1998bq}. We have computed the gravity partition function of the black hole solution and evaluated two-point functions of bulk scalar field on the black hole background. In the main context, we presented the analysis mainly from the dual CFT as it is rather non-trivial. In this appendix, we explain its bulk counterparts, which may be obtained quite straightforwardly by analytically continuing the analysis done in \cite{Kraus:2011ds,Kraus:2012uf} for the case of AdS$_3$.

\subsection{Higher-spin black holes}
\label{app:hslambdabh}

As usual, we start by reviewing the known results for AdS$_3$ case.
The massless sector of the Prokushkin-Vasiliev theory can be described by Chern-Simons gauge theory with the action \eqref{CSaction}, but the gauge algebra is given by an infinite dimensional one denoted by $\mathfrak{hs}[\lambda]$.
The generators of the algebra are 
$V_m^s$ with $s\geq0, |m|<s$, and the commutation relations may be written as
\begin{align}
    [V_m^s , V_n^t] = \sum_u g_u^{st}(m,n;\lambda) V_{m+n}^{s + t- u} \, ,
\end{align}
see also \eqref{comm}.
The explicit expression of the structure constants may be found in \cite{Pope:1989sr,Gaberdiel:2011wb}. It is also convenient to define the lone star product 
\begin{align}
V_m^s \star V_n^t = \frac{1}{2} \sum_u g_u^{st}(m,n;\lambda) V_{m+n}^{s + t- u} \, . 
\end{align}

The black hole solutions of the Chern-Simons gravity with negative cosmological constant were obtained in \cite{Kraus:2011ds}. We consider the gauge configuration of the form \eqref{gaugeE} with \eqref{gaugeE2}. The gauge configuration corresponding to the BTZ black hole is given by (see \eqref{solution})
\begin{align}
    a_z = V_1^2 - \frac{2 \pi \mathcal{L}^\text{AdS}}{k} V_{-1}^2 \, , \quad 
    a_{\bar z} = 0 \, .
\end{align}
On the other hand, the ansatz for the gauge configuration with higher spin charges is considered as
\begin{align}
\begin{aligned}
	&a_z=V_1^2-\frac{2 \pi {\cal L}^{\text{AdS}}}{k}  V_{-1}^2-N(\lambda)\frac{\pi{\cal W}^{\text{AdS}}}{2k}V_{-2}^3+J_{\text{AdS}}\,,\\
	&a_{\bar z} = - \mu^{\text{AdS}} N(\lambda)\left( a_z\star a_z-\frac{2\pi{\cal L}^{\text{AdS}}}{3k}(\lambda^2-1)\right) \, .
\end{aligned}
\end{align}
Here we use
\begin{align} \label{Nlambda}
N(\lambda)=\sqrt{\frac{20}{\lambda^2-4}}
\end{align}
and 
\begin{align}
J_{\text{AdS}}=J_{\text{AdS}}^{(4)}V_{-3}^4+J_{\text{AdS}}^{(5)}V_{-4}^5+\cdots
\end{align}
where $J^{(s)}_\text{AdS}$ are spin-$s$ charges.
We can check that the gauge configuration solves the equations of motion $[a_z , a_{\bar{z}}] = 0$.

As in the SL(3) case, we require that the eigenvalues of holonomy matrix are the same as that of the BTZ black hole. Since the holonomy matrix is computed as in \eqref{hsholonomyAdS}, we have
\begin{align} \label{omegaads0}
	\Omega_{\text{BTZ}}=i\beta^{\text{AdS}}\left( V_1^2-\frac{2 \pi {\cal L}^{\text{AdS}}}{k}  V_{-1}^2\right)
\end{align}
for the BTZ black hole and 
\begin{align}
	\Omega=i\beta^{\text{AdS}}a_z-2\pi\alpha^{\text{AdS}} N(\lambda)\left( a_z\star a_z-\frac{2\pi{\cal L}^{\text{AdS}}}{3k}(\lambda^2-1)\right)
\label{omegaads}
\end{align}
for the ansatz of higher-spin black hole.
We thus consider the holonomy constraints
\begin{align}
	\text{tr}(\Omega^n)=\text{tr}(\Omega_{\text{BTZ}}^n)\,,\quad n=2,3,\cdots \, .
\end{align}
We treat the effect of spin-3 charge perturbatively. This means that we solve the constraints perturbatively in $\alpha^\text{AdS}$. Then we can express charges $\mathcal{L}^\text{AdS},\mathcal{W}^\text{AdS}$ in terms of parameters $k$ and $\tau^{\text{AdS}}$. Integrating an equation in \eqref{intZ}, we arrive at the expression of the partition function in \eqref{AdSPF}.

Let us turn to the case with positive cosmological constant. 
We use the gauge configuration of the form \eqref{gaugedS}.
For the black hole without higher-spin charges, the gauge configuration is (see \eqref{soldS})
\begin{align}
 	&a_z=V_1^2+\frac{2 \pi {\cal L}}{\kappa}V_{-1}^2 \, , \quad a_{\bar z} = 0 \, .
\end{align}
For the black hole with higher-spin charges, 
we use the ansatz
\begin{align} \label{adshs}
\begin{aligned} 
	&a_z=V_1^2+\frac{2 \pi {\cal L}}{\kappa}  V_{-1}^2+iN(\lambda)\frac{\pi{\cal W}}{2k}V_{-2}^3+J\,,\\
	&a_{\bar z} = \mu N(\lambda)\left( a_z\star a_z + \frac{2\pi{\cal L}}{3\kappa}(\lambda^2-1)\right) \, .
\end{aligned}
\end{align}
Here $N(\lambda)$ is given in \eqref{Nlambda} and 
\begin{align}
   J=J^{(4)}V_{-3}^4+J^{(5)}V_{-4}^5+\cdots\,.
\end{align}
We can check that this ansatz reduces to \eqref{dSconfig} with the SL(3) case. 
The holonomy matrices become
\begin{align}
    &\Omega_{\text{dSBH}}=\beta\left( V_1^2+\frac{2 \pi {\cal L}}{\kappa}  V_{-1}^2\right)
\label{eq:omega12}
\end{align}
for the black hole without higher-spin charges and
\begin{align}
	&\Omega=\beta a_z+2\pi\alpha N(\lambda)\left( a_z\star a_z+\frac{2\pi{\cal L}}{3\kappa}(\lambda^2-1)\right)
\label{eq:omega2}
\end{align}
for the black hole with higher-spin charges.
Here we also use the same constructions as the negative cosmological constant case:
\begin{align}
	\text{tr}(\Omega^n)=\text{tr}(\Omega_{\text{dSBH}}^n)\,,\quad n=2,3,\cdots
\end{align}
The lower even-$n$ traces are
\begin{align}
\begin{aligned}
	&\text{tr}(\Omega_{\text{dSBH}}^2)=-8\pi^2\,,\\
	&\text{tr}(\Omega_{\text{dSBH}}^4)=\frac{8\pi^4}{5}(3\lambda^2-7)\,,\\
	&\text{tr}(\Omega_{\text{dSBH}}^6)=-\frac{8\pi^6}{7}(3\lambda^4-18\lambda^2+31)\,,
\end{aligned}
\label{omegads}
\end{align}
and all odd-$n$ traces vanish. Solving the constraint equations \eqref{omegads} perturbatively in $\alpha$, we obtain the following solutions:
\begin{align}
\begin{aligned}
	{\cal L}&=-\frac{\kappa}{8\pi\tau^2}+\frac{5\kappa}{6\pi\tau^6}\alpha^2-\frac{50\kappa}{3\pi\tau^{10}}\frac{\lambda^2-7}{\lambda^2-4}\alpha^4+\frac{2600\kappa}{27\pi\tau^{14}}\frac{5\lambda^4-85\lambda^2+377}{(\lambda^2-4)^2}\alpha^6\\
	&\qquad-\frac{68000\kappa}{81\pi\tau^{18}}\frac{20\lambda^6-600\lambda^4+6387\lambda^2-23357}{(\lambda^2-4)^3}\alpha^8+\cdots \,,\\
	{\cal W}&=-\frac{\kappa}{3\pi\tau^5}\alpha+\frac{200\kappa}{27\pi\tau^9}\frac{\lambda^2-7}{\lambda^2-4}\alpha^3-\frac{400\kappa}{9\pi\tau^{13}}\frac{5\lambda^4-85\lambda^2+377}{(\lambda^2-4)^2}\alpha^5\\
	&\qquad+\frac{32000\kappa}{81\pi\tau^{17}}\frac{20\lambda^6-600\lambda^4+6387\lambda^2-23357}{(\lambda^2-4)^3}\alpha^7+\cdots\,.
\end{aligned}
\label{dSLW}
\end{align}
In addition, we find 
\begin{align}
\begin{aligned}
	J^{(4)}&=\frac{35}{9\tau^8}\frac{1}{\lambda^2-4}\alpha^2-\frac{700}{9\tau^{12}}\frac{2\lambda^2-21}{(\lambda^2-4)^2}\alpha^4+\frac{2800}{9\tau^{16}}\frac{20\lambda^4-480\lambda^2+3189}{(\lambda^2-4)^3}\alpha^6+\cdots\,,\\
	J^{(5)}&=\frac{100\sqrt{5}}{9\tau^{11}}\frac{1}{(\lambda^2-4)^{3/2}}\alpha^3-\frac{400\sqrt{5}}{27\tau^{15}}\frac{44\lambda^2-635}{(\lambda^2-4)^{5/2}}\alpha^5+\cdots,\\
	J^{(6)}&=\frac{14300}{81\tau^{14}}\frac{1}{(\lambda^2-4)^2}\alpha^4+\cdots\,.
\end{aligned}
\label{dSJ}
\end{align}
Note that we have the relations
\begin{align}
	\frac{{\cal L}^{\text{AdS}}}{k}=-\frac{{\cal L}}{\kappa} \,,\quad \frac{{\cal W}^{\text{AdS}}}{k}=-i\frac{{\cal W}}{\kappa} \,,\quad J_{\text{AdS}}^{(4)}= J^{(4)},\quad J_{\text{AdS}}^{(5)}= iJ^{(5)}\,,\quad J_{\text{AdS}}^{(6)}= -J^{(6)} \, , 
\end{align}
which imply that
\begin{align}
	J_{\text{AdS}}^{(s)}= e^{(\pi i/2)s }J^{(s)} \, .
\label{eq:Jrule}
\end{align}
Integrating an equation in \eqref{intZ}, we arrive at
\begin{align} \label{ZdSBHf}
\begin{aligned}
 &\ln Z(\tau,\alpha)=\frac{i\pi\kappa}{2\tau}\left[1-\frac{4}{3}\frac{\alpha^2}{\tau^4}+\frac{400}{27}\frac{\lambda^2-7}{\lambda^2-4}\frac{\alpha^4}{\tau^8}-\frac{1600}{27}\frac{5\lambda^4-85\lambda^2+377}{(\lambda^2-4)^2}\frac{\alpha^6}{\tau^{12}}\right.\\
	&\hspace{100pt}\left.+\frac{32000}{81}\frac{20\lambda^6-600\lambda^4+6387\lambda^2-23357}{(\lambda^2-4)^3}\frac{\alpha^8}{\tau^{16}}\right]+\cdots
\end{aligned}
\end{align}
as in \eqref{ZdSBH}.

\subsection{Two-point functions}
\label{app:hsbh2pt}

In this subsection, we study the two-point functions in higher-spin black holes on dS$_3$ by extending the argument in AdS$_3$ \cite{Kraus:2012uf}. 
We again start by reviewing the known results in the case of AdS$_3$.
In the Prokushkin-Vasiliev theory, the master field $C$, which contains the bulk scalar field, satisfies the linearized equation 
\begin{align}
   dC+A\star C-C\star {\overline A}=0\,,
\end{align}
where $A$ is the gauge field of $\mathfrak{hs}[\lambda]$ Chern-Simons theory. 
The scalar field can be obtained by $\mathop{\mathrm{Tr}}[C]$, which satisfies a Klein-Gordon equation, see also \cite{Ammon:2011ua}.
According to \cite{Kraus:2012uf}, we may derive the scalar propagator by
\begin{align}
 G_\pm(\rho,\mathbf{x};\mathbf{0})=\pm\frac{\lambda}{\pi}e^{(1\pm\lambda)\rho}\text{Tr}[e^{-\Lambda_\rho}\star c_\pm \star e^{\bar{\Lambda}_\rho}]\,,
\label{sprop}
\end{align}
where $+$ and $-$ denote the standard and alternate quantization conditions, respectively. Here $c_\pm$ is the master field in the trivial gauge with $A=0$. 
The master field $C_\pm$ can be obtained by gauge transformation as
\begin{align}
    C_\pm=g^{-1}\star c_\pm \star \bar g \, .
\end{align}
Moreover, $\Lambda_\rho,\,\bar{\Lambda}_\rho$ are defined as
\begin{align}
    \Lambda_\rho=e^{-\rho V_0^2}\star (a_\mu x^\mu) \star e^{\rho V_0^2},\quad  \bar{\Lambda}_\rho=e^{\rho V_0^2}\star (\bar a_\mu x^\mu) \star e^{-\rho V_0^2} 
\end{align}
with gauge solutions $a,\,\bar{a}$.

At $\lambda=1/2$, the lone star product is known to reduce to the Moyal product. Let us introduce generators $y_\alpha\,(\alpha=1,2)$, which satisfy $[y_\alpha,y_\beta]_\ast=2i\epsilon_{\alpha\beta}$. For the function of $y_\alpha$, the Moyal product acts as the differential form
\begin{align}
    (f\ast g)(y)=\left.\exp\left[ i\epsilon_{\alpha\beta}\frac{\partial}{\partial y_\alpha}\frac{\partial}{\partial y'_\beta} \right]f(y)g(y')\right|_{y=y'}\,,
\end{align}
where $\epsilon_{\alpha\beta}$ is the anti-symmetric tensor $\epsilon_{12}=1$. 
In terms of $y_\alpha$, we can rewrite the $\mathfrak{hs}[\lambda]$ generators
\begin{align}
    V_m^s=\left(\frac{-i}{4}\right)^{s-1}y_1^{s+m-1}y_2^{s-m-1}\,.
\end{align}
The authors of \cite{Kraus:2012uf} has nicely shown that we may obtain the scalar propagator for all backgrounds with following $c_\pm$
\begin{align}
    c_-=e^{-iy_1y_2}\, ,\quad c_+=\left(\frac{1}{2i}\right)y_1\ast e^{-iy_1y_2}\ast y_2 \, ,
\label{master}
\end{align}
and the propagator becomes
\begin{align}
 G_\pm(\rho,\mathbf{x};\mathbf{0})=\pm\frac{1}{2\pi}e^{(1\pm\frac{1}{2})\rho}\text{Tr}[e^{-\Lambda_\rho}\ast c_\pm \ast e^{\bar{\Lambda}_\rho}]\,.
\end{align}
Using the solution (\ref{solution}), it leads to the correct AdS$_3$ result.
 Besides, for general value of $\lambda$, we should use $c_\pm$ as the highest weight state of $\mathfrak{hs}[\lambda]$.

Let us extend this calculation for dS$_3$ black holes. We use the gauge configuration \eqref{connection} and the solution \eqref{soldS} for the case of pure dS$_3$ and define $\Lambda_{\tilde{\rho}}$ and $\bar{\Lambda}_{\tilde{\rho}}$ as
\begin{align}
\begin{aligned}
    \Lambda_{\tilde\rho}&=e^{-\tilde\rho V_0^2}\ast \left[i\left( V_{1}^2-\frac{2\pi{\cal L}}{\kappa}V_{-1}^2\right)z\right]\ast e^{\tilde\rho V_0^2},\\
    \bar{\Lambda}_{\tilde\rho}&=e^{\tilde\rho V_0^2}\ast\left[i\left( V_{-1}^2-\frac{2\pi{\cal L}}{\kappa}V_{1}^2\right){\bar z}\right]\ast e^{-\tilde\rho V_0^2} \, ,
\end{aligned}
\end{align}
see \eqref{tilderho}.
It leads to the scalar bulk-boundary propagator at $\lambda=1/2$ as
\begin{align}
\begin{aligned}
    G_\pm(\tilde\rho,\mathbf{x};\mathbf{0})&=\pm\frac{1}{2\pi}e^{(1\pm\frac{1}{2})\tilde\rho}\text{Tr}[e^{-\Lambda_{\tilde\rho}}\ast c_\pm \ast e^{\bar{\Lambda}_{\tilde\rho}}]\\
    &=\pm\frac{1}{2\pi}\left( \frac{e^{-\tilde\rho}}{e^{-2\tilde\rho}\cosh\left( \frac{z}{2\tau} \right)\cosh\left( \frac{\bar z}{2\bar{\tau}} \right)-4\tau\bar{\tau}\sinh\left( \frac{z}{2\tau} \right)\sinh\left( \frac{\bar z}{2\bar{\tau}} \right)} \right)^{1\pm\frac{1}{2}} \, .
\end{aligned}
\label{GBTZ}
\end{align}
At $\tilde\rho\rightarrow\infty$, the two-point function can be read off as (see \eqref{dS2pt})
\begin{align} \label{bdS2pt}
	G_\pm(r,\mathbf{x};\mathbf{0})\sim\left[ -\cosh (r_+t) + \cos (r_+\phi) \right]^{-(1\pm\frac{1}{2})}\, ,
\end{align}
where $z=it+\phi$. 
At $\tilde\rho\rightarrow-\infty$, we also obtain the behavior of different boundaries as (see \eqref{dS2pts})
\begin{align} \label{bdS2pts}
	G_\pm(r,\mathbf{x};\mathbf{0})\sim\left[ \cosh (r_+t) + \cos (r_+\phi) \right]^{-(1\pm\frac{1}{2})} \, .
\end{align}
Let us turn to the case with higher-spin charges. We use the solution \eqref{adshs} as
\begin{align}
\begin{aligned}
    &e^{-(\pi i/2)V_0^2}a e^{(\pi i/2)V_0^2}=ia_z-i\mu N(\lambda)\left( a_z\star  a_z-\frac{2\pi {\cal L}}{3k}(\lambda^2-1)\right)d\bar z\,,\\
    &a_z=V_{1}^2-\frac{2\pi{\cal L}}{\kappa}V_{-1}^2-N(\lambda)\frac{\pi{\cal W}}{2\kappa}V_{-2}^3+J^{(4)}V_{-3}^4+J^{(5)}V_{-4}^5+\cdots \, .
\end{aligned}
\end{align}
Here we focus on the first order $\alpha$ correction of $G_\pm$. Note that the charges \eqref{dSLW} and \eqref{dSJ} are
\begin{align}
\begin{aligned}
    {\cal L}=-\frac{\kappa}{8\pi\tau^2}+{\cal O}(\alpha^2)\,,\quad {\cal W}=-\frac{\kappa}{3\pi\tau^5}\alpha+{\cal O}(\alpha^3)\,,\quad J^{(s)}={\cal O}(\alpha^{s-2})\,.
\end{aligned}
\end{align}
We also expand the propagator in $\alpha$ as 
\begin{align} \label{dS2ptcorr0}
    G_\pm=G_\pm^{(0)}+\sum_{n=1}^\infty G_\pm^{(n)}\,,
\end{align}
where $G_\pm^{(0)}$ is the result without any higher-spin charges in \eqref{GBTZ}.
Through complicated computation, it leads to the ratio of the scalar propagators at $\tilde\rho\rightarrow+\infty$ as
\begin{align} \label{dS2ptcorr1}
    \frac{G^{(1)}_-|_{\tilde\rho\rightarrow+\infty}}{G^{(0)}_-|_{\tilde\rho\rightarrow+\infty}}\sim\frac{\alpha}{16\tau^2}\frac{3\sin\frac{iz}{\tau}+(2+\cos\frac{iz}{\tau})(\frac{i{\bar z}}{\bar\tau}-\frac{iz}{\tau})}{\sin^2\frac{iz}{2\tau}} \, ,
\end{align}
and 
at $\tilde\rho\rightarrow-\infty$ as:
\begin{align} \label{dS2ptcorr2}
    \frac{G^{(1)}_-|_{\tilde\rho\rightarrow-\infty}}{G^{(0)}_-|_{\tilde\rho\rightarrow-\infty}}\sim\frac{\alpha}{16\tau^2}\frac{\sin\frac{iz}{\tau}+(2-\cos\frac{iz}{\tau})(\frac{i{\bar z}}{\bar\tau}-\frac{iz}{\tau})}{\cos^2\frac{iz}{2\tau}} \, .
\end{align}

\section{Some CFT calculations}
\label{app:CFTcal}

In the main context, we have analyzed partition function on $S^2$ in the presence of two heavy operators. Due to the symmetry under the combination of Weyl transformation and shift of fields, we can compute the same quantity as a partition function without any insertions of operators but on a locally $S^2$ with two conical deficits as explained in the next subsection. Here we focus on the leading contribution to the Gibbons-Hawking entropy. For the purpose, we use the Liouville description. However, if the vertex operator is of the form $V_\alpha$ with $\alpha \propto \rho$, then the same analysis applies to the Toda case as shown in subsection \ref{app:thooft}.

\subsection{Partition function of CFT on conical defect}
\label{app:cd}

The  metric of $S^2$ is given by
\begin{align}
 ds^2 = \frac{4 dz d\bar z}{|1 + z \bar z|^2}
  = \frac{4 (dr^2 + r^2 d \phi^2)}{(1 + r^2)^2}
\end{align}
with $z = r e^{i \phi}$ and $\bar z = re^{- i \phi}$.
Performing a coordinate transformation $r = 1/\tan (\theta/2)$, we find
\begin{align}
    ds^2 = d \theta ^2 + \sin ^2 \theta d\phi^2 \, .
\end{align}
The partition function is evaluated in (4.9) of \cite{Hikida:2022ltr} as
\begin{align}
    |Z_\text{CFT}|^2  \sim e^{ \frac{\pi }{3} c^{(g)} } \, .
\end{align}
Let us introduce the conical deficits at $\theta = 0 , \pi/2$ with the deficit angle $4 \pi \eta$ $(0 \leq \eta \leq 1/2)$. Performing a coordinate transformation as $z = r^{1 - 2 \eta} e^{i (1 - 2 \eta) \phi}$ and $\bar z = r^{1 - 2 \eta} e^{- i (1 - 2 \eta) \phi}$, the metric becomes
\begin{align}
 ds^2 = \frac{4 dz d\bar z}{|1 + z \bar z|^2}
  = \frac{4 (1 - 2 \eta )^2 (dr^2 + r^2 d \phi^2)}{r^{4 \eta }(1 + r^{2 - 4 \eta})^2} \, .
\end{align}
Applying another coordinate transformation
as $r^{1 - 2 \eta} = 1/\tan (\theta/2)$, then we find
\begin{align}
    ds^2 = d \theta ^2 + (1 - 2 \eta)^2 \sin ^2 \theta d\phi^2 \, .
\end{align}
The metric indeed has conical deficits at $\theta = 0 , \pi/2$ with the deficit angle $4 \pi \eta$. The period of $\phi$ is now $2 \pi (1 - 2 \eta)$, which makes the volume to be $(1 - 2 \eta)$ times that of $S^2$. From the computation in \cite{Hikida:2022ltr}, we can see that the leading contribution to the entropy is proportional to the volume, i.e.
\begin{align}
    |Z_\text{CFT}|^2 \sim e^{ \frac{\pi }{3} c^{(g)} (1 -2 \eta) } \, .
\end{align}
Using the relation \eqref{eta2E}, we reproduce $\exp S_\text{GH}$ with \eqref{BHentropy}.

\subsection{A two-point function of 't Hooft limit CFT}
\label{app:thooft}

In the main context, we deal with $\mathfrak{sl}(N)$ Toda field theory with finite $N$, which can be used to compute correlation functions of the coset \eqref{coset} with finite $N$ as shown in \cite{Creutzig:2021ykz}. However, we also need to consider the coset \eqref{coset} at the 't Hooft limit, where $N,k \to \infty$ but $\lambda$ in \eqref{thooft} finite. In order to obtain the expressions at the 't Hooft limit, we utilize the triality relation in \cite{Gaberdiel:2012ku}.
Namely, we obtain the expressions in terms of $N,c$ and then replace $N$ by $\pm \lambda$.
The relation to parameter $b$ is
\begin{align}
b = i \sqrt{1 - \frac{\lambda}{N}} \, .
\end{align}
In particular, we have
\begin{align}
 (b + b^{-1})^2 \sim - \frac{\lambda^2}{N^2} \sim - \frac{N \lambda^2}{12 (\rho,\rho)} + \mathcal{O} (N^0) \, .
\end{align}
Here we compute the two-point function
\begin{align}
\langle V_\alpha (z_1) V_{\alpha^*} (z_2) \rangle
\end{align}
with $\alpha = (1 -\sqrt{1 - 8 G_N E} ) Q$. 
The background charge $Q$ is given in \eqref{Q}.
If $\alpha$ is proportional to the Weyl vector $\rho$, then the operator does not carry any higher-spin charges. Therefore, the operator is dual to dS$_3$ black hole without any higher-spin charges. In this case, the two-point function
\eqref{2ptheavy} becomes simplified as
\begin{align} 
\left| \langle V_\alpha (z_1) V_{\alpha^*} (z_2) \rangle \right|^2 \sim e^{\frac{\pi}{3} c^{(g)} \sqrt{1 - 8 G_N E} } \, .
\end{align}
We thus reproduced the leading order expression of $\exp (S_\text{GH})$ with \eqref{BHentropy}.

\section{Wilson lines in de Sitter gravity}
\label{app:Wilson}

In this appendix, we extend the calculation of holographic entanglement entropy using Wilson line in \cite{Ammon:2013hba,deBoer:2013vca} to our dS$_3$ higher-spin gravity. See also \cite{Narayan:2015vda,Sato:2015tta,Hikida:2021ese,Hikida:2022ltr,Doi:2022iyj,Narayan:2022afv,Doi:2023zaf} for related works.

\subsection{Entanglement entropy from open Wilson lines}
\label{sec:EE}

Let us introduce the Wilson line operator with two boundary states $|U_i\rangle, |U_f\rangle$ \cite{Ammon:2013hba,deBoer:2013vca}
\begin{align}
\label{Wline}
    W(C_{ij})=\langle U_f|{\cal P} \exp\left(\int_C A \right){\cal P}\exp\left(\int_C \bar{A} \right)|U_i\rangle\,,
\end{align}
where $\cal{P}$ is the path ordering. Denoting a world line field on $s\in [s_i,s_f]$ by $U(s)$, the expectation value of Wilson line (\ref{Wline}) is given by
\begin{align}
    W(C_{ij})=\int {\cal D}U {\cal D}P \exp(-I(U,P))\,,
\end{align}
where $P$ is a canonical momentum conjugate to $U$. The action $I(U,P)$ describes an auxiliary system, which lives on the Wilson line. It is explicitly given by
\begin{align}
\label{IUP}
    I(U,P)=\int_{s_i}^{s_f} ds \left( \text{tr}\left[PU^{-1}D_s U\right]+\Sigma(s)\left(\text{tr}\left[P^2\right]-c_2\right)\right)
\end{align}
in the case of a SL$(2)$-valued $U$. Here $c_2$ is the quadratic Casimir and the covariant derivative $D_s$ is defined by
\begin{align}
   D_sU=\frac{dU}{ds}+A_sU-U\bar{A}_s\,,\quad A_s\equiv A_\mu\frac{dx^\mu}{ds}\,.
\end{align}
Note that the action (\ref{IUP}) is invariant under a local gauge transformation
\begin{align}
    U(s)\rightarrow L(s)U(s)R(s)\,,\quad P(s)\rightarrow R^{-1}(s)P(s)R(s)\,.
\end{align}
The equations of motion from $I(U,P)$ are given by
\begin{align}
\label{UEOM}
   U^{-1}D_sU+2\Sigma P=0\,,\quad \frac{dP}{ds}+[\bar{A}_s, P]=0\,,\quad \text{tr}\left[P^2\right]=c_2\,.
\end{align}
A trivial solution for $A=\bar{A}=0$ is
\begin{align}
    P(s)=P_0\,,\quad U(s)=U_0(s)\equiv u_0\exp\left( -2v (s) P_0\right) \, ,\quad \frac{dv }{ds}=\Sigma \, ,
\end{align}
where $P_0$ and $u_0$ are constant elements. Moreover, (\ref{UEOM}) leads to the on-shell action
\begin{align}
    I(U,P)_\text{on-shell}=-2c_2\int_{s_i}^{s_f}ds \Sigma(s)=-2c_2\Delta v \, ,
\end{align}
where $\Delta v=v(s_f)-v(s_i)$. 
The authors of \cite{Ammon:2013hba} showed that this on-shell action with boundary conditions $U(s_i)=U(s_f)=1$ leads to the holographic entanglement entropy with \eqref{solution} and setting $\sqrt{2c_2}=c/6$. In the following, we consider our case of dS$_3$.

In the pure dS$_3$, we can write the solution \eqref{gaugedS} in the following pure gauge form:
\begin{align}
    A=LdL^{-1}\,,\quad \bar{A}=R^{-1}dR\,,
\end{align}
where $L$ and $R$ given by
\begin{align}
    L=e^{-i\theta L_0} e^{-iL_1x^+}\,,\quad R=e^{-iL_{-1}x^-}e^{-i\theta L_0}\,.
\end{align}
The actual solution can be found by acting gauge transformation to a trivial solution as
\begin{align}
\begin{aligned}
    &U(s_i)=L(s_i)\left(u_0e^{-2 v(s_i) P_0}\right)R(s_i)\,,\\
    &U(s_f)=L(s_f)\left(u_0e^{-2v(s_f) P_0}\right)R(s_f)\,.
\end{aligned}
\end{align}
Eliminating $u_0$ with boundary conditions $U(s_i)=U(s_f)=1$, we obtain
\begin{align}
\label{Ufi}
    e^{-2\Delta v P_0}=R(s_i)L(s_i)\left(R(s_f)L(s_f)\right)^{-1}\,.
\end{align}
To evaluate $\Delta v$, we take the trace of (\ref{Ufi}) for the fundamental representation of SL(2). Then we obtain
\begin{align}
    I(U,P)_\text{on-shell}=\sqrt{2c_2}\cosh^{-1}\left( 1-\frac{(\Delta t)^2+(\Delta \phi)^2}{2}e^{2i\theta_0} \right),
\end{align}
where we set $\theta(s_f)=\theta(s_i)=\theta_0, \Delta t=t(s_f)-t(s_i), \Delta\phi=\phi(s_f)-\phi(s_i)$ with $x^\pm=it\pm\phi$. For $e^{i\theta_0}\Delta\phi=\epsilon^{-1}\Delta\phi\gg1$ and $\Delta t=0$, we obtain
\begin{align}
    S_{EE}=I(U,P)_\text{on-shell}=i\frac{c^{(g)}}{3}\log\left( \frac{\Delta\phi}{\epsilon} \right)+\frac{\pi c^{(g)}}{6}\,,
\end{align}
where we set $\sqrt{2c_2}=c/6=ic^{(g)}/6$. 
This reproduces the results in \cite{Hikida:2022ltr}.
As argued in \cite{Doi:2022iyj,Doi:2023zaf}, this quantity should be interpreted as an generalization of entanglement entropy, called pseudo entropy \cite{Nakata:2020luh}, which takes complex-valued. 
In the dS$_3$ black holes, let us consider the solution \eqref{soldS} and define the following $L$ and $R$
\begin{align}
    L=e^{-i\theta L_0} e^{-i\left(L_1-\frac{2\pi {\cal L}}{\kappa}L_{-1}\right)x^+},\quad R=e^{-i\left(L_{-1}- \frac{2\pi {\bar{\cal L}}}{\kappa}L_1 \right)x^-}e^{-i\theta L_0}\,.
\end{align}
In the same steps as above, we obtain 
\begin{align}
\label{EEdSBH}
    S_{EE}=i\frac{c^{(g)}}{6}\log\left( \frac{\kappa}{2\pi\sqrt{{\cal L}{\bar{\cal L}}}}\frac{1}{\epsilon^2}\sin\left( \sqrt{\frac{2\pi {\cal L}}{\kappa}}\Delta\phi \right)\sin\left( \sqrt{\frac{2\pi {\bar{\cal L}}}{\kappa}}\Delta\phi \right) \right)+\frac{\pi c^{(g)}}{6}
\end{align}
for $\epsilon^{-1}\Delta\phi\gg1$.
If we set ${\cal L}={\bar{\cal L}}$ and define $\beta=\pi\sqrt{\frac{\kappa}{2\pi{\cal L}}}$, the result (\ref{EEdSBH}) becomes
\begin{align}
    S_{EE}=i\frac{c^{(g)}}{3}\log\left( \frac{\beta}{\pi\epsilon}\sin\left( \frac{\pi\Delta\phi}{\beta} \right) \right)+\frac{\pi c^{(g)}}{6}\,.
\end{align}

Let us move to the higher-spin dS$_3$ black holes. According to \cite{Ammon:2013hba}, the evaluation of Wilson line for SL(3)-valued $U$ is given by the action
\begin{align}
\label{IUP3}
    I(U,P)=\int_{s_i}^{s_f} ds \left(\text{tr}\left[PU^{-1}D_s U\right]+\Sigma_2(s)\left(\text{tr}\left[P^2\right]-c_2\right)+\Sigma_3(s)\left(\text{tr}\left[P^3\right]-c_3\right)\right)\, ,
\end{align}
where we define
\begin{align}
\begin{aligned}
    &\text{tr}\left[P^2\right]\equiv P^aP^b\delta_{ab}\,,\quad \delta_{ab}=\frac{1}{2}\text{tr}\left[T_aT_b\right]\,,\\
    &\text{tr}\left[P^3\right]\equiv P^aP^bP^c h_{abc}\,,\quad h_{abc}=\text{tr}\left[T_{(a}T_bT_{c)}\right] \,,
\end{aligned}
\end{align}
with  $T_a$ as the generators of $\mathfrak{sl}(3)$. The equations of motion are given by 
\begin{align}
\label{UEOM3}
   U^{-1}D_sU+2\Sigma_2 P+3\Sigma_3P\times P=0\,,\quad \frac{dP}{ds}+[\bar{A}_s, P]=0\,,\quad \text{tr}\left[P^2\right]=c_2\,,\quad \text{tr}\left[P^3\right]=c_3
\end{align}
with the definition $P\times P=h_{abc}T^aP^bP^c$. 
These equations of motion lead to the on-shell action
\begin{align}
    I(U,P)_{\text{on-shell}}=-2c_2\Delta v_2-3\Delta v_3\,.
\end{align}
We again start from a trivial solution
\begin{align}
    U_0(s)= u_0\exp\left( -2 v(s) P_0-3 v_3(s)P_0\times P_0\right)\,,\quad \frac{d v_i}{ds}=\Sigma_i
\end{align}
and use the non-rotating solution \eqref{dSconfig} as 
\begin{align}
\begin{aligned}
    L=e^{-i\theta L_0}e^{-i(a_+x^++a_-x^-)}\,,\quad R=e^{i (\bar{a}_-x^-+\bar{a}_+x^+)}e^{-i\theta L_0}\,.
\end{aligned}
\end{align}
Here $L,R$ take values in SL(3). We consider the case of $c_2\neq 0,\, c_3=0$ in order to evaluate the entanglement entropy. Using the same steps as above, see also \cite{Ammon:2013hba, Castro:2014mza}, for $\epsilon^{-1}\Delta\phi\gg1$ and $\Delta t=0$, we obtain the following first spin-3 correction at $\mu\rightarrow0$
\begin{align}
\begin{aligned}
    &S_{EE}=i\frac{c^{(g)}}{3}\log\left(\frac{\beta}{\pi\epsilon}\sin\left( \frac{\pi\Delta\phi}{\beta} \right) \right)+\frac{\pi c^{(g)}}{6}\\
   &+i\frac{c^{(g)}}{18}\frac{4\pi^2\mu^2}{\beta^2}\sin^{-4}\left(\frac{\pi\Delta\phi}{\beta}\right)\left[\sin^2\left(\frac{\pi\Delta\phi}{\beta}\right)\left(1+5\cos\left(\frac{\pi\Delta\phi}{\beta}\right)\right)\right.\\
   &\left.-\frac{4\pi\Delta\phi}{\beta}\sin\left(\frac{2\pi\Delta\phi}{\beta}\right)\left(1+2\cos\left(\frac{2\pi\Delta\phi}{\beta}\right)\right)+6\left(\frac{2\pi\Delta\phi}{\beta}\right)^2\cos\left(\frac{2\pi\Delta\phi}{\beta}\right)\right]+\cdots\,.
\end{aligned}
\end{align}
Here we use the relation in terms of a dimensionless parameter $C$ in \eqref{CdS};
\begin{align}
    {\cal W}=\frac{4(C-1)}{C^{3/2}}{\cal L}\sqrt{\frac{2\pi{\cal L}}{\kappa}}\, ,\quad \mu=\frac{3\sqrt{C}}{4(2C-3)}\sqrt{\frac{\kappa}{2\pi{\cal L}}}\, ,\quad \frac{\mu}{\beta}=\frac{3}{4\pi}\frac{(C-3)\sqrt{4C-3}}{(2C-3)^2}\,.
\label{Crelation}
\end{align}

\subsection{Thermal entropy from Wilson loops}
Let us consider the Wilson loop and evaluate the thermal entropy. The expectation value of Wilson loop is given by \cite{Ammon:2013hba,deBoer:2013vca}
\begin{align}
\label{Wloop}
    W(C)=\mathop{\text{tr}}\left[ {\cal P} \exp\left(\oint_C A \right){\cal P}\exp\left(\oint_C \bar{A} \right)\right]=\int {\cal D}U {\cal D}P \exp(-I(U,P))\,,
\end{align}
where we take the trace for the representation of the gauge group.  The action $I(U,P)$ is same as \eqref{IUP} with the path $x^\mu(s_i)=x^\mu(s_f)$ and the boundary condition for $U(s)$ is
\begin{align}
	U(s_i)=U(s_f)\,,\quad P(s_i)=P(s_f)\,.
\end{align}
With this condition, as in the section \ref{sec:EE}, we evaluate $\Delta  v$ from 
\begin{align}
    L(s_i)\left(u_0e^{-2 v(s_i)P_0}\right)R(s_i)=L(s_f)\left(u_0e^{-2 v(s_f) P_0}\right)R(s_f)\,.
\end{align}
Note that, from $x^\mu(s_i)=x^\mu(s_f)$, we have
\begin{align}
	&L^{-1}(s_f)L(s_i)=e^{\oint d\phi  a_\phi}=e^{-2\pi a_\phi}\,,\\
    &R(s_i)R^{-1}(s_f)=e^{-\oint d\phi {\bar a}_\phi}=e^{-2\pi{\bar a}_\phi}\,,
\end{align}
where we only consider the holonomy along the $\phi$-cycle. Thus we obtain the on-shell action
\begin{align}
    I(U,P)_\text{on-shell}=-2c_2\Delta v=2\pi\sqrt{2c_2}\mathop{\text{tr}_f}\big((\Lambda_\phi-\bar{\Lambda}_\phi)J_0\big)\,,
\end{align}
where $\Lambda_\phi$ and $\bar{\Lambda}_\phi$ are the diagonal matrix of the eigenvalues of $a_\phi$ and $\bar a_\phi$. Moreover, $\mathop{\text{tr}_f}$ means that we take the trace for the fundamental representation.

Let us examine explicit examples.
In the case of dS$_3$ black holes, the thermal entropy of BTZ black holes is evaluated as
\begin{align}
	S=2\pi\sqrt{2\pi \kappa {\cal L}}+2\pi\sqrt{2\pi \kappa \bar{{\cal L}}} \, .
\end{align}
Here the diagonal matrix is 
\begin{align}
	\Lambda_\phi=\text{diag}\left(i \sqrt{\frac{2\pi{\cal L}}{\kappa}},- i\sqrt{\frac{2\pi{\cal L}}{\kappa}} \right)\,,
\end{align}
which comes from the solution \eqref{soldS}.
Let us now turn to the case of the higher-spin black hole. The thermal entropy is evaluated as
\begin{align}
	S=4\pi\sqrt{2\pi \kappa{\cal L}}\frac{\sqrt{1-\frac{3}{4C}}}{1-\frac{3}{2C}} \, .
\label{TE}
\end{align}
The diagonal matrix is
\begin{align}
	\Lambda_\phi=2i\sqrt{\frac{2\pi{\cal L}}{\kappa}}\text{diag}\left( \frac{3+C(-2+\sqrt{-3+4C})}{\sqrt{C}(-3+2C)},\frac{2}{\sqrt{C}},\frac{3-C(2+\sqrt{-3+4C})}{\sqrt{C}(-3+2C)} \right)\,,
\end{align}
which comes from the solution \eqref{dSconfig}.
Here we have used the relation \eqref{Crelation}.
Note here that the result \eqref{TE} is the dS$_3$ counterpart of AdS$_3$ one computed with the bulk Hamiltonian \cite{Perez:2013xi,deBoer:2013gz}. On the other hand, the result \eqref{hsentropy} is the dS$_3$ counterpart of AdS$_3$ one by using the boundary stress tensor \cite{Gutperle:2011kf}. According to \cite{deBoer:2013gz}, in the presence of the chemical potential $\mu$, these results disagree, see also \cite{Ammon:2013hba}.


\providecommand{\href}[2]{#2}\begingroup\raggedright\endgroup

\end{document}